\documentclass[twocolumn,showpacs,prd,preprintnumbers,superscriptaddress,nofootinbib,10pt,aps]{revtex4-2}

\usepackage{amsmath}
\usepackage{graphicx}
\usepackage{natbib}
\usepackage{siunitx}
\usepackage{hyperref}
\usepackage{isotope}
\usepackage{enumitem}
\usepackage[mathlines]{lineno}
\usepackage{booktabs}
\usepackage{multirow}
\usepackage{tabularx}
\usepackage{makecell} % thead
\usepackage[dvipsnames]{xcolor}
\usepackage{listofitems}
\usepackage{fp}
\usepackage{longtable}
\usepackage{orcidlink}
\usepackage{footnote}

\begin{document}

%\preprint{APS/123-QED}

\title{\texorpdfstring{Measurement of neutron production in atmospheric neutrino interactions\\at Super-Kamiokande}{Measurement of neutron production in atmospheric neutrino interactions at Super-Kamiokande}}

\newcommand{\AFFicrr}{\affiliation{Kamioka Observatory, Institute for Cosmic Ray Research, University of Tokyo, Kamioka, Gifu 506-1205, Japan}}
\newcommand{\AFFkashiwa}{\affiliation{Research Center for Cosmic Neutrinos, Institute for Cosmic Ray Research, University of Tokyo, Kashiwa, Chiba 277-8582, Japan}}
\newcommand{\AFFipmu}{\affiliation{Kavli Institute for the Physics and
Mathematics of the Universe (WPI), The University of Tokyo Institutes for Advanced Study,
University of Tokyo, Kashiwa, Chiba 277-8583, Japan }}
\newcommand{\AFFmad}{\affiliation{Department of Theoretical Physics, University Autonoma Madrid, 28049 Madrid, Spain}}
\newcommand{\AFFubc}{\affiliation{Department of Physics and Astronomy, University of British Columbia, Vancouver, BC, V6T1Z4, Canada}}
\newcommand{\AFFbu}{\affiliation{Department of Physics, Boston University, Boston, MA 02215, USA}}
\newcommand{\AFFuci}{\affiliation{Department of Physics and Astronomy, University of California, Irvine, Irvine, CA 92697-4575, USA }}
\newcommand{\AFFcsu}{\affiliation{Department of Physics, California State University, Dominguez Hills, Carson, CA 90747, USA}}
\newcommand{\AFFcnm}{\affiliation{Institute for Universe and Elementary Particles, Chonnam National University, Gwangju 61186, Korea}}
\newcommand{\AFFduke}{\affiliation{Department of Physics, Duke University, Durham NC 27708, USA}}
\newcommand{\AFFgifu}{\affiliation{Department of Physics, Gifu University, Gifu, Gifu 501-1193, Japan}}
\newcommand{\AFFgist}{\affiliation{GIST College, Gwangju Institute of Science and Technology, Gwangju 500-712, Korea}}
\newcommand{\AFFuh}{\affiliation{Department of Physics and Astronomy, University of Hawaii, Honolulu, HI 96822, USA}}
\newcommand{\AFFicl}{\affiliation{Department of Physics, Imperial College London , London, SW7 2AZ, United Kingdom }}
\newcommand{\AFFkek}{\affiliation{High Energy Accelerator Research Organization (KEK), Tsukuba, Ibaraki 305-0801, Japan }}
\newcommand{\AFFkobe}{\affiliation{Department of Physics, Kobe University, Kobe, Hyogo 657-8501, Japan}}
\newcommand{\AFFkyoto}{\affiliation{Department of Physics, Kyoto University, Kyoto, Kyoto 606-8502, Japan}}
\newcommand{\AFFliv}{\affiliation{Department of Physics, University of Liverpool, Liverpool, L69 7ZE, United Kingdom}}
\newcommand{\AFFmiyagi}{\affiliation{Department of Physics, Miyagi University of Education, Sendai, Miyagi 980-0845, Japan}}
\newcommand{\AFFnagoya}{\affiliation{Institute for Space-Earth Environmental Research, Nagoya University, Nagoya, Aichi 464-8602, Japan}}
\newcommand{\AFFkmi}{\affiliation{Kobayashi-Maskawa Institute for the Origin of Particles and the Universe, Nagoya University, Nagoya, Aichi 464-8602, Japan}}
\newcommand{\AFFpol}{\affiliation{National Centre For Nuclear Research, 02-093 Warsaw, Poland}}
\newcommand{\AFFsuny}{\affiliation{Department of Physics and Astronomy, State University of New York at Stony Brook, NY 11794-3800, USA}}
\newcommand{\AFFokayama}{\affiliation{Department of Physics, Okayama University, Okayama, Okayama 700-8530, Japan }}
\newcommand{\AFFosaka}{\affiliation{Department of Physics, Osaka University, Toyonaka, Osaka 560-0043, Japan}}
\newcommand{\AFFox}{\affiliation{Department of Physics, Oxford University, Oxford, OX1 3PU, United Kingdom}}
\newcommand{\AFFqmul}{\affiliation{School of Physics and Astronomy, Queen Mary University of London, London, E1 4NS, United Kingdom}}
\newcommand{\AFFregina}{\affiliation{Department of Physics, University of Regina, 3737 Wascana Parkway, Regina, SK, S4SOA2, Canada}}
\newcommand{\AFFseoul}{\affiliation{Department of Physics and Astronomy, Seoul National University, Seoul 151-742, Korea}}
\newcommand{\AFFsheff}{\affiliation{Department of Physics and Astronomy, University of Sheffield, S3 7RH, Sheffield, United Kingdom}}
\newcommand{\AFFshizuokasc}{\affiliation{Department of Informatics in
Social Welfare, Shizuoka University of Welfare, Yaizu, Shizuoka, 425-8611, Japan}}
\newcommand{\AFFstfc}{\affiliation{STFC, Rutherford Appleton Laboratory, Harwell Oxford, and Daresbury Laboratory, Warrington, OX11 0QX, United Kingdom}}
\newcommand{\AFFskk}{\affiliation{Department of Physics, Sungkyunkwan University, Suwon 440-746, Korea}}
\newcommand{\AFFtodai}{\affiliation{Department of Physics, University of Tokyo, Bunkyo, Tokyo 113-0033, Japan }}
\newcommand{\AFFtit}{\affiliation{Department of Physics, Institute of Science Tokyo, Meguro, Tokyo 152-8551, Japan }}
\newcommand{\AFFtus}{\affiliation{Department of Physics, Faculty of Science and Technology, Tokyo University of Science, Noda, Chiba 278-8510, Japan }}
\newcommand{\AFFtriumf}{\affiliation{TRIUMF, 4004 Wesbrook Mall, Vancouver, BC, V6T2A3, Canada }}
\newcommand{\AFFtokai}{\affiliation{Department of Physics, Tokai University, Hiratsuka, Kanagawa 259-1292, Japan}}
\newcommand{\AFFtsinghua}{\affiliation{Department of Engineering Physics, Tsinghua University, Beijing, 100084, China}}
\newcommand{\AFFynu}{\affiliation{Department of Physics, Yokohama National University, Yokohama, Kanagawa, 240-8501, Japan}}
\newcommand{\AFFllr}{\affiliation{Ecole Polytechnique, IN2P3-CNRS, Laboratoire Leprince-Ringuet, F-91120 Palaiseau, France }}
\newcommand{\AFFbari}{\affiliation{ Dipartimento Interuniversitario di Fisica, INFN Sezione di Bari and Universit\`a e Politecnico di Bari, I-70125, Bari, Italy}}
\newcommand{\AFFnapoli}{\affiliation{Dipartimento di Fisica, INFN Sezione di Napoli and Universit\`a di Napoli, I-80126, Napoli, Italy}}
\newcommand{\AFFroma}{\affiliation{INFN Sezione di Roma and Universit\`a di Roma ``La Sapienza'', I-00185, Roma, Italy}}
\newcommand{\AFFpadova}{\affiliation{Dipartimento di Fisica, INFN Sezione di Padova and Universit\`a di Padova, I-35131, Padova, Italy}}
\newcommand{\AFFkeio}{\affiliation{Department of Physics, Keio University, Yokohama, Kanagawa, 223-8522, Japan}}
\newcommand{\AFFwinnipeg}{\affiliation{Department of Physics, University of Winnipeg, MB R3J 3L8, Canada }}
\newcommand{\AFFkcl}{\affiliation{Department of Physics, King's College London, London, WC2R 2LS, UK }}
\newcommand{\AFFwarwick}{\affiliation{Department of Physics, University of Warwick, Coventry, CV4 7AL, UK }}
\newcommand{\AFFral}{\affiliation{Rutherford Appleton Laboratory, Harwell, Oxford, OX11 0QX, UK }}
\newcommand{\AFFwu}{\affiliation{Faculty of Physics, University of Warsaw, Warsaw, 02-093, Poland }}
\newcommand{\AFFbcit}{\affiliation{Department of Physics, British Columbia Institute of Technology, Burnaby, BC, V5G 3H2, Canada }}
\newcommand{\AFFtohoku}{\affiliation{Department of Physics, Faculty of Science, Tohoku University, Sendai, Miyagi, 980-8578, Japan }}
\newcommand{\AFFicise}{\affiliation{Institute For Interdisciplinary Research in Science and Education, ICISE, Quy Nhon, 55121, Vietnam }}
\newcommand{\AFFilance}{\affiliation{ILANCE, CNRS - University of Tokyo International Research Laboratory, Kashiwa, Chiba 277-8582, Japan}}
\newcommand{\AFFibs}{\affiliation{Center for Underground Physics, Institute for Basic Science (IBS), Daejeon, 34126, Korea}}
\newcommand{\AFFglasgow}{\affiliation{School of Physics and Astronomy, University of Glasgow, Glasgow, Scotland, G12 8QQ, United Kingdom}}
\newcommand{\AFFoecu}{\affiliation{Media Communication Center, Osaka Electro-Communication University, Neyagawa, Osaka, 572-8530, Japan}}
\newcommand{\AFFminn}{\affiliation{School of Physics and Astronomy, University of Minnesota, Minneapolis, MN  55455, USA}}
\newcommand{\AFFsilesia}{\affiliation{August Che\l{}kowski Institute of Physics, University of Silesia in Katowice, 75 Pu\l{}ku Piechoty 1, 41-500 Chorz\'{o}w, Poland}}
\newcommand{\AFFtoyama}{\affiliation{Faculty of Science, University of Toyama, Toyama City, Toyama 930-8555, Japan}}
\newcommand{\AFFbmcc}{\affiliation{Science Department, Borough of Manhattan Community College / City University of New York, New York, New York, 1007, USA.}}

\AFFicrr
\AFFkashiwa
\AFFmad
\AFFbmcc
\AFFbu
\AFFbcit
\AFFuci
\AFFcsu
\AFFcnm
\AFFduke
\AFFllr
\AFFgifu
\AFFgist
\AFFglasgow
\AFFuh
\AFFibs
\AFFicise
\AFFicl
\AFFbari
\AFFnapoli
\AFFpadova
\AFFroma
\AFFilance
\AFFkeio
\AFFkek
\AFFkcl
\AFFkobe
\AFFkyoto
\AFFliv
\AFFminn
\AFFmiyagi
\AFFnagoya
\AFFkmi
\AFFpol
\AFFsuny
\AFFokayama
\AFFoecu
\AFFox
\AFFral
\AFFseoul
\AFFsheff
\AFFshizuokasc
\AFFsilesia
\AFFstfc
\AFFskk
\AFFtohoku
\AFFtokai
%\AFFtokyo
\AFFtodai
\AFFipmu
\AFFtit
\AFFtus
\AFFtoyama
\AFFtriumf
\AFFtsinghua
\AFFwu
\AFFwarwick
\AFFwinnipeg
\AFFynu

\author{S.~Han} 
%\altaffiliation{Present Address: Department of Physics, Kyoto University, Kyoto, Kyoto 606-8502, Japan}
%\AFFkashiwa
\AFFkyoto

%%%%%%%%%%%%%%%%%%%%%%%%%%%%%%%%%%%%%%%%%%%%%%%%%%%%%%%%%%%%%%%%%%%%
%ICRR
\author{K.~Abe}
\AFFicrr
\AFFipmu
\author{S.~Abe}
\author{Y.~Asaoka}
\AFFicrr
\AFFipmu
\author{C.~Bronner}
\author{M.~Harada}
\AFFicrr
\author{Y.~Hayato}
\AFFicrr
\AFFipmu
\author{K.~Hiraide}
\AFFicrr
\AFFipmu
\author{K.~Hosokawa}
\AFFicrr
\author{K.~Ieki}
\author{M.~Ikeda}
\AFFicrr
\AFFipmu
% ...

%%%%%%%%%%%%%%%%%%%%%%%%%%%%%%%%%%%%%%%%%%%%%%%%%%%%%%%%%%%%%%%%%%%%%

\collaboration{The Super-Kamiokande Collaboration}
\noaffiliation

\date{\today}

\begin{abstract}
%We present measurements of total neutron production from atmospheric neutrino interactions in water, analyzed as a function of the electron-equivalent visible energy over a range of 30 MeV to 10 GeV. These results are based on 4,270 days of data collected by Super-Kamiokande, including 564 days with 0.011 wt\% gadolinium added to enhance neutron detection. The measurements are compared to predictions from neutrino event generators combined with various hadron-nucleus interaction models, which consist of an intranuclear cascade model and a nuclear de-excitation model. We observe significant variations in the predictions depending on the choice of hadron-nucleus interaction model. We discuss key factors that contribute to describing our data, such as in-medium effects in the intranuclear cascade and the accuracy of statistical evaporation modeling.

We present measurements of total neutron production from atmospheric neutrino interactions in water, analyzed as a function of electron-equivalent visible energy over a range of 30 MeV to 10 GeV. These results are based on 4,270 days of data collected by Super-Kamiokande, including 564 days with 0.011 wt\% gadolinium added to enhance neutron detection. Neutron signal selection is based on a neural network trained on simulation, with its performance validated using an Am/Be neutron point source. The measurements are compared to predictions from neutrino event generators combined with various hadron-nucleus interaction models, which include an intranuclear cascade model and a nuclear de-excitation model. We observe significant variations in the predictions depending on the choice of hadron-nucleus interaction model. We discuss key factors that contribute to describing our data, such as in-medium effects in the intranuclear cascade and the accuracy of statistical evaporation modeling.

%, which simulate intranuclear hadron transport and subsequent nuclear de-excitation.

%\textcolor{blue}{Among the tested models, our data favor those predicting fewer secondary neutrons for nucleon projectiles and more for pion projectiles.}

%The Liège model (INCL++) coupled with Geant4 Precompound model showed better agreement with our observations across the entire energy range, compared to the widely used Geant4 Bertini cascade model and its variants. 

\end{abstract}

\maketitle

\section{Introduction}
\label{sec:intro}

A large fraction of neutrino experiments rely on nuclear targets, yet significant uncertainties remain regarding the influence of nucleon correlations on interaction cross sections and particle kinematics. Modeling these ``nuclear effects'' is particularly critical for GeV-scale neutrino experiments aiming to measure neutrino oscillation parameters, including CP violation and mass ordering. 

Outgoing hadrons serve as valuable probes of these effects, with recent advancements in neutrino detectors enabling precise measurements of hadron multiplicities and kinematics. For instance, proton measurements in neutrino-argon interactions have revealed discrepancies between observed and predicted kinematic distributions \cite{uboone}. However, detecting neutrons in tracking detectors is challenging due to the limited detection efficiency from the small $(n, p)$ reaction cross section \cite{mv_neutron_2019, uboone_neutron}. This issue is crucial, as inaccuracies in estimating the ``missing energy'' carried by neutrons can bias key measurements, such as the Dirac CP phase \cite{missing_energy_bias_to_cp}.

Neutrons with kinetic energies of a few MeV or lower tend to thermalize and can be detected via the radiative neutron capture $(n, \gamma)$ reactions with well-defined timescales and energy signatures, enabling clean signal selection with virtually no energy threshold. Historically, this made neutrons effective tags for antineutrino charged-current (CC) interactions (e.g., $\bar{\nu}_e p \rightarrow e^+ n$) compared to neutrino interactions (e.g., $\nu_e n \rightarrow e^- p$) \cite{reines_cowan}. Neutron tagging remains relevant today, for instance, in atmospheric neutrino oscillation analyses, where it enhances sensitivity to both neutrino mass ordering and CP violation, by preventing the cancellation of opposite-sign effects in neutrino and antineutrino oscillation probabilities \cite{sk_osc_2024}. It also helps suppress atmospheric neutrino backgrounds in searches for rare events, such as proton decay (e.g., $p \rightarrow e^+ \pi^0$ \cite{sk_pdk_2020}), which in many cases is not expected to produce neutrons, or inverse beta decay ($\bar{\nu}_e p \rightarrow e^+ n$) induced by supernova $\bar{\nu}_e$, which emits only one neutron. 

\begin{figure}
\includegraphics[width=0.99\columnwidth]{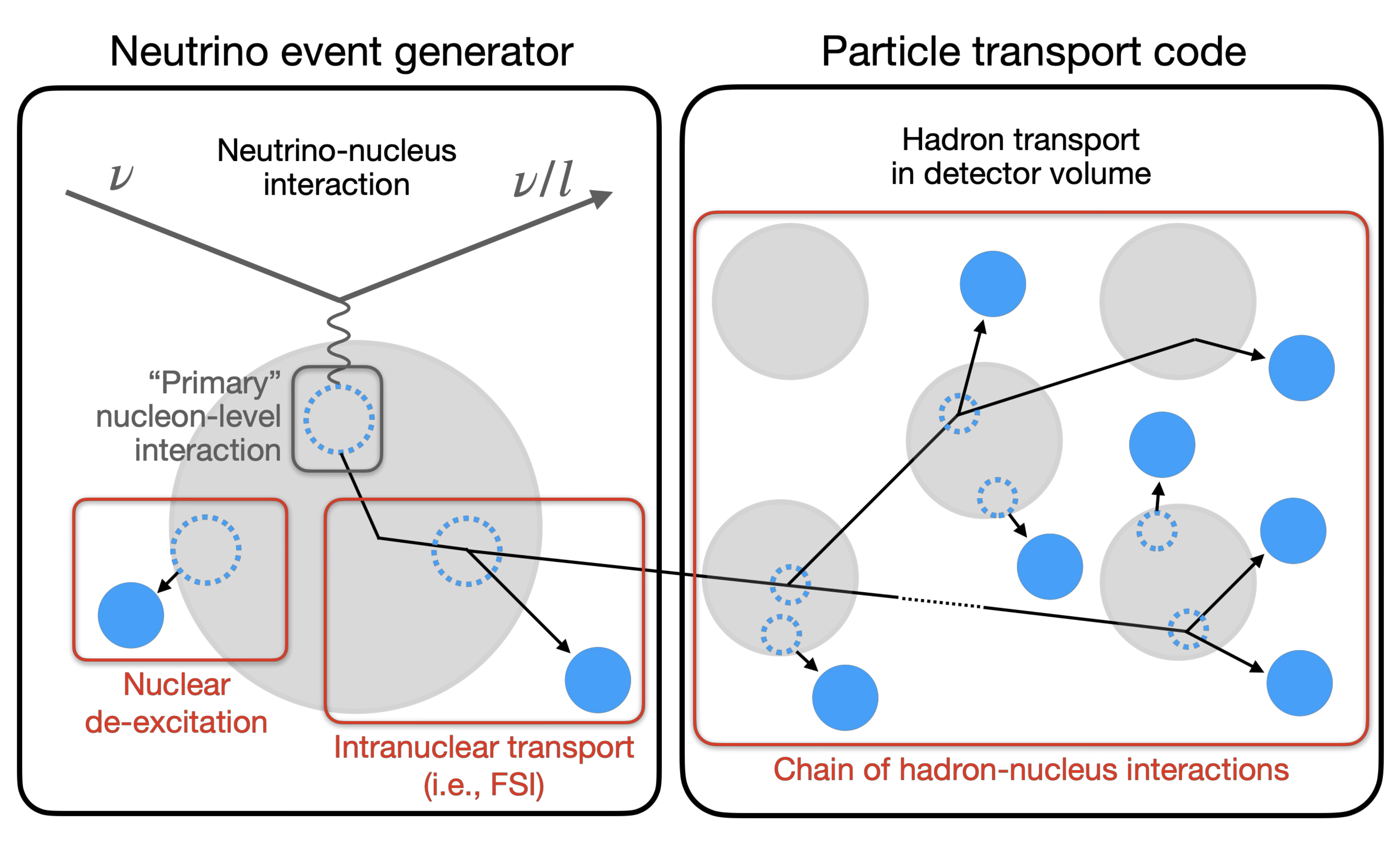}
\caption{Schematic illustration of nucleon production mechanisms in a typical GeV-scale neutrino interaction simulation. Black arrows indicate nucleon trajectories; solid blue circles represent detectable nucleons ejected from nuclei, while dashed blue circles denote resulting nucleon holes.}
\label{fig:ngpern}
\end{figure}

Accurate prediction of detectable neutrons is essential and requires well-constrained uncertainties. The modeling approach commonly adopted by GeV-scale neutrino experiments is illustrated in Figure \ref{fig:ngpern}. Neutrino event generators sample outgoing hadrons from the initial neutrino interaction, either at the nucleon or quark level, and subsequently model intranuclear hadron transport (often referred to as final-state interactions, or FSI), followed by nuclear de-excitation. Similarly, particle transport codes such as Geant4 \cite{geant4_1, geant4_2, geant4_3} simulate hadron transport and nuclear de-excitation to describe downstream hadron interactions within the detector.

Accurately modeling secondary neutron production is particularly important. For hadron transport in the $O(0.1\text{--}1)$ GeV energy range, Intranuclear Cascade (INC) models \cite{inc_review} are commonly used. Subsequent nuclear de-excitation involves an evaporation process that releases neutrons with kinetic energies of a few MeV and also contributes significantly to total neutron production. Variations in how these models account for nuclear effects often lead to significant discrepancies in predictions \cite{iaea_benchmark}. Several studies have measured neutron production from atmospheric or artificial neutrino interactions using water (T2K \cite{t2k_neutron_2019}), heavy water (SNO \cite{sno_neutron_2019}), and hydrocarbon (MINERvA \cite{mv_neutron_2019,mv_neutron_2023}, KamLAND \cite{kl_ncqe}) as target materials. Several of these studies \cite{t2k_neutron_2019, mv_neutron_2019, mv_neutron_2023} reported deficits in observed neutron signals compared to predictions from neutrino event generators, with the discrepancies often attributed to the inaccuracy of hadron transport models.

%These models approximate the process as a series of binary collisions between a hadron projectile and a quasi-free nucleon within the target nucleus. 

In this paper, we present a measurement of neutron production following atmospheric neutrino interactions in water. Using data collected by the Super-Kamiokande (SK) detector from 2008 to 2022, we evaluated the average multiplicity of $(n, \gamma)$ reactions (``neutron captures'') as a function of electron-equivalent visible energy, a calorimetric proxy for neutrino momentum transfer. The results were compared with predictions from various models relevant to secondary neutron production. This study focuses on an event sample with visible energy greater than 30 MeV, distinct from our previous neutron measurement \cite{sakai} targeting neutral-current quasi-elastic (NCQE) events with visible energy below 30 MeV.

This paper is structured as follows. Section II provides a brief overview of the SK detector. The selection process for atmospheric neutrino events and neutron signals, along with the estimation of selection performance, are detailed in Sections III and IV. Section V outlines the methodology for determining the average $(n,\gamma)$ multiplicity per visible energy bin and the associated systematic uncertainties. Section VI introduces the interaction models used for generating predictions. Finally, Sections VII and VIII present a comparison of observations with predictions and discuss the implications of the results.

\begin{table*}[ht!]
\caption{\label{tab:skphase} SK operational phases and neutron-related characteristics. SK-IV, V, VI data were used in this analysis.}
\begin{ruledtabular}
\begin{tabular}{lcccccc}
 &&Livetime&Gd concentration\footnote{Based on the amount of dissolved Gd.}&\multicolumn{2}{c}{Expected $(n,\gamma)$ fraction\footnote{Based on the evaluated thermal $(n,\gamma)$ reaction cross sections and uncertainties of ENDF/B-VII.1 \cite{endf7}.}}&$(n,\gamma)$ time constant\footnote{Weighted mean of all Am/Be neutron source measurements, explained in Section \ref{sec:ambe}.}\\
 Phase&Years&[days]&[wt\%]&$\text{H}(n,\gamma)$ [\%]
&Gd$(n,\gamma)$ [\%]&[µs]\\ \hline 
 SK-I-III&1996--2008&2805.9&-&$>$99.9&-&No data \\
 \hline
 SK-IV&2008--2018&3244.4&-&$>$99.9&-&204.8 $\pm$ \phantom{1}9.8 \phantom{[12]}\\
 SK-V&2019--2020 &461.0&-&$>$99.9&-&199.8 $\pm$ 10.2 \phantom{[12]}\\
 SK-VI&2020--2022 &564.4\footnote{Excludes earlier runs which showed signs of non-uniform Gd concentration, i.e., varying time constant by position.}&0.0110 $\pm$ 0.0001 \cite{sk_1st_gdloading}&$56.1\pm1.5$&$43.9\mp1.5$&116.2 $\pm$ \phantom{1}2.3 \phantom{[12]}\\
 \hline

 SK-VII-VIII &2022-present&-&0.0332 $\pm$ 0.0002 \cite{sk_2nd_gdloading}&$29.7\pm0.7$&$70.3\mp0.7$& \phantom{1}61.8 $\pm$ \phantom{1}0.1 \cite{sk_2nd_gdloading}\\

\end{tabular}
\end{ruledtabular}
\end{table*}

%This issue is critical with GeV-scale neutrino interactions, where neutrons often carry substantial energy.

%Neutrons can also be detected via the radiative neutron capture $(n, \gamma)$ reaction, which has no practical energy threshold. 

%Outgoing hadrons serve as valuable probes of these effects, and advancements in neutrino detectors have enabled precise measurements of hadron multiplicities and kinematics. For instance, proton measurements in neutrino-argon interactions have revealed discrepancies between observed and predicted kinematic distributions \cite{uboone}. Detecting neutrons in tracking detectors, however, is challenging since the detection efficiency is limited by the small $(n, p)$ reaction cross section \cite{mv_neutron_2019, uboone_neutron}. This issue is critical as misestimating the neutrino energy carried by neutrons, often called ``missing energy,'' can bias key measurements such as the Dirac CP phase \cite{missing_energy_bias_to_cp}.
\section{The Super-Kamiokande Detector}
\label{sec:detector}

Super-Kamiokande (SK) \cite{sk_detector} is an underground water Cherenkov detector located in Gifu, Japan. It consists of two optically separated, concentric cylindrical volumes: the inner detector (ID) containing 32.5 ktons of water and equipped with 11,129 inward-facing photomultiplier tubes (PMTs) and the outer detector (OD) serving as a cosmic-ray veto. The detector registers a PMT signal with a pulse height greater than 0.25 photoelectron-equivalent charge as a ``hit.'' If the number of ID or OD PMT hits within a 200-ns sliding time window ($N_\text{200-ns}$) exceeds a given threshold, an event trigger is issued. The details of the detector can be found in \cite{sk_detector, sk_calib}.

Charged particles, namely electrons and muons produced by charged-current neutrino interactions, are identified through Cherenkov radiation. The radiation is projected onto the PMTs as a characteristic ring pattern that depends on the particle type and energy. This ring pattern serves as the basis for particle reconstruction. Neutrons are indirectly identified via Compton-scattered electrons resulting from $(n,\gamma)$ reactions. In pure water, most occur on $\isotope[1]{H}$, emitting a single 2.2 MeV $\gamma$-ray. With the recent addition of gadolinium (Gd), a large fraction of neutrons are expected to be captured by Gd isotopes, resulting in a total $\gamma$-radiated energy of around 8 MeV. 

\begin{figure}
\includegraphics[width=0.99\columnwidth]{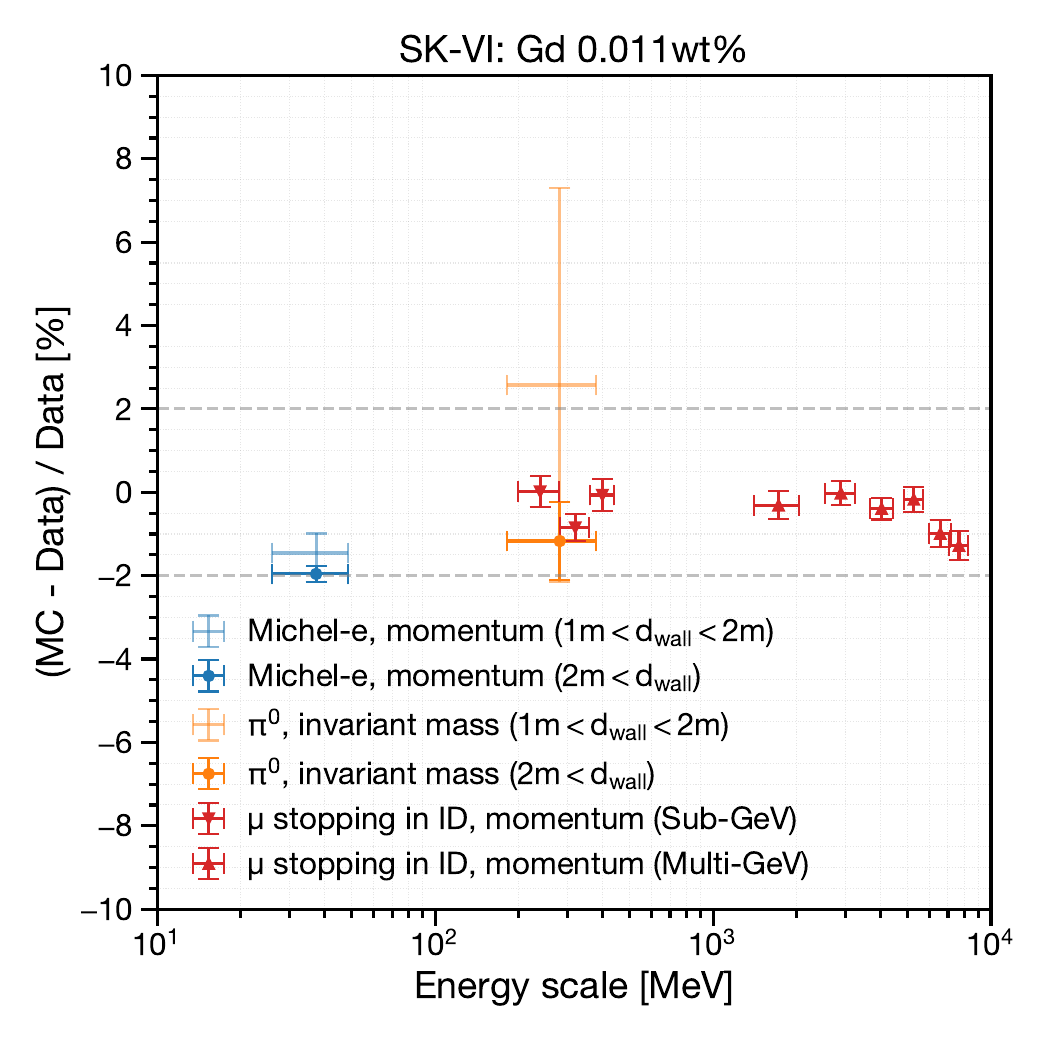}
\caption{Comparison of energy reconstruction performance between Monte Carlo (MC) simulation and observed data for naturally occurring particles in the Gd-loaded SK-VI phase. The variable $d_\text{wall}$
  represents the distance (in meters) from the reconstructed vertex to the nearest ID photodetector wall. The fiducial volume for this study is defined as $d_\text{wall}>1$ m.}
\label{fig:escale}
\end{figure}

The $O(1)$ MeV signal identification performance is significantly influenced by variations in detector characteristics. Parameters such as individual PMT gain, timing properties, quantum efficiency, and optical absorption and scattering in water are continuously monitored using cosmic-ray muons and light sources \cite{sk_calib}. Additionally, the uncertainty in Cherenkov ring energy reconstruction (described in Section \ref{sec:evreco}) is evaluated over a wide energy range using naturally occurring particles, including cosmic-ray muons, Michel electrons, and neutral pions produced in neutral-current (NC) atmospheric neutrino interactions in water. Figure \ref{fig:escale} illustrates the agreement between data and simulation in energy reconstruction for the Gd-loaded SK-VI phase, which is mostly within 2\% across the $O(10\text{--}10^4)$ MeV range and consistent with the pure water phase results reported in the recent atmospheric neutrino oscillation analysis \cite{sk_osc_2024}.

The detector has operated through eight different phases. Neutron detection began with the fourth phase SK-IV, following the electronics upgrade \cite{sk_qbee} that allowed extended event recording up to 535 µs after certain ID triggers. This has enabled analysis of delayed neutron captures that occur with a time scale of $O(10\text{--}100)$ µs following an atmospheric neutrino interaction. Between SK-IV and SK-V, in 2018, the detector underwent refurbishment, during which malfunctioning PMTs were replaced. The later phases, SK-VI, SK-VII, and SK-VIII involved the dissolution of Gd$_2$(SO$_4$)$_3$ into the water volume to enhance neutron detection efficiency \cite{sk_1st_gdloading,sk_2nd_gdloading}. Table \ref{tab:skphase} summarizes the relevant operational conditions. 
\begin{figure*}[ht!]
\includegraphics[width=0.99\linewidth]{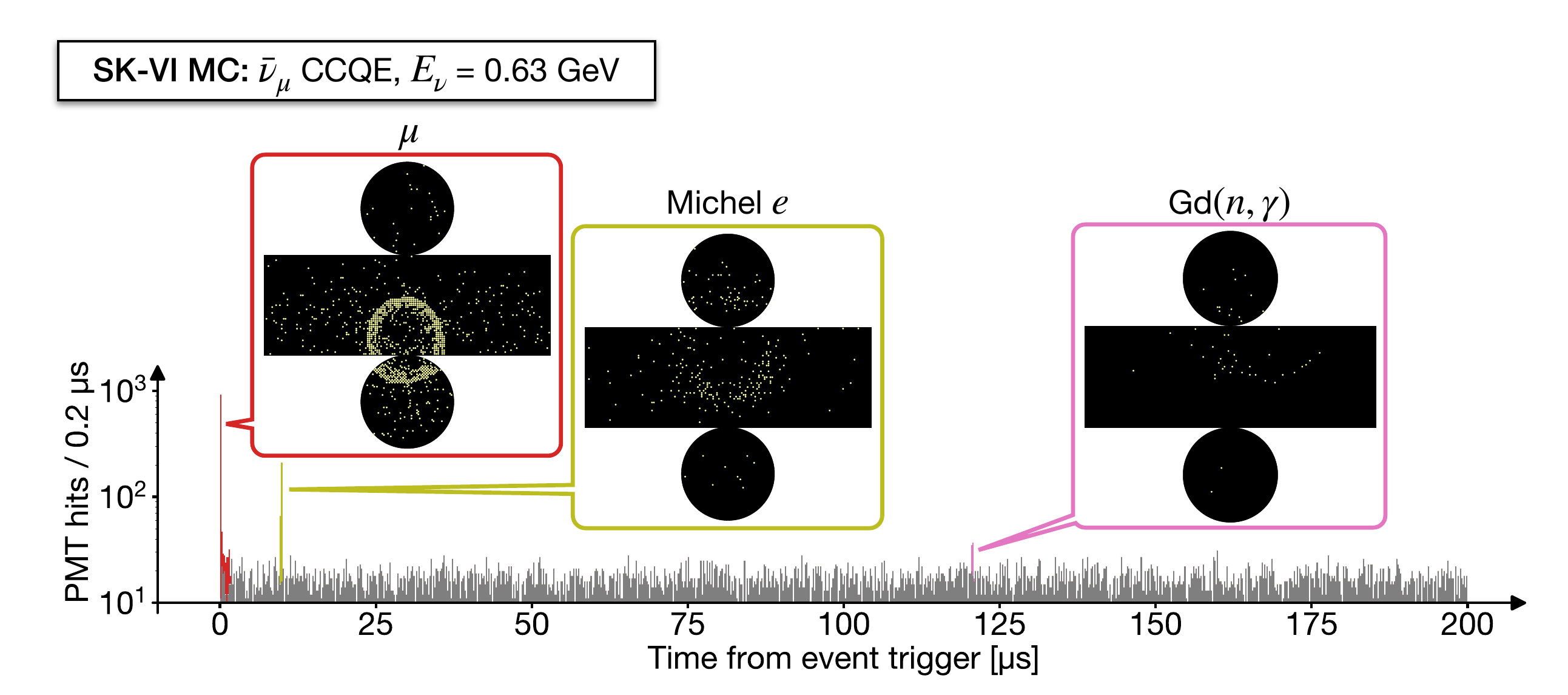}
\caption{\label{fig:evdisp} A PMT hit time distribution for a typical $\bar{\nu}_\mu$ charged-current (CC) quasi-elastic (QE) interaction with a muon and a neutron as final states. The zero is set at the event trigger. The event displays feature the ``prompt'' muon signal (red) and the two types of ``delayed'' coincident signals --- Michel electrons from muon decay (olive) and neutron captures on Gd (pink). The gray bars represent randomly recorded background PMT hits. This event was simulated in the SK-VI configuration.}
\end{figure*}

\section{Atmospheric neutrino events}
\label{sec:simulation}

\subsection{Event selection}
\label{sec:reduc}

%Following the electronics upgrade in 2008, any integrated PMT signal with over 0.25 photoelectron-equivalent charge is digitized and registered as a ``hit''. If the number of hits within a 200-ns sliding time window ($N_{200}$) exceeds a threshold, an event trigger is issued, and a data block spanning [-5, 35] µs surrounding the trigger is recorded as an event.

%The typical event rate of a low-energy trigger with $N_{200}\geq47$ (roughly corresponding to an 8 MeV electron) is $O(10)$ Hz, or around $10^6$ events per day. Most of these event triggers are due to cosmic-ray muons and low-energy radioactivity, which can be reduced through OD veto and cuts on ID charge and reconstructed event vertex. Event triggers resulting from PMT discharges, which often present recurring patterns, are identified and rejected by a dedicated algorithm that searches for such repetitive patterns in the data. The process is detailed in Ref. [].

We followed a typical selection process for atmospheric neutrino interactions that are fully contained within the ID, similar to previous studies conducted at SK \cite{sk_osc_2024}. All events were required to pass the ID trigger with the threshold $N_\text{200-ns}\geq58$ PMT hits—roughly corresponding to a 10 MeV electron—followed by the extended event window of 535 µs for neutron detection. Background events from cosmic-ray muons, radioactivity, and neutrino interactions with exiting particles were reduced using OD veto and ID charge cuts.

Selected events were reconstructed as described in Section \ref{sec:evreco}. To further reject low-energy backgrounds, we required that the reconstructed vertex be more than 1 m away from the ID tank wall (defining the fiducial volume with 27.2 kton of water) and that the visible energy be larger than 30 MeV. The remaining background contamination, mainly due to cosmic-ray muons stopping in the ID and PMT discharges, was estimated to be below 0.2\%, based on visual inspection \cite{sk_pdk_2020}.

\begin{figure}
\includegraphics[width=.9\columnwidth]{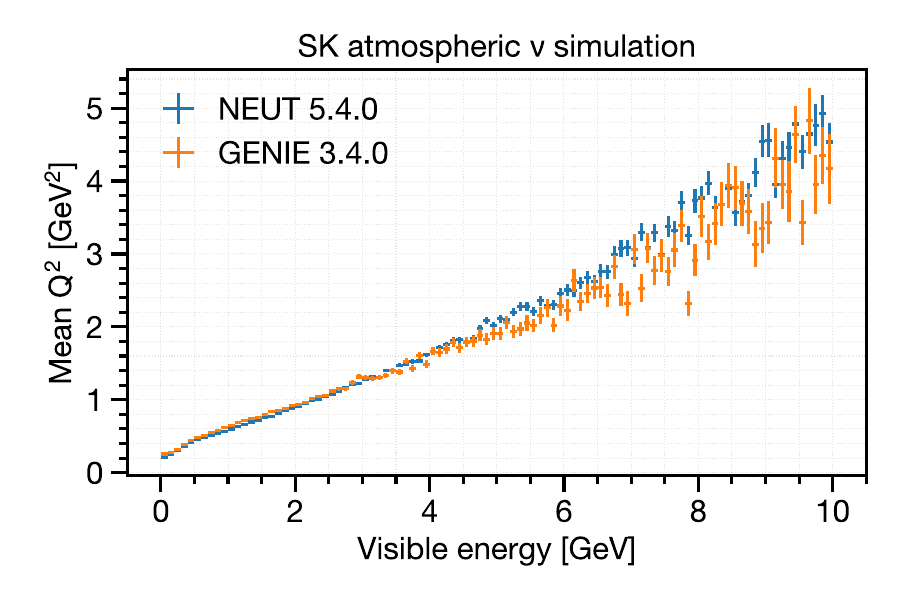}
\caption{\label{fig:evis_vs_qw} Average squared momentum transfer ($Q^2$) per visible energy bin for simulated atmospheric neutrino events at SK, compared between two different neutrino event generators.  NEUT 5.4.0 (blue) uses the baseline setup described in Section \ref{sec:defsimsetup}, while GENIE 3.4.0 \cite{genie_main,genie_manual,genie_v3tune} uses the ``hN'' setup as described in Section \ref{sec:models}.}
\end{figure}

\begin{figure*}[ht!]
\includegraphics[width=\linewidth]{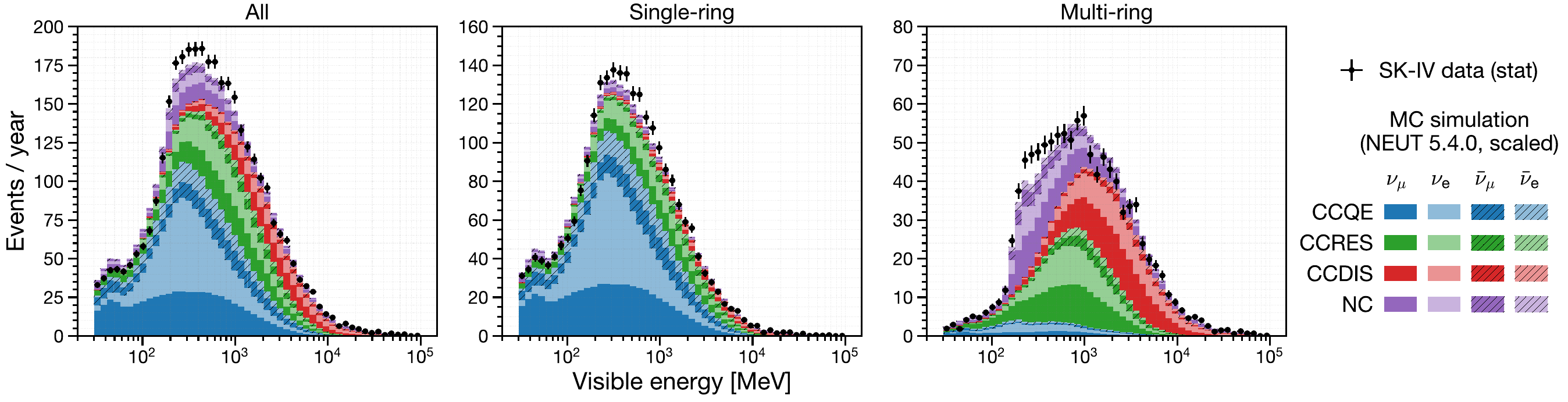}
\caption{\label{fig:intbd} Observed and simulated visible energy distributions for selected atmospheric neutrino interactions, for the longest SK-IV phase: all events (left), events with one reconstructed Cherenkov ring (middle), and events with two or more rings (right). CCQE events dominate the single-ring sample, while CCRES and CCDIS events dominate the multi-ring sample. NC events in the multi-ring sample are mostly RES or DIS interactions. The simulation setup is detailed in Section \ref{sec:defsimsetup}. Simulated distributions are scaled to match the SK-IV data to account for the uncertainty in absolute neutrino flux normalization.}
\end{figure*}

%while $\Delta$ resonance (labels with 1$\pi$ or 1$x$) and DIS events are dominant in the multi-ring sample.}

%with the baseline simulation setup as described in Section %\ref{sec:defsimsetup},

\subsection{Reconstruction of prompt Cherenkov rings}
\label{sec:evreco}

Figure \ref{fig:evdisp} shows a simulated PMT hit time distribution of a typical $\bar{\nu}_\mu$ charged-current (CC) quasi-elastic (QE) event followed by a Michel electron and a Gd$(n,\gamma)$ reaction, along with the corresponding event displays. For the ``prompt'' radiation due to charged particles (namely, electrons and muons) produced via the primary neutrino interaction, we followed the Cherenkov ring reconstruction process \cite{apfit} as applied in previous SK analyses \cite{sk_pdk_2020,sk_osc_2024}.

The visible energy of an event is defined as the sum of the reconstructed kinetic energies of all Cherenkov rings, assuming each ring originates from an electron. This calorimetric measure serves as a reliable proxy for neutrino momentum transfer, as demonstrated by the positive correlation shown in Figure \ref{fig:evis_vs_qw}.

\subsection{Baseline simulation setup}
\label{sec:defsimsetup}

The simulation of atmospheric neutrino events involves a convolution of the atmospheric neutrino flux, neutrino event generator, and detector simulation.

To determine the incoming neutrino kinematics and event rate, we used atmospheric flux calculations for $\nu_e$, $\nu_\mu$, $\bar{\nu}_e$, and $\bar{\nu}_\mu$ at the detector site, assuming no oscillations, as provided by Honda \textit{et al.}\ \cite{honda_2011} for $[10^2, 10^7]$ MeV range. Neutrinos with energies below 100 MeV, as well as the $\nu_\tau$ and $\bar{\nu}_\tau$ interactions, were neglected as their overall presence in data is expected to be small.

Neutrino interactions in water were modeled using neutrino event generators \cite{pdg_nu_ev_gen_rev}, which compute the cross sections for each interaction channel and sample outgoing particle kinematics. NEUT 5.4.0 \cite{neut} was used as the baseline event generator. It covers major interaction channels, including QE scattering with single (1p1h) or double (2p2h) nucleon knockout, single pion production due to $\Delta$ resonance (RES), and multiple pion production via deep inelastic scattering (DIS). The models used are consistent with the latest SK atmospheric neutrino oscillation analysis \cite{sk_osc_2024}. The transport of hadronic final states within the target nucleus (FSI) was modeled separately from the primary neutrino interaction. The transport of low-momentum ($<$ 500 MeV/c) pions was based on Salcedo et al.\ \cite{salcedo_oset}, with parameters tuned to fit external pion-nucleus scattering data \cite{neut_pi_fsi_tune}. For higher energy nucleons and pions, FSIs were modeled using the INC approach, which includes elastic scattering and single/double pion production. The cross sections for reactions with free target nucleons were sourced from Bertini \cite{bertini} for nucleon projectiles and from external pion scattering data for high-momentum pion projectiles \cite{pdperio}. The de-excitation of an oxygen target was modeled using tabulated occupation probabilities for nucleon energy states \cite{160_occ_prob} and branching ratios for knockout of each state \cite{16o_knockout_br}.

The subsequent particle transport in water and detector responses were simulated using GCALOR \cite{gcalor} coupled with GEANT 3.21 \cite{geant3}. Propagation of nucleons and charged pions below 10 GeV was simulated based on the Bertini cascade model \cite{bertini_validation}. For cross sections, GCALOR uses ENDF/B-VI \cite{endf6} for low-energy (below 20 MeV, or 195 MeV/c in momentum) neutrons, and Bertini-based tabulation \cite{bertini_validation} for the remaining nucleons and charged pions. The transport of low-momentum pions was separately modeled using the NEUT pion FSI routine to ensure consistency between NEUT and GEANT.

For the transport of low-energy neutrons in the Gd-loaded SK-VI phase, we imported reaction cross sections from the Geant4.10.5.p01 \cite{geant4_1,geant4_2,geant4_3} NeutronHP model \cite{neutronhp} based on ENDF/B-VII.1 \cite{endf7}, replacing ENDF/B-VI for neutron energies below 20 MeV. Additionally, we modeled the $\gamma$-cascade resulting from neutron captures on $^{155/157}{\text{Gd}}$ using the ANNRI-Gd model \cite{annrigd_157,annrigd_155}.

The characteristics of individual PMTs and the optical parameters in water were adjusted to align with calibration data obtained from light sources and through-going cosmic-ray muons. To accurately account for detector noise, randomly recorded PMT hits were included as background, represented as gray bars in Figure \ref{fig:evdisp}.

500 years of atmospheric neutrino events were simulated for each SK phase and processed as described in Sections \ref{sec:evreco} and \ref{sec:reduc}. The events in the final sample were weighted based on the standard three-flavor oscillation probability in matter \cite{barger}, using the oscillation parameters fitted with reactor constraints in the previous SK analysis \cite{mjiang} and the PREM model \cite{prem} for Earth's matter density. Corrections for atmospheric neutrino flux accounting for solar activity were also applied.

Figure \ref{fig:intbd} shows the visible energy distributions of the simulated event sample. Single-ring events exhibit a higher fraction of QE interactions and a smaller fraction of DIS compared to multi-ring events. For the same visible energy, multi-ring events are expected to produce more secondary neutrons than single-ring events, due to a higher fraction of DIS interactions.

\section{Neutron signal selection}
\label{sec:interp}

Neutrons from atmospheric neutrino interactions in the SK ID are mostly captured by $\isotope[1]{H}$ or $\isotope[155/157]{Gd}$ within $O(100)$ µs, with capture ratios summarized in Table \ref{tab:skphase}. The resulting $O(1)$ MeV $\gamma$-rays scatter electrons, producing faint Cherenkov rings on the ID wall (hereafter referred to as the ``neutron signal''; see Figure \ref{fig:evdisp}). 

Since fast neutrons, $\gamma$-rays, and scattered electrons travel only a few tens of centimeters, the associated Cherenkov photons can be reliably treated as originating from a single vertex. This enables time-of-flight (ToF) corrections to suppress random PMT hits from dark noise. The vertex can either be assumed \textit{a priori}—e.g., using the reconstructed lepton vertex—for better resolution and signal efficiency at the cost of potential bias, or reconstructed independently from the lower energy neutron signal (Appendix \ref{appendix:phaseconsistency}), yielding lower resolution and efficiency but greater robustness to bias and kinematic uncertainties. Residual Michel electrons from muon decays are reduced using timing and energy cuts.

\subsection{Signal selection algorithm}
\label{sec:ntagalgo}

The signal selection algorithm is based on Ref.\ \cite{sk4_ntag}, and consists of candidate search and classification stages. %The first stage searches for signal candidates, by triggering on the number of PMT hits within a small time window,  The second stage further reduces the remaining PMT dark noise and Michel electron contamination within the selected candidates, by applying a neural network and cuts on distinguishing features, as illustrated in Figure \ref{fig:simfeatures} and \ref{fig:ecut}.

%In the first candidate search stage, we first subtract expected photon time-of-flight (ToF) from the individual PMT hit times, assuming all PMT hits are due to Cherenkov photons generated from the neutrino interaction vertex. Then, we slided a time window of 14 ns width on the ToF-corrected PMT hit times to trigger on the number of included PMT hits.

%All PMT hit times after correcting photon time-of-flight, assuming all photons are created at the reconstructed neutrino vertex, are arranged in a single array in a decreasing order. A sliding time window of 14 ns, which has been designed to optimize the signal-to-noise ratio specifically for $^1H(n,\gamma)$ reactions. 

In the candidate search stage, PMT hit times are corrected for photon ToF from an \textit{a priori} vertex given by the reconstructed lepton vertex. A 14-ns sliding time window is applied, with thresholds of 5 hits for SK-IV/V (pure water) and 7 for SK-VI (Gd-loaded), optimized for signal efficiency \cite{han_mthesis}. Overlapping candidates within 50 ns are resolved by selecting the one with the most PMT hits. In SK-IV/V, the search window was set to [18, 534] µs from the event trigger, following earlier analyses \cite{sk4_ntag, sk_osc_2024, sk_pdk_2020} designed to avoid PMT afterpulses ($<$15 µs) and Michel electrons. For SK-VI, it was extended to [3, 534] µs to capture more of the faster neutron captures with Gd, aided by improved candidate classification.

 %534 µs is to have 1 µs buffer at the end of the event.

%This stage reduces random coincidences of PMT dark hits.

%For overlapping candidates within 50 ns, only the candidate with the largest number of PMT hits is selected.
%This algorithm is highly dependent on the reconstructed neutrino vertex, works poor on neutron captures far from reconstructed neutrino vertex. Any candidate is dismissed if the signal time deviates by the minimum peak separation from the original time, or if the new hit count is lower than the threshold.

%For the sixth phase of Super-Kamiokande with Gd, the number of signal hits is large enough for a good independent vertex reconstruction (resolution $\sim$1 m) so that we can add additional steps to make algorithm works evenly well for all possible neutron capture vertices. BONSAI is a maximum likelihood vertex fitter. For a given set of PMT hit times (ToF-corrected, residual), it finds the vertex that maximizes the likelihood calculated based on LINAC $e$ calibration data. (Reference) The BONSAI vertex and signal time are reconstructed using the hits registered within a timespan of -0.5 to 1 microseconds. This timeframe accommodates all possible vertices in the tank. The time window is then reset based on a range of -7 to 7 nanoseconds from the newly reconstructed signal time.

In the candidate classification stage, we extract features of each candidate and use a neural network to classify each candidate into signal and noise based on input features. These features characterize the signal energy, the background hit level, timing spread assuming the vertex, correlation between the input vertex and the hit PMT positions, correlation to the known properties of PMT noise, and angular correlation among hit PMTs relative to the Cherenkov cone opening angle. 

The major changes from the original algorithm \cite{sk4_ntag} include a simplified algorithm, a reduced set of features, and a heuristically tuned neural network architecture. These modifications aim to reduce performance bias between the data and the simulation that is used to train the neural networks.

\begin{figure*}[ht!]
\includegraphics[width=0.85\linewidth]{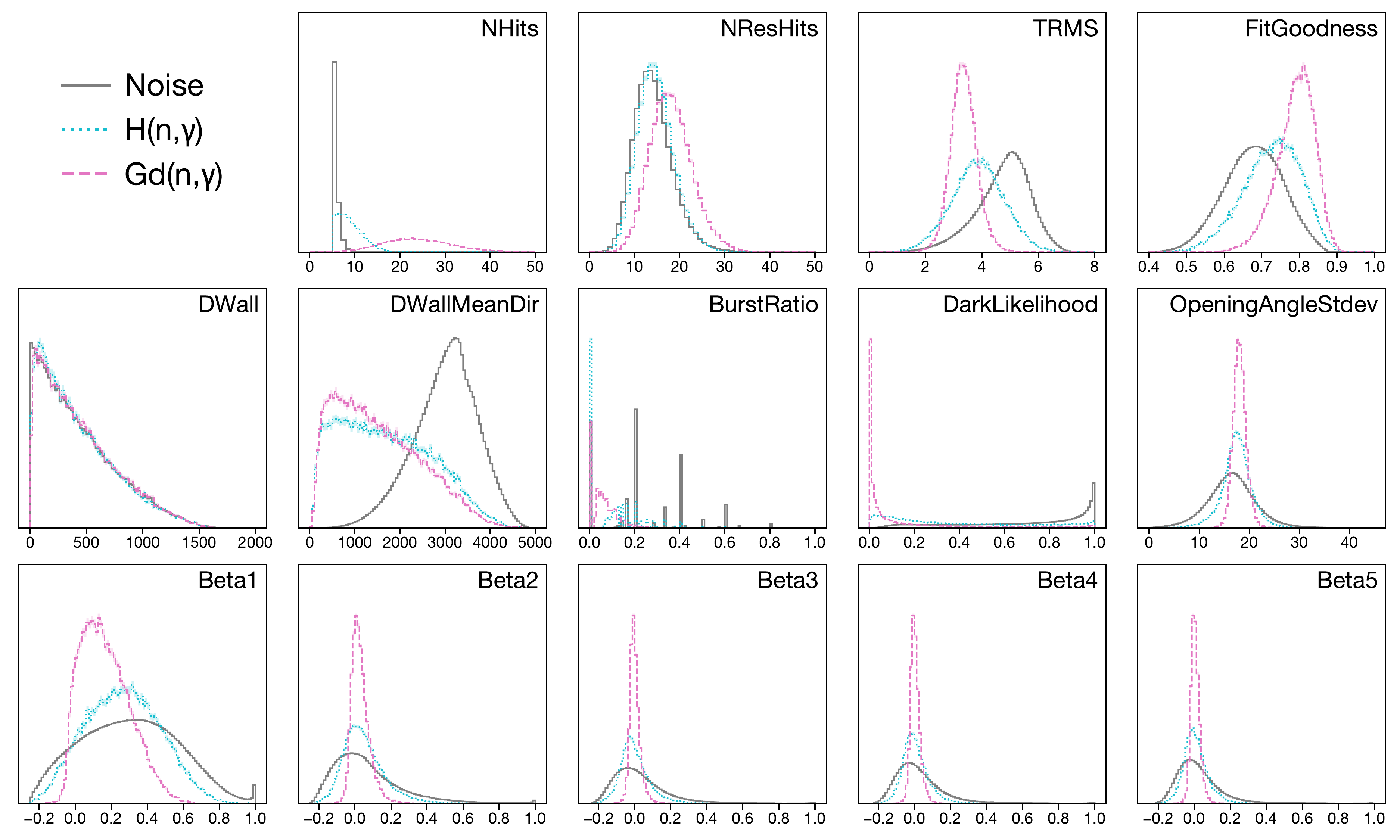}
\caption{\label{fig:simfeatures} The features (area-normalized) of neutron signals and noise from the thermal neutron MC simulation used for training the neural network for SK-VI phase.}
\end{figure*}

Here, we provide the definition and unit of each feature used for the classification of signal candidates, along with their expected distributions as shown in Figure \ref{fig:simfeatures}: %

\begin{itemize}
    \item \texttt{NHits}

    The number of selected PMT hits within the 14-ns sliding time window.

    \item \texttt{NResHits}

    The number of PMT hits within $[-100,+100]$ ns from the center of the 14-ns sliding time window, minus \texttt{NHits}.
    
    \item \texttt{TRMS} [ns]

    The root mean square (RMS) of the ToF-corrected time distribution of the selected PMT hits.
    
    \item \texttt{FitGoodness}

    The normalized likelihood of the ToF-corrected time distribution of the selected PMT hits, given the \textit{assumed} signal vertex and the Gaussian PMT timing resolution of 5 ns. 

    \item \texttt{DWall} [cm]

    The distance from the \textit{assumed} signal vertex to the nearest tank wall. 

    \item \texttt{DWallMeanDir} [cm]

The shorter of the radial and vertical distances from the \textit{assumed} signal vertex to the tank wall, calculated along the average direction of the unit vectors connecting the vertex to each hit PMT.
    
    \item \texttt{BurstRatio}

    The ratio of the selected PMTs with a preceding hit within 10 µs, which are likely caused by scintillation within the irradiated PMT glass.
    
    \item \texttt{DarkLikelihood}

    The normalized log likelihood ratio based on measured individual PMT dark rates, given by:

    \begin{equation}
        \texttt{DarkLikelihood} = \sigma\bigg(\log \prod_{i=1}^\texttt{NHits} \frac{r_i}{\langle r\rangle}\bigg)
    \end{equation}
    where $\sigma$ represents the sigmoid function, $r_i$ is the dark rate of the $i^\textrm{th}$ PMT, and $\langle r\rangle$ is the average dark rate of all ID PMTs.
    
    \item \texttt{OpeningAngleStdev} [deg]

The standard deviation of the opening angles of cones formed by every possible combination of three hit PMTs and the \textit{assumed} signal vertex.
    
    \item \texttt{Beta(k)}, \texttt{k}$\in\{1,2,3,4,5\}$

    \begin{equation}
    \texttt{Beta(k)} = \frac{2}{\texttt{NHits}(\texttt{NHits}-1)}\sum_{\substack{i\neq j}}P_\texttt{k}(\cos{\theta_{ij}})
\end{equation}

where $P_\texttt{k}$ is the $k^\text{th}$ Legendre polynomial and $\theta_{ij}$ is the opening angle between the assumed signal vertex and the $i^\text{th}$ and $j^\text{th}$ hit PMTs.

\end{itemize}

\begin{figure}[htbp!]
\includegraphics[width=0.79\linewidth]{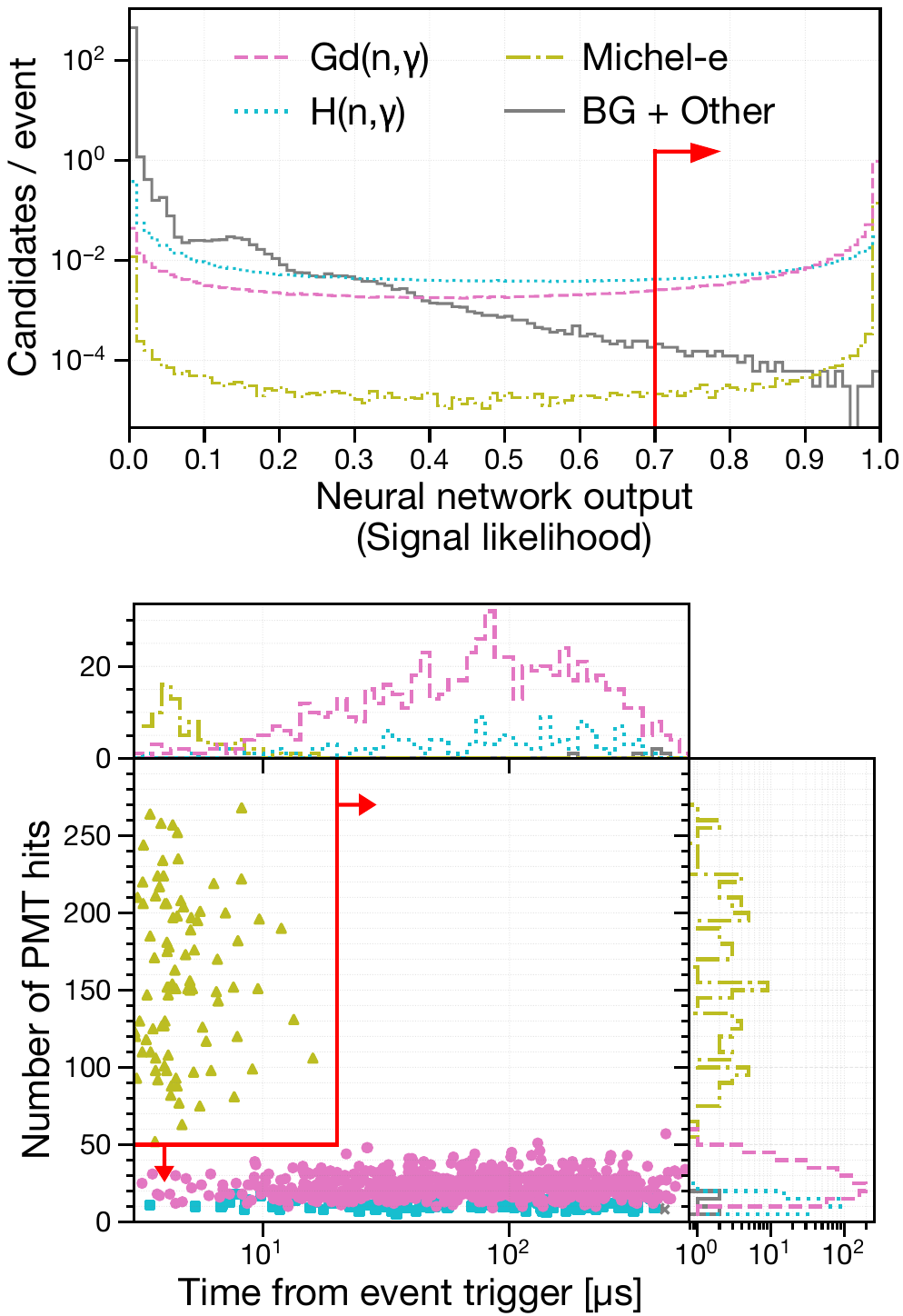}
\caption{\label{fig:ecut}The neural network response to neutron signals and backgrounds in the SK-VI atmospheric neutrino simulation (top), and the time versus energy distribution of signal candidates passing the neural network selection (bottom). Red arrows indicate the corresponding cut points.}
\end{figure}

For training the neural network, we utilized the features and labels of first-stage candidates from a particle-gun simulation of thermal neutrons, with vertices randomized within the ID. Each first-stage candidate, as shown in Figs.\ \ref{fig:simfeatures} and \ref{fig:ecut}, was labeled as signal if it was triggered within 50 ns of the simulated $(n,\gamma)$ reaction, and otherwise as background. Of the training dataset, 80\% was used for updating the neural network weights, while the remaining 20\% was reserved for validation.

We implemented a feed-forward fully connected neural network using Keras 2.6.0 \cite{keras}. The network consisted of an input layer with 14 features, followed by three dense layers, each comprising 128 ReLU-activated nodes with a 50\% dropout rate, and a single sigmoid output node. Weights and biases were initialized following He \textit{et al.}\ \cite{he_normal} and optimized by minimizing the binary cross-entropy loss iteratively on minibatches of size 2,048 using the Adam optimizer \cite{adam}. The initial learning rate was set to 0.0001. Training was stopped when signal efficiency on the validation set showed no improvement for 5 consecutive epochs. A neural network was trained for each SK phase: SK-IV, SK-V, and SK-VI.

Candidates with a neural network output greater than 0.7 were classified as signals, while those with a large number of PMT hits ($\texttt{NHits}>50$) and occurring earlier than the typical neutron capture timescale ($<20$ µs) were identified as Michel electrons and excluded. The effectiveness of this Michel electron rejection is illustrated in Figure \ref{fig:ecut}. When applied to cosmic-ray muons decaying within the ID, the selection achieved an efficiency of 98.4$\pm$1.3\% and a purity of 98.7$\pm$0.5\%.

\subsection{Signal selection performance on calibration data}
\label{sec:ambe}

An Am/Be neutron source with a measured total intensity of 236.8$\pm$5.0 neutrons/s \cite{ambe_intensity} was used to obtain calibration data for estimating the signal detection performance. The first excited state of the alpha-absorbed $\isotope[9]{Be}$, with a roughly 60\% branching ratio, emits a fast neutron and a 4.44 MeV $\gamma$-ray simultaneously. This source was encapsulated with Bismuth Germanate ($\text{Bi}_4 \text{Ge}_3 \text{O}_{12}$, BGO) crystals so that the 4.44 MeV $\gamma$'s can induce scintillation. The setup was deployed in various positions within the ID, and events were recorded for 30 minutes to 1 hour. Events with a trigger charge yield corresponding to the 4.44 MeV $\gamma$-ray scintillation were regarded as the single neutron control sample. 

\begin{figure}[b]
\includegraphics[width=0.99\columnwidth]{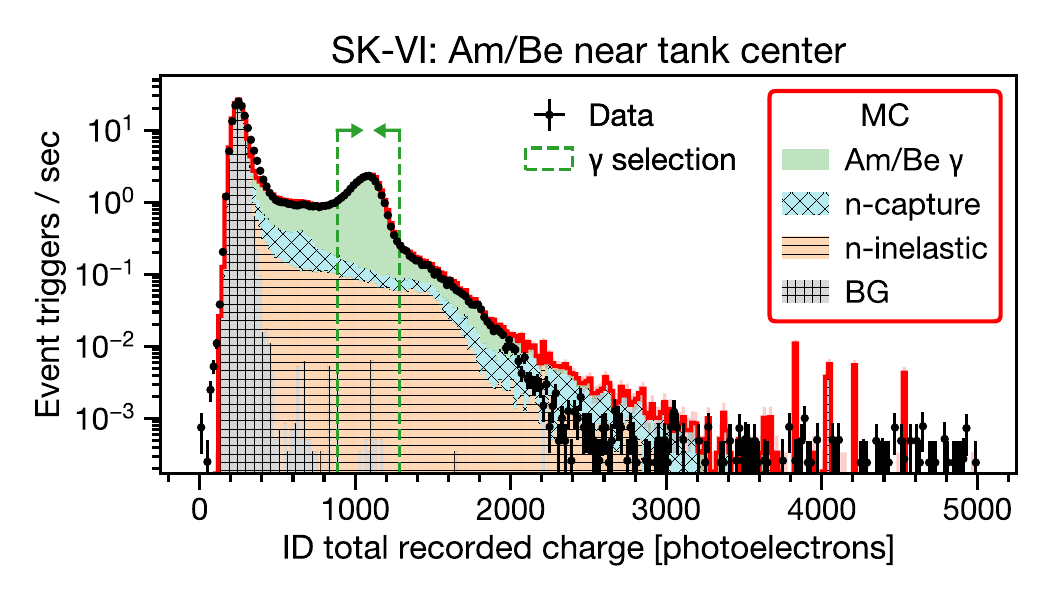}
\caption{\label{fig:qismsk} Distribution of recorded charge within the time window $[-0.5, 1.0]$ µs relative to event triggers, measured with an Am/Be neutron source positioned near the ID tank center. The black points represent SK-VI data, while the red line shows the simulated prediction. The green dashed arrows indicate the selection window for 4.44 MeV $\gamma$-ray-induced scintillation. Roughly 95\% of the selected events are attributed to $\gamma$'s from the Am/Be source (``Am/Be $\gamma$''), with the remaining 5\% arising from ambient neutron captures (``n-capture'') and neutrons inelastically producing charged particles in the scintillator (``n-inelastic''). The contribution from background events (``BG'') is negligible. The pink shades indicate MC statistical errors.}
\end{figure}

The observed light yield distribution was compared with dedicated simulation, as shown in Figure \ref{fig:qismsk}. The simulation accounts for continuous source activity and pile-up, by reorganizing the simulated detector response to Am/Be neutron emission on a single global time axis, based on the measured total neutron intensity and the estimated branching ratios to each excited state of alpha-absorbed $\isotope[9]{Be}$. As shown in Figure \ref{fig:qismsk}, this simulation accurately models event triggers due to ambient neutron captures and neutron inelastic interactions within scintillator crystal. The contamination of such unwanted event triggers in the 4.44 MeV $\gamma$-ray event selection was estimated to be at a few percent level.

Within the selected events in the single neutron control sample, signal candidates were obtained following the algorithm described in Section \ref{sec:ntagalgo}, with the \textit{assumed} signal vertex set at the source position.

%The obtained calibration data was compared with MC simulation.

\begin{figure}
\includegraphics[width=0.99\columnwidth]{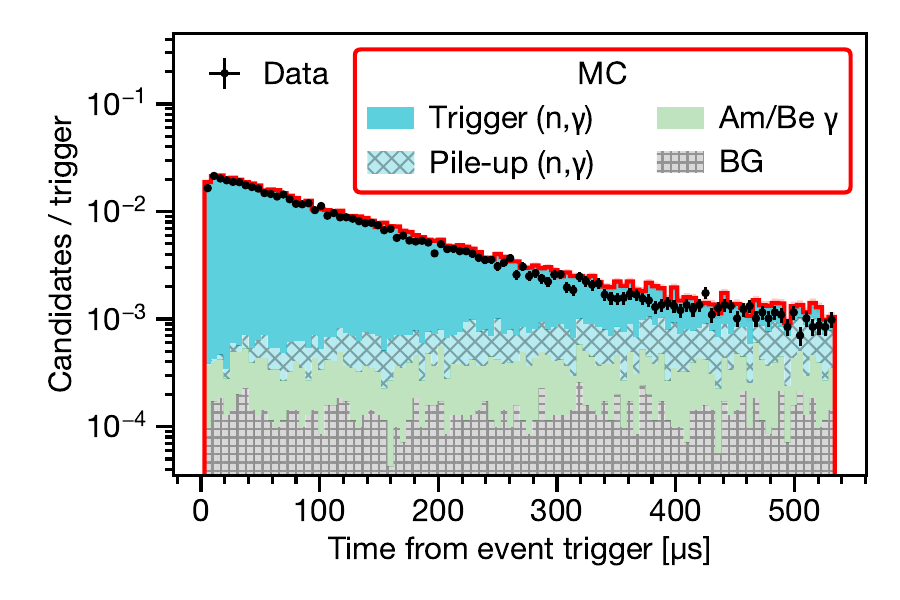}
\caption{\label{fig:fittdist} Exponential decrease of the selected neutron signal candidates as a function of the time from the selected event triggers with the Am/Be neutron source positioned near the ID tank center, in the SK-VI phase. The label ``Trigger $(n,\gamma)$'' indicates captures of neutrons produced within 350 ns from the event trigger, while the label ``Pile-up $(n,\gamma)$'' indicates captures of piled-up neutrons without such correlation to the event trigger. The pink shades indicate MC statistical errors.}
\end{figure}

\begin{figure*}
\includegraphics[width=\linewidth]{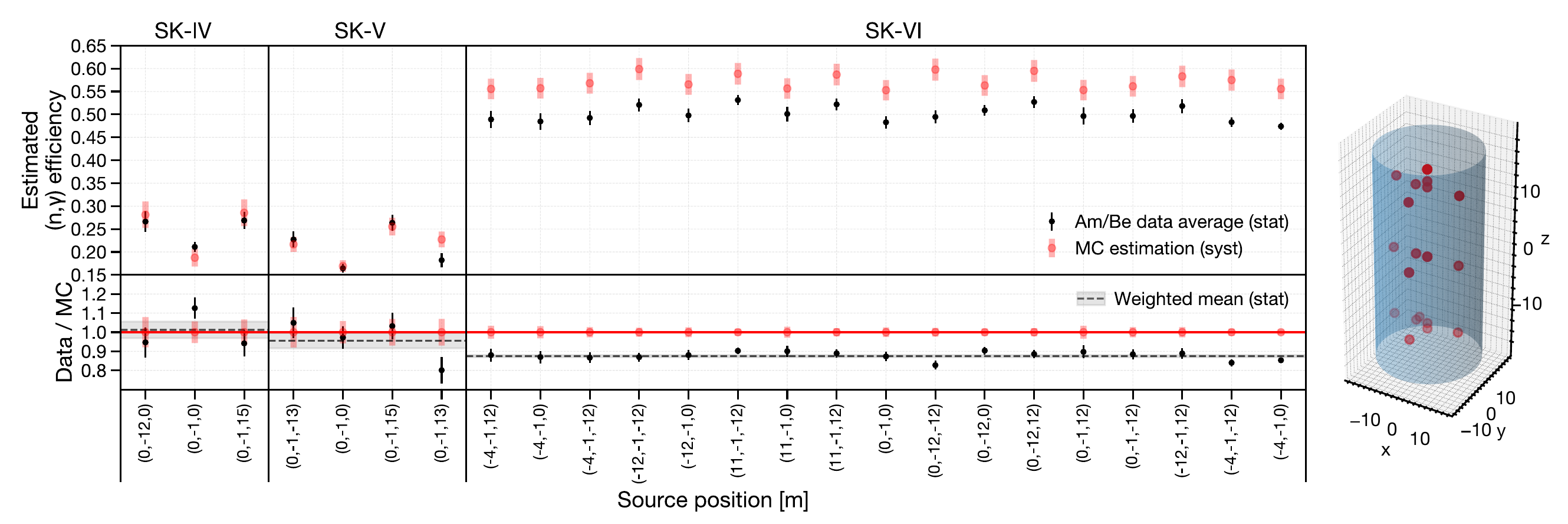}
\caption{\label{fig:pospvx} Estimated $(n,\gamma)$ signal selection efficiencies for each calibration position within the ID tank (red markers in the schematic on the right). For SK-VI positions with multiple measurements taken on different dates, the reported efficiency is the average of those measurements. Systematic uncertainties considered in the MC simulation are summarized in Table \ref{tab:ambe_mcunc}.}
\end{figure*}

Figure \ref{fig:fittdist} shows an example time distribution of the selected signal candidates. Such distributions of the time $t$ were fitted with a function $f$ of the form:

\begin{equation}
f(t) = A (1-e^{-t/\tau_\text{thermal}})e^{-t/\tau_\text{capture}} + B
\label{eq:tau}
\end{equation}
where the normalization constant $A$, the background constant $B$, the neutron thermalization time scale $\tau_\text{thermal}$ (set to 0 for SK-IV/V assuming a velocity-independent neutron capture rate), and the neutron capture time constant $\tau_\text{capture}$ are free parameters. The measured $\tau_\text{thermal}$ in SK-VI was $4.71\pm0.04$ µs. The signal efficiency was evaluated as the number of identified signals per selected event triggers, corrected by the constant background term $B$. Figure \ref{fig:pospvx} shows the estimated neutron detection efficiencies for various source positions in the ID.

The major sources of systematic uncertainty are summarized in Table \ref{tab:ambe_mcunc}. In the pure water phases (SK-IV and V), the dominant source of uncertainty lies in the potential bias caused by the calibration setup, such as the unwanted event triggers or time correlation of false positives to event triggers, often leading to an overestimation of the background constant $B$. The size of this uncertainty was conservatively estimated by comparing the true and estimated signal efficiencies from the simulations and quantifying the fluctuation within each SK phase. In the Gd-loaded phase (SK-VI), the dominant source of uncertainty is in the fraction and the $\gamma$ emission model of the Gd$(n,\gamma)$ reaction. The size of these uncertainties is estimated based on the evaluated thermal neutron capture cross section uncertainties in ENDF/B-VII.1, as well as variations in the estimated signal efficiency when using an alternative photon strength function to describe the Gd continuum $\gamma$-cascade in the ANNRI-Gd model \cite{annrigd_157}.
%The impact on the estimated neutron detection efficiency was evaluated by varying parameters within uncertainties or testing different models. 

%due to the uncertainty in thermal neutron capture cross sections and in the $\gamma$ emission model in the Gd$(n,\gamma)$ reaction, which was evaluated by comparing the use of single Lorentzian (SLO) and enhanced generalized Lorentzian (EGLO) models for the description of quasi-continuum levels of \isotope[156/158]{Gd}^*$.

\begin{table}[ht!]
\caption{Major sources of systematic uncertainty in the neutron detection efficiency estimated with the Am/Be neutron source.}
\centering
\begin{ruledtabular}
\begin{tabular}{lrrr}
Source                       & \multicolumn{1}{c}{SK-IV} & \multicolumn{1}{c}{SK-V} & \multicolumn{1}{c}{SK-VI} \\ \hline
Am/Be neutron characterization             & 0.5\%                      & 0.9\%                     & 0.5\%                                               \\ %\hline
Detector response & 2.2\%                      & 3.3\%                     & 1.2\%                                              \\ %\hline
Bias due to calibration setup            & 6.9\%                      & 4.6\%                     & 1.1\%                                             \\ %\hline
Gd$(n,\gamma)$ fraction       & \multicolumn{1}{c}{-}                          & \multicolumn{1}{c}{-}                         & 2.1\%        \\ %\hline
Gd$(n,\gamma$) $\gamma$ emission model    & \multicolumn{1}{c}{-}                          & \multicolumn{1}{c}{-}                         & 2.6\%                                             \\ \hline
Total                        & 7.3\%                      & 5.7\%                     & 3.8\%         \\ %\hline
\end{tabular}
\end{ruledtabular}
\label{tab:ambe_mcunc}
\end{table}

The discrepancy between observed and predicted signal efficiencies in SK-VI (Figure \ref{fig:pospvx}) likely arises from an overestimation of the Gd$(n,\gamma)$ fraction in Geant4 \cite{han_thesis, hino_g4}. For example, the Gd$(n,\gamma)$ fraction estimated using SK-VI Am/Be data was $(44\pm3)$\% \cite{han_thesis}, consistent with the $(43.9\pm1.5)$\% evaluated with thermal neutron capture cross sections in ENDF/B-VII.1 \cite{endf_paper}, yet both were lower than the 52\% predicted by Geant4.10.5.p01 NeutronHP using the same cross sections. This problem with Geant4 is attributed to its treatment of hydrogen as free rather than bound in water molecules, leading to an underestimation of the competing $\isotope[1]{H}(n,\gamma)$ reaction \cite{hino_g4}.

To address this discrepancy, we applied a correction based on the weighted mean of Am/Be data-to-MC efficiency ratios across all source positions. Signal efficiencies in the pure water phases with limited calibration data were further corrected to ensure consistency in the overall neutron counts with the well-calibrated Gd-loaded phase, and the final correction factors were $0.90\pm0.12$ (SK-IV), $0.94\pm0.04$ (SK-V), and $0.88\pm0.01$ (SK-VI) (see Appendix \ref{appendix:phaseconsistency} for details). 

The measured $\tau_\text{capture}$ in Equation \ref{eq:tau} were $200.4\pm3.7$ \textmu s (pure water, SK-IV/V) and $116.9\pm0.3$ \textmu s (Gd-loaded, SK-VI), consistent with ENDF/B-VII.1 predictions of $204.7\pm5.3$ \textmu s and $114.9\pm2.5$ \textmu s.
%Geant4 predicted $112.4$ \textmu s for the Gd-loaded phase.  

%Figure \ref{fig:gdfrac_simple} compares the MC-simulated Gd$(n,\gamma)$ fraction $r_\text{Gd}$ with the analytically evaluated fraction, assuming completely thermalized neutrons:

%begin{equation}
%    r_\text{Gd} \approx 1 - r_\text{H} \approx 1 - \frac{n_\text{H} g_\text{H}(T) \sigma_\text{H}(v_\text{thermal})}{\sum_i n_i g_i (T) \sigma_i(v_\text{thermal})}
%    \label{eq:gdfrac}
%\end{equation}

%Here, for the $i^\text{th}$ isotope, $n_i$ is the number density, $g_i(T)$ is the Westcott $g$-factor for temperature $T$, and $\sigma_i(v_\text{thermal})$ is the neutron capture cross section evaluated at thermal neutron speed $v_\text{thermal} = 2200$ m/s.

%\begin{figure}
%\includegraphics[width=\linewidth]{figures/4_neutron_selection/gdfrac_magnified.pdf}
%\caption{\label{fig:gdfrac_simple} Predicted Gd$(n,\gamma)$ fraction in Gd-loaded water as a function of Gd concentration, compared with SK-VI Am/Be data (black). The red dots represent predictions by MC simulation using the NeutronHP model in Geant4.10.05.p01 with neutron cross sections from ENDF/B-VII.1, while the blue line and shades represent evaluation of Equation \ref{eq:gdfrac} based on the evaluated thermal $(n,\gamma)$ cross sections and uncertainties in ENDF/B-VII.1.}
%\end{figure}
\section{\texorpdfstring{$(n,\gamma)$ multiplicity estimation}{Neutron capture multiplicity estimation}}
\label{sec:mult_est}

The average multiplicity of total $(n,\gamma)$ reactions is computed as the average of the expected number of $(n,\gamma)$ reactions estimated on an event-by-event basis, as follows:

\begin{equation}
\langle N \rangle = \bigg\langle \frac{N^\text{detected}_i - N^\text{BG}_i}{\epsilon_i} \bigg\rangle
\label{nmult_eq}
\end{equation}

Here, $N^\text{detected}_i$ is the count of detected signals, $N^\text{BG}_i$ is the estimated number of false positives, and $\epsilon_i$ is the estimated signal detection efficiency of the $i^\text{th}$ event. 

Accurate estimation of $N_i^\text{BG}$ and $\epsilon_i$ is crucial. While Am/Be calibration data provides a basis for these estimates, additional factors in atmospheric neutrino events—such as neutrino vertex reconstruction accuracy and larger neutron kinetic energy—may significantly impact the signal selection performance. To better account for these effects in the calculation of $N_i^\text{BG}$ and $\epsilon_i$, we trained Generalized Additive Models (GAMs) \cite{gam} on the baseline atmospheric neutrino simulations. Using pyGAM 0.9.0 \cite{pygam}, a total of six linear GAMs were constructed across the three SK phases, for two output metrics: signal selection efficiency and false positive rate.

Using the baseline MC simulation generated with NEUT 5.4.0, GAMs were fitted to the simulated distributions of each metric, averaged within bins in a five-dimensional feature space. The features consist of reconstructed neutrino event variables, including visible energy, Cherenkov ring multiplicity, the particle type of the most energetic ring, and the radial and vertical displacements of the neutrino interaction vertex. No assumptions were made regarding feature correlations, and appropriate smoothing was applied to mitigate overfitting. The 1$\sigma$ prediction interval was determined based on the MC statistical uncertainties within each bin.

\begin{figure}[ht!]
\centering
\includegraphics[width=0.94\columnwidth]{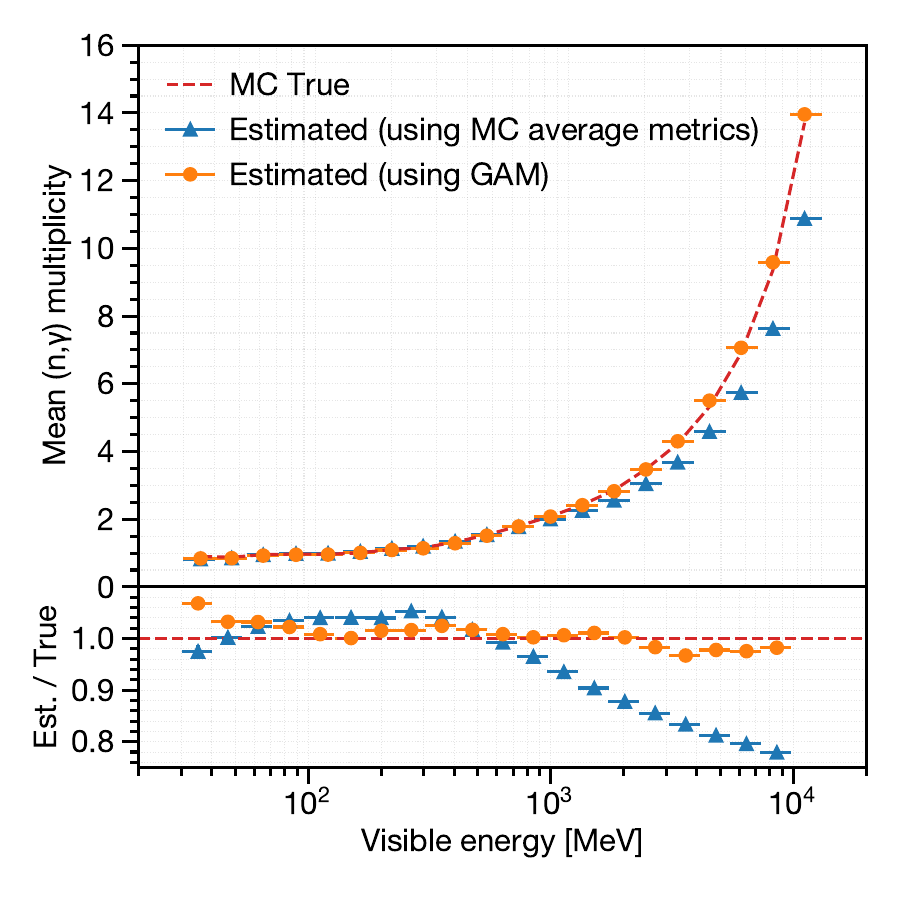}
\caption{\label{fig:gam_est_perf} The true (dashed lines) and estimated average $(n,\gamma)$ multiplicity (circle markers: using GAM, triangle markers: using the overall average signal efficiency and false positive rate obtained from the given simulation) as a function of visible energy, for the test simulations of atmospheric neutrino events produced with NEUT 5.1.4.}
\end{figure}

 To evaluate uncertainty in its performance, trained GAMs were deliberately tested on unseen simulation data produced with NEUT 5.1.4, which predicts roughly 10\% lower $(n,\gamma)$ multiplicity overall. Figure \ref{fig:gam_est_perf} shows the performance of the trained GAM in estimating the true average signal multiplicity per visible energy bin. By reconstructing $N_i^\text{BG}$ and $\epsilon_i$ on an event-by-event basis, the GAM helps reduce potential biases in the results, particularly in the multi-GeV bins.

The following major systematic uncertainties affecting signal counting were evaluated on a bin-by-bin basis for each operational phase and data subsample:

\begin{enumerate}[label={(\arabic*)}, leftmargin=12pt]
    \item Overall signal efficiency scale

This includes calibration uncertainties (Section \ref{sec:ambe}) and phase-dependent variations (to be explained in Appendix \ref{appendix:phaseconsistency}).
    
    \item Signal selection performance modeling (GAM)

This is quantified as the difference between true and estimated signal multiplicities in simulations. The ratio is also used to correct the estimated distribution.
    
    \item Neutrino event reconstruction

    This assumes 2\% visible energy resolution (Figure \ref{fig:escale}), with ring-counting errors accounted for single- and multi-ring events.
\end{enumerate}

\begin{figure}[ht!]
\includegraphics[width=0.9\columnwidth]{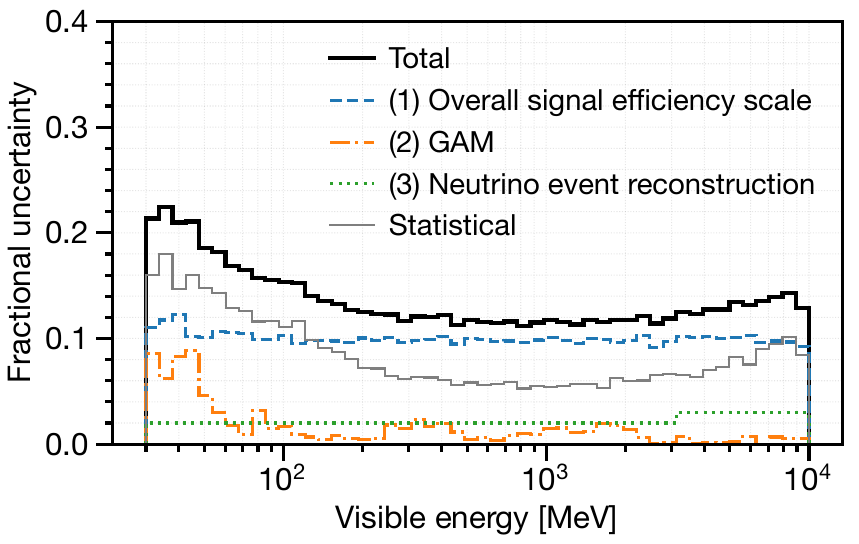}
\caption{\label{fig:fracerr} Fractional uncertainties assigned to the estimated average $(n,\gamma)$ multiplicity per bin for the full data sample.}
\end{figure}

Figure \ref{fig:fracerr} shows fractional uncertainties per visible energy bin for the full data sample. The largest uncertainty lies in the overall signal efficiency scale and statistics.

\section{Tested models}
\label{sec:models}

%Figure \ref{fig:n_source} shows the sources of neutron production in the baseline SK-IV simulation as a function of visible energy, offering insight into the origins of neutron production across different neutrino energy regions. The number of outgoing neutrons from neutrino-nucleus interactions is expected to be nearly constant regardless of visible energy. In events with lower visible energies, the momentum of the outgoing neutrons is small, so each outgoing neutron corresponds almost directly to a single $(n,\gamma)$ reaction. At higher energies, the outgoing hadrons gain more energy, leading to a linear increase in secondary neutron production. This suggests that the $(n,\gamma)$ multiplicity in lower visible energies is sensitive to the modeling of low-energy nucleon transport within the nucleus, while the linear increase is primarily determined by the choice of secondary hadron-nucleus interaction model.

% \begin{figure}[b]
%    \centering
%    \includegraphics[width=0.85\columnwidth]{figures/7_results/n_source.pdf}
%    \caption{Mean multiplicity of outgoing neutrons from the primary neutrino interaction (blue) and subsequent FSIs (green) and resulting total $(n,\gamma)$ reactions (red), in the baseline SK-IV simulation setup.}
%    \label{fig:n_source}
%\end{figure}

To compare with data, we generated predictions for the average $(n,\gamma)$ multiplicity as a function of neutrino event visible energy using various neutrino event generators and hadron-nucleus interaction models. Six generator configurations were tested: \texttt{NEUT 5.4.0}, \texttt{NEUT 5.6.3}, \texttt{GENIE hA}, \texttt{GENIE hN}, \texttt{GENIE BERT}, and \texttt{GENIE INCL}. The \texttt{GENIE} setups use GENIE 3.4.0 with G\_18a\_10$x$\_02\_11b physics tunes \cite{genie_v3tune}, where $x \in \{\text{a}, \text{b}, \text{c}, \text{d}\}$ corresponds to the FSI model: INTRANUKE/hA \cite{genie_manual} (\texttt{hA}), INTRANUKE/hN \cite{genie_manual} (\texttt{hN}), the Geant4 Bertini cascade model \cite{g4_bertini} (\texttt{BERT}), and the Liège INC model \cite{g4_incl} (\texttt{INCL}).

\texttt{NEUT 5.4.0} follows the setup described in Section~\ref{sec:defsimsetup}, while \texttt{NEUT 5.6.3} includes a modified nuclear binding energy, removing roughly 10\% of QE interactions in which the struck nucleon falls below the revised threshold. The \texttt{GENIE} setups share QE and single-pion production models with NEUT but differs in FSI and hadronization. Except for \texttt{GENIE hA}, all FSI models use the full INC approach. Also, the Liège INC model in \texttt{GENIE} is coupled with ABLA07 \cite{abla07} for nuclear de-excitation. 

Secondary hadron-nucleus interaction models were tested using SK detector simulations with GEANT 3.21 and Geant4.10.5.p01, in six configurations: \texttt{SK-IV/V default}, \texttt{SK-VI default}, \texttt{G3 GCALOR}, \texttt{G4 BERT}, \texttt{G4 BERT\_PC}, and \texttt{G4 INCL\_PC}. The \texttt{SK-VI} setup differs from \texttt{SK-IV/V} in its use of the Geant4 NeutronHP and Bertini cascade models for neutron tracking below 20 MeV and $\mu^-$ captures, respectively. \texttt{G3 GCALOR} relies entirely on GCALOR within GEANT 3.21, while the baseline setups use the NEUT pion FSI routine. Geant4-based models adopt ENDF/B-VII.1 for low-energy neutron transport, with \texttt{G4 BERT} and \texttt{G4 BERT\_PC} using the Bertini cascade, and \texttt{G4 INCL\_PC} using the Liège INC model. Configurations with the \texttt{PC} suffix employ the Geant4 Precompound model \cite{g4preco} for nuclear de-excitation, while \texttt{G4 BERT} uses a simpler native model \cite{g4_bertini}. GEANT-3-based models rely on Bertini-based tabulation \cite{bertini_validation} for hadron-nucleus cross sections (excluding low-energy neutrons), whereas Geant4-based models use Glauber parameterization \cite{glauber}.

\begin{figure}
\includegraphics[width=0.9\columnwidth]{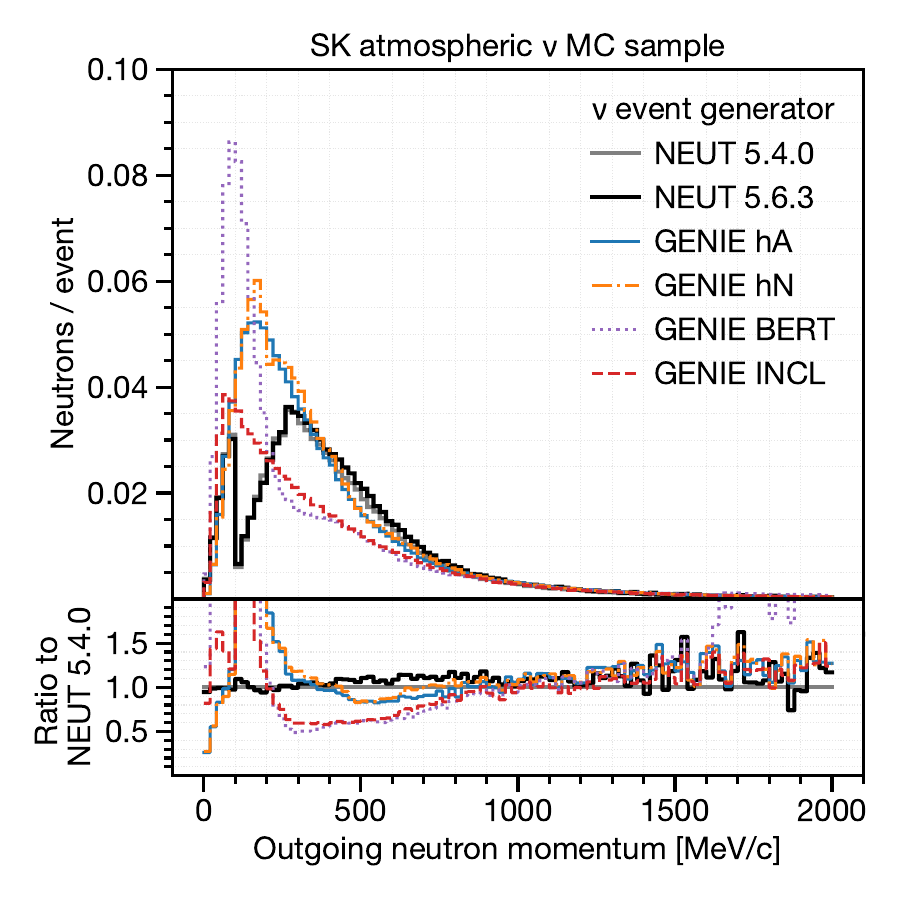}
\caption{\label{fig:out_n_mom} Comparison of neutrino event generator options with different FSI models: Outgoing neutron momentum distribution per selected atmospheric neutrino event.}
\end{figure}

\begin{figure}
\includegraphics[width=0.92\columnwidth]{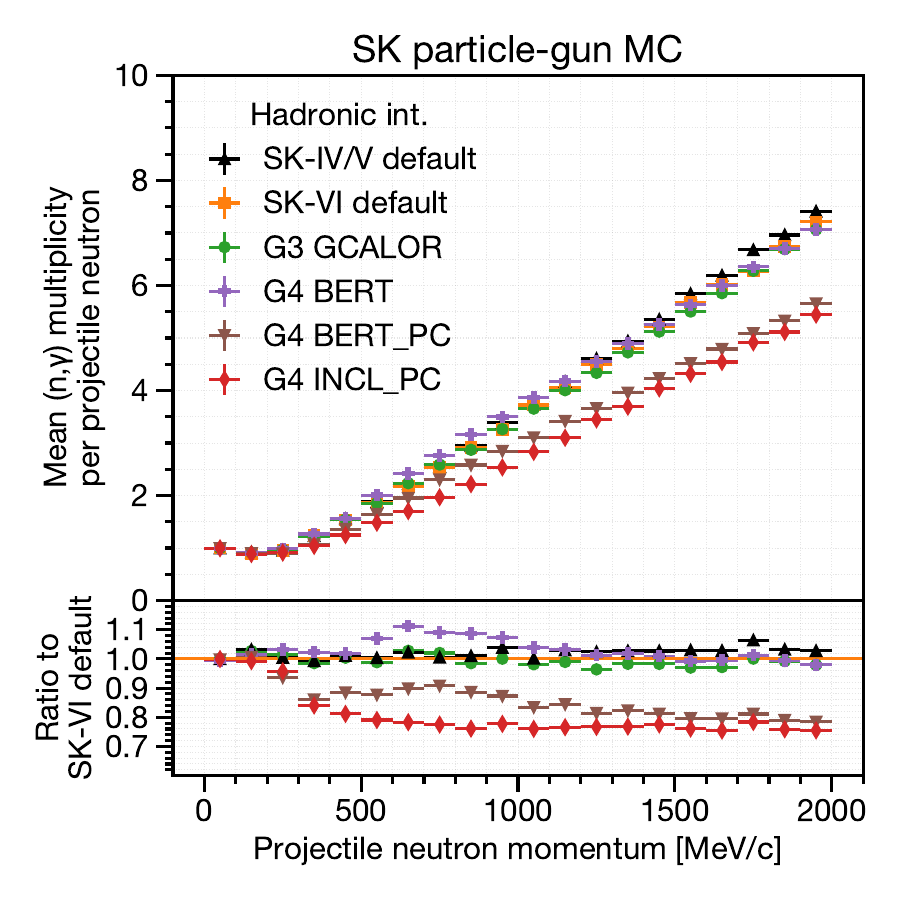}
\caption{\label{fig:n_pgun} Comparison of secondary hadron-nucleus interaction models: Average $(n,\gamma)$ multiplicity per projectile neutron momentum bin, based on simulations using a single-neutron particle-gun setup in water.}
\end{figure}

\begin{figure}
\includegraphics[width=0.92\columnwidth]{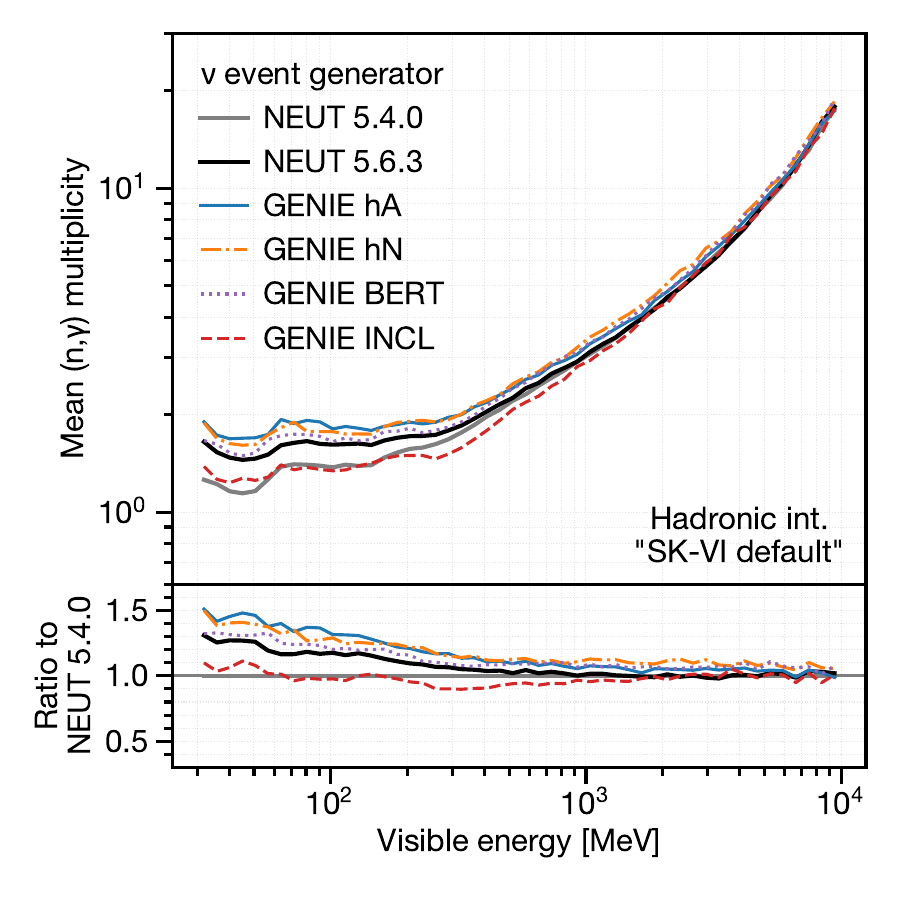}
\caption{\label{fig:pifsi_comp} Comparison of the predicted average $(n,\gamma)$ multiplicity predicted by the neutrino event generator options paired with \texttt{SK-VI default}.}
\end{figure}

\begin{figure}
\includegraphics[width=0.9\columnwidth]{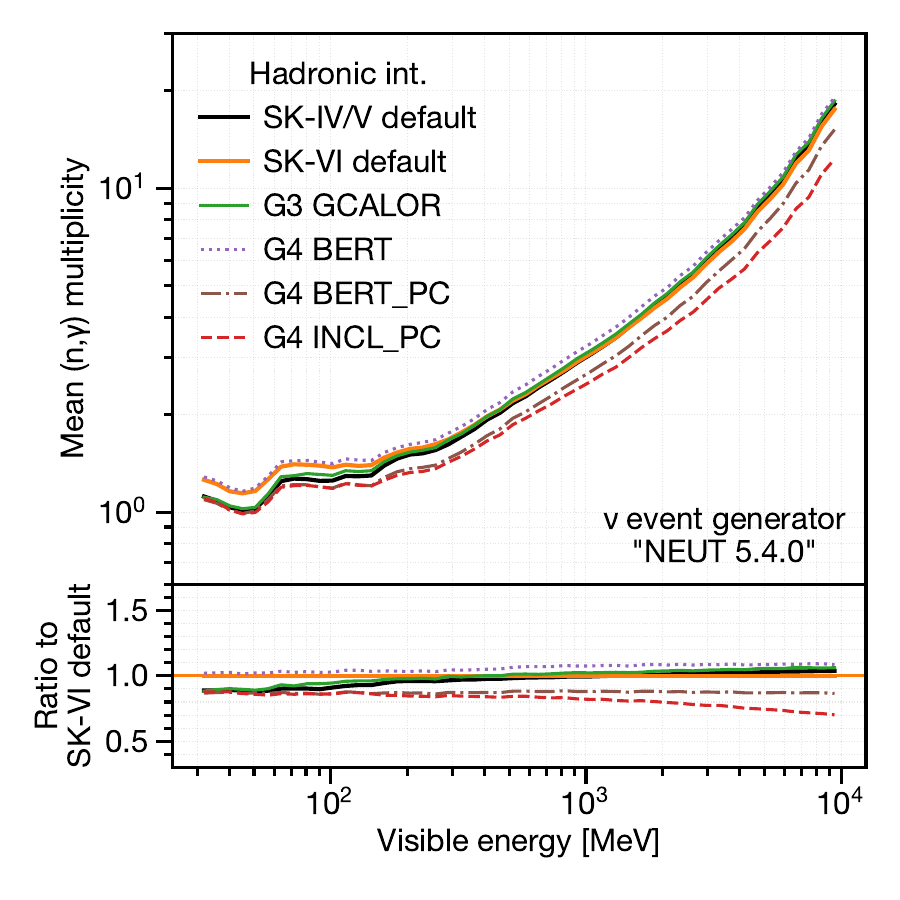}
\caption{\label{fig:si_comp} Comparison of the predicted average $(n,\gamma)$ multiplicity predicted by the secondary hadron-nucleus interaction models paired with \texttt{NEUT 5.4.0}.}
\end{figure}

%\begin{figure}
%\centering
%\includegraphics[width=0.85\linewidth]{figures/6_models/modelcomp_xylog.pdf}
%\caption{\label{fig:model_pi_si_comp} Comparison of the neutrino event generator options (top) and secondary hadron-nucleus interaction models (bottom) on the predicted average $(n,\gamma)$ multiplicity. In the top figure, each neutrino event generator option is paired with the baseline secondary interaction model (\texttt{SK-VI default}). In the bottom figure, each model is paired with the baseline neutrino event generator option (\texttt{NEUT 5.4.0}).}
%\end{figure}

Figure \ref{fig:out_n_mom} compares outgoing neutron momentum distributions predicted by the six neutrino event generator options, highlighting notable differences below 1 GeV/c, which are further discussed in Section \ref{sec:discussion}.

Figure \ref{fig:n_pgun} shows the average signal multiplicities as a function of projectile neutron momentum for the six secondary hadron-nucleus interaction models, with variations of up to 40\%.

For each neutrino event generator, a full MC simulation equivalent to 50 years of atmospheric neutrino exposure was generated using the baseline detector model (\texttt{SK-VI default}). To account for variations in secondary hadron-nucleus interactions without additional full simulations, we precomputed average signal multiplicities for neutron-producing projectiles ($n$, $p$, $\pi^\pm$, and $\mu^-$) in water up to 10 GeV/c, based on particle-gun MC simulations. %Figure \ref{fig:n_pgun} shows an example tabulation for neutron projectiles.

Predictions were made by convolving projectile momentum distributions with momentum-to-multiplicity tables. For each event generator, ratios relative to the baseline detector simulation model (\texttt{SK-VI default}) were used to scale the signal multiplicities obtained from the full MC simulation. In configurations using either \texttt{NEUT 5.4.0} or \texttt{NEUT 5.6.3} with \texttt{SK-IV/V default}, the discrepancy between this prediction method and the 500-year exposure full MC simulation was within 5\% across all visible energy range and data subsamples.

Figures \ref{fig:pifsi_comp} and \ref{fig:si_comp} show the model predictions. In visible energy range above a few hundred MeV, the average $(n,\gamma)$ multiplicities are expected to increase linearly with visible energy. Differences across neutrino event generator configurations are more pronounced at visible energies below 1 GeV, while variability among secondary hadron-nucleus interaction models remains relatively constant. \texttt{GENIE INCL} predicts fewer neutrons than the other FSI models, as already suggested by Figure \ref{fig:out_n_mom}. \texttt{NEUT 5.4.0} predicts fewer neutrons than \texttt{NEUT 5.6.3}, due to a larger QE fraction. Additionally, the Geant4 Bertini cascade model with the Precompound model (\texttt{G4 BERT\_PC}) predicts fewer neutrons than when using the native de-excitation model (\texttt{G4 BERT}).

%As mentioned earlier, the Liège INC model resulted in the least neutron production, both when used in the neutrino event generator (coupled with ABLA07) and in the detector simulator (coupled with the Geant4 Precompound model).

\begin{table*}[htbp!]
    \caption{Summary of the number of atmospheric neutrino events (``$\nu$ events'') and detected neutron signals (``$n$ signals'') in the final sample. $\langle N \rangle_\text{overall}$ is the unbinned application of Equation \ref{nmult_eq}, followed by signal efficiency scale corrections with the factors given in Section \ref{sec:ambe}. These factors are specifically chosen to ensure consistency of $\langle N \rangle_\text{overall}$ across the three SK phases (see Appendix~\ref{appendix:phaseconsistency}). Errors for the observed $\langle N \rangle_\text{overall}$ are listed as statistical first, followed by systematic uncertainty. Other errors are statistical only. The bottom two rows show the expected $\langle N \rangle_\text{overall}$ and the true overall $(n,\gamma)$ multiplicity extracted from the baseline full MC simulations (\texttt{NEUT 5.4.0} with \texttt{SK-IV/V default} for SK-IV/V, and \texttt{SK-VI default} for SK-VI).}

\begin{ruledtabular}
    \centering
    \begin{tabular}{lccc}
         & SK-IV & SK-V & SK-VI  \\ \hline
      $\nu$ events & 29,942 & 4,231 & 5,203 \\
     Events/day & $9.23\pm0.05$ & $9.18\pm0.14$ & $9.22\pm0.13$  \\
     $n$ signals & 15,705 & 2,035 & 5,752   \\
     $n$ signals/event & $0.525\pm0.004$ & $0.481\pm0.011$ & $1.106\pm0.015$ \\
     Observed $\langle N \rangle_\text{overall}$& $2.49\pm0.03\pm0.23$ & $2.49\pm0.10\pm0.11$ & $2.49\pm0.06\pm0.05$ \\
     \hline
      Expected $\langle N \rangle_\text{overall}$ & 2.83 & 2.84 & 2.85 \\ 
    True $(n,\gamma)$ multiplicity & 2.85 & 2.85 & 2.86 \\
   
     %& 2.11$\pm$0.30 \\
    \end{tabular}
    \label{tab:evsummary}
\end{ruledtabular}
\end{table*}

%\begin{ruledtabular}
%    \centering
%    \begin{tabular}{lcccc}
%         & SK-IV & SK-V & SK-VI & SK-VI (Reference)  \\ \hline
%      $\nu$ events & 29,942 & 4,231 & 5,203 & Same as SK-VI\\
%     Events/day & $9.23\pm0.05$ & $9.18\pm0.14$ & $9.22\pm0.13$ & Same as SK-VI \\
%     $n$ signals & 15,705 & 2,035 & 5,752 & 4,412  \\
%     $n$ signals/event & $0.525\pm0.004$ & $0.481\pm0.011$ & $1.106\pm0.015$ & $0.848\pm0.017$\\
%     Observed $\langle N \rangle_\text{overall}$& $2.21\pm0.03\pm0.11$ & $2.46\pm0.10\pm0.11$ & $2.50\pm0.06\pm0.05$ & $2.49\pm0.06\pm0.05$ \\
%     \hline
%      Expected $\langle N \rangle_\text{overall}$ & 2.83 & 2.84 & 2.85 & 2.87 \\ 
%    True $(n,\gamma)$ multiplicity & 2.85 & 2.85 & 2.86 & Same as SK-VI \\
%   
%     %& 2.11$\pm$0.30 \\
%    \end{tabular}
%    \label{tab:evsummary}
%\end{ruledtabular}
%\end{table*}

\section{Results}
\label{sec:results}

Table \ref{tab:evsummary} summarizes the number of atmospheric neutrino events and detected neutron signals in the final data sample. The overall average $(n,\gamma)$ multiplicity per neutrino event, $\langle N \rangle_\text{overall}$, is estimated using Equation \ref{nmult_eq} for all selected atmospheric neutrino events without binning.

\begin{figure}[t]
\includegraphics[width=0.9\columnwidth]{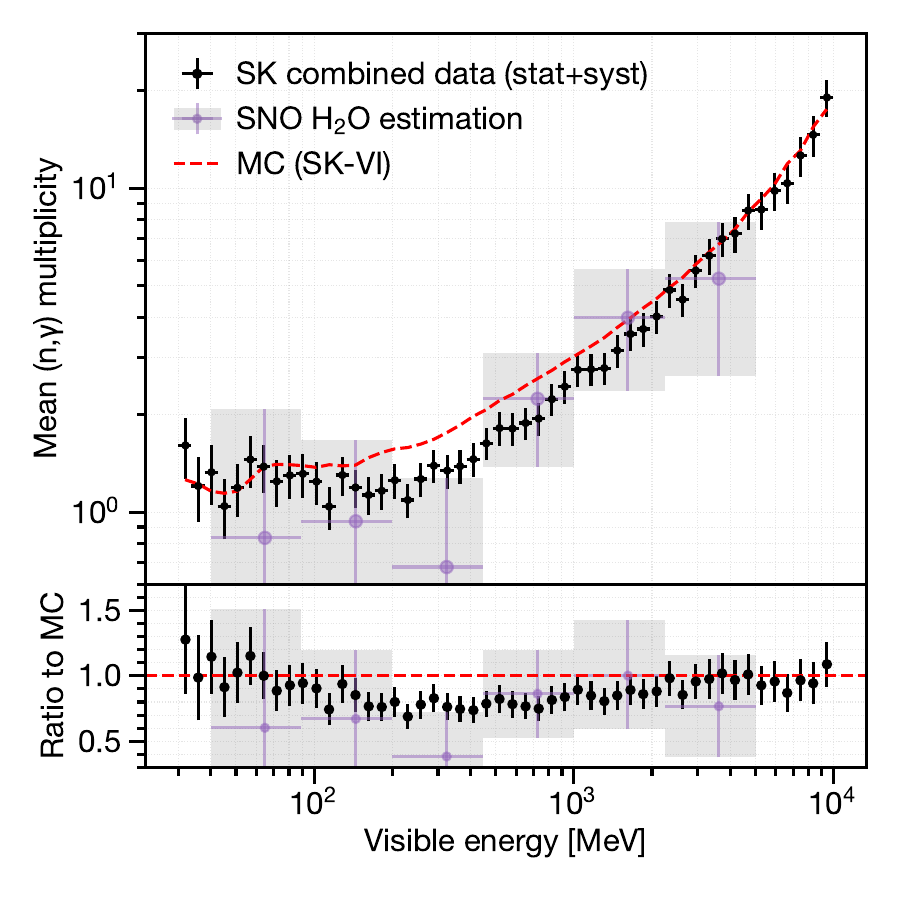}
\caption{\label{fig:result_sno} Average $(n,\gamma)$ multiplicity in atmospheric neutrino events as a function of their visible energy. Black data points represent combined SK data with statistical and systematic uncertainties. Purple crosses with shades are estimates for pure water based on SNO measurements \cite{sno_neutron_2019} using a $\text{D}_2\text{O}$ target. The red dashed line represents the true average $(n,\gamma)$ multiplicity from the baseline SK-IV MC simulation.}
\end{figure}

The combined data was compared with the baseline simulation and neutron production estimates in water from SNO \cite{sno_neutron_2019}, as shown in Figure \ref{fig:result_sno}. For visible energies above 200 MeV—where outgoing lepton Cherenkov rings are well reconstructed—a linear relationship between visible energy and the average $(n,\gamma)$ multiplicity was observed, as expected. While the data aligned well with the SNO estimate, it was 10–30\% lower than the baseline predictions in the sub-GeV energy range. A similar deficit was reported by SNO in heavy water \cite{sno_neutron_2019}, where their data was compared to GENIE 2.10.2 (hA FSI model) coupled with the Geant4 Bertini cascade model.

\begin{figure*}
\includegraphics[width=\linewidth]{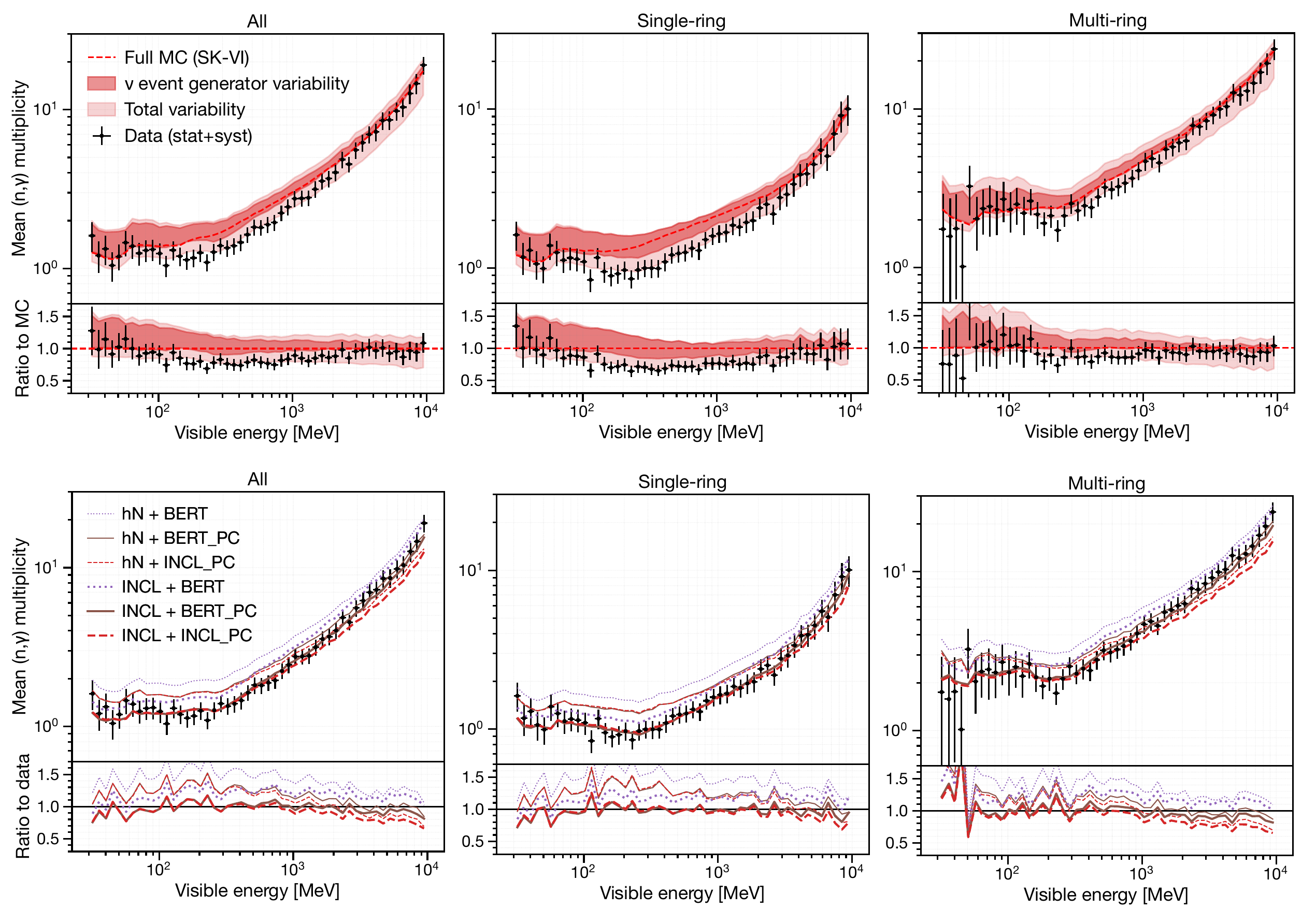}
\caption{\label{fig:nmult_comp} Comparison of various model predictions with the combined data, showing the average $(n,\gamma)$ multiplicity as a function of atmospheric neutrino event visible energy, for all events (left), single-ring events (middle), and multi-ring events (right). The black data points represent the combined data, including both statistical and systematic uncertainties. In the top panel, the combined data are compared to the prediction intervals of the models. Thicker shaded regions represent the range of predictions obtained by varying only the neutrino event generator options while keeping the secondary hadron-nucleus interaction model fixed to the baseline (\texttt{SK-VI default}). Lighter shaded regions indicate the broader range of predictions from all model combinations. The bottom panel shows corresponding plots for specific model choices that are based on GENIE and Geant4, including the one (\texttt{GENIE INCL} + \texttt{G4 BERT\_PC}) that shows good agreement with the data (see Table \ref{tab:chisq}).}

% The dashed red line represents the prediction from the full SK-VI MC simulation using the default setup. The thick red shaded area shows the range of predictions when varying the neutrino event generator options, while keeping the secondary spallation model fixed to the SK-VI default. The light red shaded area shows the range of predictions from all tested model combinations.}}
\end{figure*}

Predictions from various model combinations, as described in Section \ref{sec:models}, were compared with the combined data across different subsamples, as shown in Figure \ref{fig:nmult_comp}. The bottom panel of Figure \ref{fig:nmult_comp} contrasts the data with predictions from selected FSI models (\texttt{GENIE hN} and \texttt{INCL}) and secondary interaction models (\texttt{G4 BERT}, \texttt{G4 BERT\_PC}, and \texttt{G4 INCL\_PC}). The observed deficit in the [0.1, 0.4] GeV range was primarily found in the single-ring sample, which is expected to be dominated by CCQE interactions (see Figure \ref{fig:intbd}). This deficit was only reproduced when using models that predict fewer neutrons than the baseline for both FSI and secondary interactions, such as \texttt{GENIE INCL} and \texttt{G4 BERT\_PC}.

In the multi-ring sample at a few-GeV energies, there was significant variation in the predicted ``slope" of the average $(n,\gamma)$ multiplicity as a function of visible energy. In particular, \texttt{G4 INCL\_PC} predicted a smaller slope compared to \texttt{G4 BERT} and \texttt{G4 BERT\_PC}.

%Among all tested combinations, \texttt{GENIE INCL} and \texttt{G4 BERT\_PC} provide the best agreement with data, showing the lowest chi-squared statistic when considering statistical errors only.

%\begin{figure*}[htb!]
%\includegraphics[width=0.8\linewidth]{figures/7_results/chisq_modelcomp.pdf}
%\caption{\label{fig:chisq_modelcomp} Comparison of combined reduced $\chi^2$ values (Equation \ref{eq:chisq}) for single-ring and multi-ring data across various model predictions. Colors indicate different secondary hadron-nucleus interaction models. The unhatched area represents the normalization-free $\chi^2$ ($s$ minimized within $[0,2]$ without penalty), while the total height, including the hatched area, represents $\chi^2$ with a 10\% normalization constraint. The hatched area shows the contribution from this constraint.}
%\end{figure*}

The goodness of fit between model predictions and data for the entire visible energy range of [0.03, 10] GeV was quantified using the following chi-square ($\chi^2$) definition, considering only the two most dominant sources of uncertainty: statistical uncertainty and signal efficiency scale uncertainty.

\begin{equation}
    \chi^2 = \sum_i^{N_\text{bins}} \frac{(sO_i-E_i)^2}{\sigma_{\text{stat},i}^2}
    \label{eq:chisq}
\end{equation}
 Here, $i$ is the bin number, $O_i$ and $E_i$ represent the observed and expected average $(n,\gamma)$ multiplicity, respectively, and $\sigma_{\text{stat},i}$ is the statistical uncertainty in the $i$-th bin. The total number of bins, $N_\text{bins}$, is 20 per subsample for both single-ring and multi-ring events, resulting in 40 bins in total. The binning was optimized to ensure a sufficient number of events per bin so that the average $(n,\gamma)$ multiplicity follows a normal distribution. The normalization scale $s$ was allowed to float within $[0,2]$. Two $\chi^2$ calculations were performed: (1) a normalization-free $\chi^2$, where $s$ was minimized without constraint, and (2) a constrained $\chi^2$, where $s$ was minimized with a penalty term $(s-1)^2/\sigma_\text{scale}^2$ and an additional constraint $|s-1|<\sigma_\text{scale}$, with $\sigma_\text{scale}$ set to 10\%. 

 \begin{table*}
     \caption{Reduced $\chi^2$ (see Equation \ref{eq:chisq}) and $p$-values for each model combination. The signal efficiency scale $s$ was either freely minimized or constrained to 10\%. The fit has 38 fixed degrees of freedom. $p$-values assume Gaussian bin errors without bin correlations.}
    \centering
    \renewcommand{\arraystretch}{1.06}
    \begin{ruledtabular}
    \begin{tabular}{ccccccccc}
        \multirow{2}{*}{} & \multirow{2}{*}{$\nu$ generator} & \multirow{2}{*}{Secondary int.} & \multicolumn{2}{c}{Minimizing $s$} & \multicolumn{2}{c}{Reduced $\chi^2$} & \multicolumn{2}{c}{$p$-value} \\
        \cline{4-9}
        & & & Free $s$ & Constrained $s$ & Free $s$ & Constrained $s$ & Free $s$ & Constrained $s$ \\
        \hline \hline
        1 & \texttt{NEUT 5.4.0} & \texttt{SK-IV/V default} & 0.75 & 0.90 & 2.76 & 11.66 & $<10^{-4}$ & $<10^{-4}$ \\
        2 & \texttt{NEUT 5.4.0} & \texttt{SK-VI default} & 0.74 & 0.90 & 3.64 & 14.38 & $<10^{-4}$ & $<10^{-4}$ \\
        3 & \texttt{NEUT 5.4.0} & \texttt{G3 GCALOR} & 0.73 & 0.90 & 2.14 & 13.47 & $<10^{-4}$ & $<10^{-4}$ \\
        4 & \texttt{NEUT 5.4.0} & \texttt{G4 BERT} & 0.70 & 0.90 & 2.14 & 20.76 & $<10^{-4}$ & $<10^{-4}$ \\
        5 & \texttt{NEUT 5.4.0} & \texttt{G4 BERT\_PC} & 0.84 & 0.90 & 2.73 & 3.68 & $<10^{-4}$ & $<10^{-4}$ \\
        6 & \texttt{NEUT 5.4.0} & \texttt{G4 INCL\_PC} & 0.89 & 0.90 & 5.08 & 5.11 & $<10^{-4}$ & $<10^{-4}$ \\
        7 & \texttt{NEUT 5.6.3} & \texttt{SK-IV/V default} & 0.72 & 0.90 & 2.92 & 15.94 & $<10^{-4}$ & $<10^{-4}$ \\
        8 & \texttt{NEUT 5.6.3} & \texttt{SK-VI default} & 0.71 & 0.90 & 4.40 & 19.97 & $<10^{-4}$ & $<10^{-4}$ \\
        9 & \texttt{NEUT 5.6.3} & \texttt{G3 GCALOR} & 0.71 & 0.90 & 2.24 & 18.30 & $<10^{-4}$ & $<10^{-4}$ \\
        10 & \texttt{NEUT 5.6.3} & \texttt{G4 BERT} & 0.67 & 0.90 & 2.53 & 27.88& $<10^{-4}$ & $<10^{-4}$ \\
        11 & \texttt{NEUT 5.6.3} & \texttt{G4 BERT\_PC} & 0.81 & 0.90 & 3.38 & 5.87 & $<10^{-4}$ & $<10^{-4}$ \\
        12 & \texttt{NEUT 5.6.3} & \texttt{G4 INCL\_PC} & 0.86 & 0.90 & 6.38 & 6.86 & $<10^{-4}$ & $<10^{-4}$ \\
        13 & \texttt{GENIE hA} & \texttt{SK-IV/V default} & 0.73 & 0.90 & 3.17 & 14.83& $<10^{-4}$ & $<10^{-4}$ \\
        14 & \texttt{GENIE hA} & \texttt{SK-VI default} & 0.67 & 0.90 & 6.71 & 30.43& $<10^{-4}$ & $<10^{-4}$ \\
        15 & \texttt{GENIE hA} & \texttt{G3 GCALOR} & 0.71 & 0.90 & 2.47 & 17.69 & $<10^{-4}$ & $<10^{-4}$ \\
        16 & \texttt{GENIE hA} & \texttt{G4 BERT} & 0.65 & 0.90 & 3.73 & 36.46 & $<10^{-4}$ & $<10^{-4}$ \\
        17 & \texttt{GENIE hA} & \texttt{G4 BERT\_PC} & 0.77 & 0.90 & 5.29 & 10.91 & $<10^{-4}$ & $<10^{-4}$ \\
        18 & \texttt{GENIE hA} & \texttt{G4 INCL\_PC} & 0.80 & 0.90 & 9.64 & 12.93 & $<10^{-4}$ & $<10^{-4}$ \\
        19 & \texttt{GENIE hN} & \texttt{SK-IV/V default} & 0.71 & 0.90 & 1.75 & 16.50 & 0.0030 & $<10^{-4}$ \\
        20 & \texttt{GENIE hN} & \texttt{SK-VI default} & 0.66 & 0.90 & 4.32 & 31.29 & $<10^{-4}$ & $<10^{-4}$ \\
        21 & \texttt{GENIE hN} & \texttt{G3 GCALOR} & 0.69 & 0.90 & 1.48 & 20.77 & 0.028 & $<10^{-4}$ \\
        22 & \texttt{GENIE hN} & \texttt{G4 BERT} & 0.63 & 0.90 & 2.29 & 40.19 & $<10^{-4}$ & $<10^{-4}$ \\
        23 & \texttt{GENIE hN} & \texttt{G4 BERT\_PC} & 0.76 & 0.90 & 3.13 & 10.10 & $<10^{-4}$ & $<10^{-4}$ \\
        24 & \texttt{GENIE hN} & \texttt{G4 INCL\_PC} & 0.79 & 0.90 & 6.48 & 10.34 & $<10^{-4}$ & $<10^{-4}$ \\
        25 & \texttt{GENIE BERT} & \texttt{SK-IV/V default} & 0.75 & 0.90 & 1.22 & 9.79 & 0.17 & $<10^{-4}$ \\
        26 & \texttt{GENIE BERT} & \texttt{SK-VI default} & 0.70 & 0.90 & 3.31 & 21.35 & $<10^{-4}$ & $<10^{-4}$ \\
        27 & \texttt{GENIE BERT} & \texttt{G3 GCALOR} & 0.73 & 0.90 & 1.21 & 13.69 & 0.17 & $<10^{-4}$ \\
        28 & \texttt{GENIE BERT} & \texttt{G4 BERT} & 0.67 & 0.90 & 1.79 & 27.70 & 0.0020 & $<10^{-4}$ \\
        29 & \texttt{GENIE BERT} & \texttt{G4 BERT\_PC} & 0.80 & 0.90 & 2.56 & 6.06 & $<10^{-4}$ & $<10^{-4}$ \\
        30 & \texttt{GENIE BERT} & \texttt{G4 INCL\_PC} & 0.83 & 0.90 & 5.60 & 7.12 & $<10^{-4}$ & $<10^{-4}$ \\
        31 & \texttt{GENIE INCL} & \texttt{SK-IV/V default} & 0.87 & 0.90 & 0.84 & 1.21 & 0.74 & 0.18 \\
        32 & \texttt{GENIE INCL} & \texttt{SK-VI default} & 0.80 & 0.90 & 1.33 & 4.59 & 0.082 & $<10^{-4}$ \\
        33 & \texttt{GENIE INCL} & \texttt{G3 GCALOR} & 0.83 & 0.90 & 1.36 & 2.75 & 0.067 & $<10^{-4}$ \\
        34 & \texttt{GENIE INCL} & \texttt{G4 BERT} & 0.76 & 0.90 & 0.87 & 8.55 & 0.71 & $<10^{-4}$ \\
        35 & \texttt{GENIE INCL} & \texttt{G4 BERT\_PC} & 0.92 & 0.92 & 0.95 & 0.96 & 0.56 & 0.53 \\
        36 & \texttt{GENIE INCL} & \texttt{G4 INCL\_PC} & 0.97 & 0.97 & 2.74 & 2.74 & $<10^{-4}$ & $<10^{-4}$ \\
    \end{tabular} 
    \end{ruledtabular} 
    \label{tab:chisq}
    \end{table*}

Table \ref{tab:chisq} compares the reduced $\chi^2$ values across different model combinations. Among the neutrino event generator options, \texttt{GENIE INCL} showed the lowest $\chi^2$ for both free and constrained $s$. Among the secondary interaction models, \texttt{G4 BERT\_PC} provided the best agreement with data under the 10\% normalization constraint. The combination of these two models yielded the best overall agreement under the same constraint. Other secondary interaction models such as \texttt{SK-IV/V default} showed reasonable agreement when coupled with \texttt{GENIE INCL} or when normalization was allowed to float. In contrast, \texttt{G4 INCL\_PC} showed higher unconstrained $\chi^2$ values, due to the shallower slope prediction for the multi-ring events. 

Model performance, separated into contributions from target nucleus FSI and secondary interactions, is further visualized using effective metrics in Appendix \ref{appendix:slopeintercept}.

\section{Discussion}
\label{sec:discussion}

%\begin{figure}
%    \centering
%    \includegraphics[width=0.9\columnwidth]{figures/7_results/sec_n_mom.pdf}
%    \caption{Momentum distributions of all secondary neutron tracks produced in particle-gun simulations of neutron propagation in water, as described in Section \ref{sec:models}, using selected secondary hadron-nucleus interaction models based on Geant4. The initial projectile neutron momentum was uniformly distributed in the [0, 10] GeV/c range.}
%    \label{fig:sec_n_mom}
%\end{figure}

Overall, our data favor models that predict relatively lower neutron production for single-ring sub-GeV events and higher neutron production for multi-ring multi-GeV events. In this section, we examine key features of these models that influence neutron production predictions, focusing on Figure \ref{fig:out_n_mom}, which illustrates differences in neutron momentum distributions across the tested FSI models. By qualitatively comparing the features of these models, we discuss potential factors contributing to the observed discrepancies in predictions.

\subsection{Nuclear de-excitation}

Nuclear de-excitation models are crucial for predicting total neutron production, as they govern low-energy neutron emission per nuclear interaction. This is evident when comparing \texttt{G4 BERT} and \texttt{G4 BERT\_PC}, which differ only in their de-excitation treatment. For instance, the inaccuracy of the Geant4 Bertini cascade model prediction is indicated by the distinct large neutron peak below 250 MeV/c in \texttt{GENIE BERT} (Figure \ref{fig:out_n_mom}), compared to the other models such as ABLA07 in \texttt{GENIE INCL} or the native de-excitation model in \texttt{NEUT}. (The \texttt{GENIE hA} and \texttt{hN} options omit nuclear de-excitation entirely.) Coupling the Geant4 Precompound model with the Geant4 Bertini cascade model significantly reduces neutron production across all visible energy ranges, leading to much better agreement with data (Figure \ref{fig:nmult_comp}). 

In the Geant4 Bertini cascade model, neutron emission follows the Weisskopf statistical evaporation model \cite{weisskopf}. A key input determining neutron emission probability is the inverse reaction (neutron absorption) cross section. The inaccuracy of the Bertini model and its variants in predicting neutron evaporation may stem from its use of the simplified cross section parameterization by Dostrovsky \cite{dostrovsky}. In comparison, the Geant4 Precompound model uses a more refined parameterization based on the nuclear optical model potential \cite{chatterjee}, fitted to a broader dataset \cite{kalbach}, resulting in lower evaporation rates particularly for low-energy neutrons. Additionally, the Geant4 Precompound model features tuned level density parameters and allows Fermi breakup in nuclei up to mass 16 \cite{g4preco}, including oxygen, further reducing isolated neutrons that contribute to $(n,\gamma)$ signals.

%\subsection{Pauli blocking}
\subsection{Nuclear in-medium effects}

A key distinction among the FSI cascade models is their treatment of Pauli blocking, which prevents nucleons from occupying the same quantum state, as dictated by the Pauli exclusion principle. For instance, \texttt{NEUT 5.4.0}, \texttt{NEUT 5.6.3} and \texttt{GENIE BERT} implement ``strict'' Pauli blocking, which forbids all nucleon scattering below the predefined oxygen Fermi momentum around 225 MeV/c, modeling the nucleus as a degenerate Fermi gas. This approach results in a characteristic dip in the outgoing neutron momentum distributions for \texttt{NEUT 5.4.0} and \texttt{NEUT 5.6.3} within the corresponding energy range, as shown in Figure \ref{fig:out_n_mom}. The Liège INC model applies a probabilistic approach for nucleon collisions throughout the cascade (except for the initial collision), accounting for nucleon holes in the surrounding phase space volume \cite{incl4, incl_pb}. This results in smoother neutron distributions near the Fermi momentum, as seen in Figure \ref{fig:out_n_mom}. In contrast, the \texttt{GENIE hA} and \texttt{hN} models omit Pauli blocking entirely, leading to significantly higher predicted neutron production than \texttt{NEUT 5.4.0}, \texttt{NEUT 5.6.3}, and \texttt{GENIE INCL}.

%\subsection{Ouclear in-medium considerations}

In the 0.2--1 GeV/$c$ range, the number of neutrons predicted by the Liège INC and Geant4 Bertini cascade models is lower than that from the \texttt{GENIE hA}, \texttt{hN}, and \texttt{NEUT} nucleon FSI models, as illustrated in Figure\ref{fig:out_n_mom}. One contributing factor may be the inclusion of cluster formation—such as deuterons, tritons, and $\alpha$ particles—among cascade products, which reduces the number of isolated neutrons contributing to the observable signal. Ref.\ \cite{ershova} also identifies cluster formation as a major mechanism behind the reduction of outgoing protons in the Liège INC model. Additional in-medium effects, considered either explicitly or implicitly in the two models—such as short-range nucleon repulsion and local nuclear density reduction after each collision—further reduce the number of nucleon-nucleon collisions during the cascade.

%The same Ref.\ \cite{ershova} suggests that protons are likely to be absorbed inside the nucleus in the Liège INC model at the similar rate those lost by cluster formation. Ref.\ \cite{ershova} has suggested that the Liège INC model has shown lower proton transparency than the generator FSI model (implemented in NuWro \cite{nuwro}), while The correlation of nuclear momentum  transmission and reflection  of outgoing nucleons at the nuclear surface is determined by the nuclear potential which in turn affects nucleon transparency. 

%In these events, secondary hadron-nucleus interactions have a smaller role in determining neutron multiplicity due to the suppression of secondary interactions by limited hadron momentum.

\subsection{\texorpdfstring{$\pi^\pm$ production}{Charged pion production}}

Accurate modeling of pion-nucleus interactions is crucial for predicting neutron production in multi-ring events with energetic pions (see Figures \ref{fig:n_contribution} and \ref{fig:n_source_by_particle} in Appendix \ref{appendix:n_source}). The difference in predicted neutron production in the few-GeV multi-ring sample between the Liège INC model (\texttt{G4 INCL\_PC}) and the Geant4 Bertini cascade model (\texttt{G4 BERT\_PC}), as observed in Figure \ref{fig:nmult_comp}, can be attributed to differences in pion production cross sections for pion projectiles. Our results are consistent with those of Ref.\ \cite{mancusi}, which report that the Liège INC model tends to predict lower inclusive $\pi^\pm$ production cross sections on light nuclei (mass number $A < 20$) compared to the Geant4 Bertini cascade model. In addition, the pion potential within the nucleus differs between the two models. In the Liège INC model, the pion potential is generally deeper \cite{incl_pion} than the constant 7 MeV potential used in the Geant4 Bertini cascade model \cite{g4_bertini}, which impacts pion transparency. For instance, Ref.\ \cite{dytman_validation} demonstrated that the Liège INC model shows lower pion transparency for pions with momenta below 300 MeV/c compared to the GENIE hA, hN, or NEUT pion FSI models. Also, using the NEUT pion FSI routine for low-energy $\pi^\pm$ transport in water slightly improves agreement with the data (e.g., \texttt{SK-IV/V default} vs.\ \texttt{G3 GCALOR}).

\subsection{Other considerations}

%Variations in low-energy neutron reaction cross section datasets (e.g., ENDF/B-VI used in \texttt{SK-IV/V default} vs.\ ENDF/B-VII.1 used in \texttt{SK-VI default}) result in a 5--10\% difference in the average $(n,\gamma)$ multiplicities for visible energy below 300 MeV, as shown in Figure \ref{fig:si_comp}. 

Neutrons produced from $\mu^-$ captures contribute 5--10\% of the average $(n,\gamma)$ multiplicity for visible energies below 300 MeV, as shown in Figure~\ref{fig:si_comp} (e.g., \texttt{SK-VI default}, which includes $\mu^-$ capture simulation, vs.\ \texttt{SK-IV/V default}, which does not).

Our model predictions do not account for interactions of $\nu_\tau$ and $\bar{\nu}_\tau$ from oscillations, nor neutrinos with energies below 100 MeV. However, their contributions are expected to be smaller than our uncertainty budget in the upper and lower ends of the visible energy range, where their effects become relevant, respectively.

%\clearpage

\begin{table*}
\caption{Summary of interaction models for final-state hadrons from neutrino–nucleus interactions used in the neutrino event generators, including their component choices and the resulting minimum reduced $\chi^2$ values when compared to data (with a 10\% constraint on the signal efficiency scale). Footnotes highlight notable features of specific models or components.}
\centering
\renewcommand{\arraystretch}{1.15}
\begin{ruledtabular}
\begin{tabular}{lccc}
Model                               & Hadron transport               & Nuclear de-excitation          & Minimum $\chi^2$ (coupled model) \\ \hline
\texttt{NEUT 5.4.0}                 & GCALOR Bertini\footnotemark[1] & Refs.\ \cite{160_occ_prob, 16o_knockout_br}                           & 3.68 (\texttt{G4 BERT\_PC})      \\
\texttt{NEUT 5.6.3}\footnotemark[2] & GCALOR Bertini\footnotemark[1] & Refs.\ \cite{160_occ_prob, 16o_knockout_br}                           & 5.87 (\texttt{G4 BERT\_PC})      \\
\texttt{GENIE hA}                   & INTRANUKE/hA\footnotemark[3]                   & -                              & 10.91 (\texttt{G4 BERT\_PC})     \\
\texttt{GENIE hN}                   & INTRANUKE/hN\footnotemark[3]                   & -                              & 10.10 (\texttt{G4 BERT\_PC})     \\
\texttt{GENIE BERT}                 & Geant4 Bertini\footnotemark[4] & Geant4 Bertini\footnotemark[5] & 6.06 (\texttt{G4 BERT\_PC})      \\
\texttt{GENIE INCL}                 & Geant4 INCL++\footnotemark[4]\footnotemark[6]  & ABLA07                         & 0.96 (\texttt{G4 BERT\_PC})     
\end{tabular}
\end{ruledtabular}

% Define the unified footnotes
\footnotetext[1]{$\pi^\pm$ inelastic scattering below 500 MeV/$c$ are modeled based on Refs.\ \cite{salcedo_oset, neut_pi_fsi_tune}.}
\footnotetext[2]{Predicts a lower fraction of QE interactions compared to \texttt{NEUT 5.4.0}, due to a modified treatment of nuclear binding energy.}
\footnotetext[3]{Pauli blocking is not included.}
\footnotetext[4]{Includes cluster formation and nucleon correlations that effectively reduce collisions, in contrast to GCALOR Bertini.}
\footnotetext[5]{Predicts larger neutron evaporation compared to the NEUT model or ABLA07.}
\footnotetext[6]{Includes a probabilistic treatment of Pauli blocking and predicts smaller $\pi$-induced $\pi^\pm$ production than Geant4 Bertini.}

\label{tab:summary_fsi}
\end{table*}

\begin{table*}
\caption{Summary of secondary hadron–nucleus interaction models used in the detector simulator, including their component choices and the resulting minimum reduced $\chi^2$ values when compared to data (with a 10\% constraint on the signal efficiency scale). Footnotes highlight notable features of specific models or components.}
\centering
\renewcommand{\arraystretch}{1.15}
\begin{ruledtabular}
\begin{tabular}{lccccc}
Model             & Hadron transport      & Nuclear de-excitation     & $\mu^-$ capture  & Minimum $\chi^2$ (coupled model) \\ \hline
\texttt{SK-IV/V default}   & GCALOR Bertini\footnotemark[1]    &GCALOR Bertini  & -                          & 1.21 (\texttt{GENIE INCL})      \\
\texttt{SK-VI default}  & GCALOR Bertini\footnotemark[1]   &GCALOR Bertini & Geant4 Bertini                           & 4.59 (\texttt{GENIE INCL})      \\
\texttt{G3 GCALOR}    & GCALOR Bertini   &GCALOR Bertini  & -                             & 2.75 (\texttt{GENIE INCL})     \\
\texttt{G4 BERT}   & Geant4 Bertini\footnotemark[2]     &Geant4 Bertini   & Geant4 Bertini                           & 8.55 (\texttt{GENIE INCL})     \\
\texttt{G4 BERT\_PC}  & Geant4 Bertini\footnotemark[2]   &Geant4 Precompound\footnotemark[3] & Geant4 Bertini & 0.96 (\texttt{GENIE INCL})      \\
\texttt{G4 INCL\_PC}   & Geant4 INCL++\footnotemark[2]\footnotemark[4]   & Geant4 Precompound\footnotemark[3]  & Geant4 Bertini                        & 2.74 (\texttt{GENIE INCL})     
\end{tabular}
\end{ruledtabular}

% Define the unified footnotes
\footnotetext[1]{$\pi^\pm$ inelastic scattering below 500 MeV/$c$ are modeled based on Refs.\ \cite{salcedo_oset, neut_pi_fsi_tune}.}
\footnotetext[2]{Includes cluster formation and nucleon correlations that effectively reduce collisions, in contrast to GCALOR Bertini.}
\footnotetext[3]{Predicts smaller neutron evaporation compared to Geant4 Bertini.}
\footnotetext[4]{Includes a probabilistic treatment of Pauli blocking and predicts smaller $\pi$-induced $\pi^\pm$ production than Geant4 Bertini.}
%\footnotetext[3]{Includes in-medium effects compared to GCALOR Bertini or INTRANUKE/hA or hN}
%\footnotetext[4]{Too much statistical neutron evaporation}
\label{tab:summary_si}
\end{table*}

Tables \ref{tab:summary_fsi} and \ref{tab:summary_si} summarize the components of each tested model and their agreement with the data. The differences in model components indicate that accurately describing the data requires both moderate evaporation of low-energy neutrons (as modeled in NEUT, ABLA07, and Geant4 Precompound) and suppression of medium-energy neutrons in the [0.3, 1] GeV/$c$ range via in-medium effects such as Pauli blocking, cascade step correlations, and cluster formation. Among the tested neutrino event generator configurations, only \texttt{GENIE INCL} satisfies both conditions as seen in Figure \ref{fig:out_n_mom}. For secondary hadron–nucleus interaction models, \texttt{G4 BERT\_PC} and \texttt{G4 INCL\_PC} meet these requirements, with \texttt{G4 BERT\_PC} performing better at higher energies due to its relatively larger $\pi^\pm$ production. These features make \texttt{GENIE INCL} and \texttt{G4 BERT\_PC} the most successful model combinations for reproducing the data. Additionally, the native de-excitation routine in the Geant4 Bertini cascade model significantly overpredicts neutron evaporation; thus, caution is advised when using it for neutron multiplicity predictions.

\section{Conclusions and prospects}
\label{sec:summary}

Accurately modeling neutron production in neutrino interactions is essential for characterizing incoming neutrinos, which is crucial for advancing precision measurements of neutrino oscillation parameters and rare event searches involving neutron tagging. Recent studies have suggested potential inaccuracies in the modeling of secondary hadron-nucleus interactions.

This paper reports a measurement of total neutron production following atmospheric neutrino interactions within the water volume of the Super-Kamiokande (SK) detector. Atmospheric neutrino events were binned by their electron-equivalent ``visible energy,'' a semi-calorimetric proxy for neutrino momentum transfer, in the range of [0.03, 10] GeV. The dominant systematic uncertainty was the roughly 10\% uncertainty in the overall signal efficiency scale. The average neutron capture multiplicity in each visible energy bin was compared against predictions from various combinations of neutrino event generators and secondary hadron-nucleus interaction models. 

Our data provides strong discriminative power for evaluating these models, with predictions varying by up to 50\%. Two key observations were made: a reduction in neutron production for sub-GeV single-ring events, where secondary hadron interactions are minimal, and a nearly linear increase in neutron production with visible energy, especially in the multi-GeV region. The first observation highlights the need for moderate neutron evaporation and consideration of nuclear in-medium effects. The second is more sensitive to pion production in cascade models, which become important at higher energies.

This study highlights the crucial role of hadron-nucleus interaction models in accurately predicting total neutron production from neutrino interactions. The observed discrepancy with the evaporation model \cite{dostrovsky} commonly used in Bertini-based cascade models aligns with previous studies, including neutron measurements with NCQE selections \cite{sakai} at SK and secondary $\gamma$-ray measurements using Germanium detectors with a neutron beam on a water target \cite{ashida_n16o, tano_n16o, e525}. Our data also supports the selection of specific models, such as \texttt{GENIE INCL} and \texttt{G4 BERT\_PC}, which reduce the uncertainty in total neutron production from atmospheric neutrino events to around 10\%. These models may help explain the deficits in total neutron capture signals reported by T2K \cite{t2k_neutron_2019} and SNO \cite{sno_neutron_2019}, and offer insights into the low-energy neutron signal deficits observed in MINERvA \cite{mv_neutron_2019}. Lastly, this work will improve the use of neutron tagging for identifying atmospheric antineutrino events at SK, enhancing sensitivity to neutrino mass ordering and searches for proton decays and the diffuse supernova neutrino background.

The data and model predictions to produce Figure \ref{fig:nmult_comp} and Table \ref{tab:chisq} are openly available \cite{data_release}.

\begin{acknowledgments}
We gratefully acknowledge the cooperation of the Kamioka Mining and Smelting Company.
The Super-Kamiokande experiment has been built and operated from funding by the 
Japanese Ministry of Education, Culture, Sports, Science and Technology; the U.S.
Department of Energy; and the U.S. National Science Foundation. Some of us have been 
supported by funds from the National Research Foundation of Korea (NRF-2009-0083526,
NRF-2022R1A5A1030700, NRF-2202R1A3B1078756) funded by the Ministry of Science, 
Information and Communication Technology (ICT); the Institute for 
Basic Science (IBS-R016-Y2); and the Ministry of Education (2018R1D1A1B07049158,
2021R1I1A1A01042256, 2021R1I1A1A01059559, RS-2024-00442775);
the Japan Society for the Promotion of Science; the National
Natural Science Foundation of China under Grants No.12375100; the Spanish Ministry of Science, 
Universities and Innovation (grant PID2021-124050NB-C31); the Natural Sciences and 
Engineering Research Council (NSERC) of Canada; the Scinet and Westgrid consortia of
Compute Canada; 
the National Science Centre (UMO-2018/30/E/ST2/00441 and UMO-2022/46/E/ST2/00336) 
and the Ministry of  Science and Higher Education (2023/WK/04), Poland;
the Science and Technology Facilities Council (STFC) and
Grid for Particle Physics (GridPP), UK; the European Union's 
Horizon 2020 Research and Innovation Programme under the Marie Sklodowska-Curie grant
agreement no.754496; H2020-MSCA-RISE-2018 JENNIFER2 grant agreement no.822070, H2020-MSCA-RISE-2019 SK2HK grant agreement no. 872549; 
and European Union's Next Generation EU/PRTR  grant CA3/RSUE2021-00559; 
the National Institute for Nuclear Physics (INFN), Italy.

\end{acknowledgments}

\appendix

\section{\texorpdfstring{$\langle N \rangle_\text{overall}$ consistency across SK phases}{Overall average neutron capture multiplicity consistency across SK phases}}
\label{appendix:phaseconsistency}

When applying a signal efficiency correction based on the weighted mean of Am/Be data-to-MC efficiency ratios—as shown in Figure~\ref{fig:pospvx} and described in Section~\ref{sec:ambe}—the efficiency-corrected estimate of total neutron production $\langle N \rangle_\text{overall}$ was $2.21 \pm 0.03\text{ (stat)} \pm 0.11\text{ (syst)}$ in SK-IV, lower than the values observed in the other two phases: $2.46 \pm 0.10 \pm 0.11$ for SK-V and $2.50 \pm 0.06 \pm 0.05$ for SK-VI. This discrepancy may be due to unaccounted systematic uncertainties in SK-IV, which had the longest data-taking period (nearly 10 years) but only a limited amount of calibration data to constrain the entire duration. In contrast, SK-VI—with a much shorter runtime (about 2 years)—periodically took calibration data across various positions within the detector volume. Thus, the signal efficiency scale for SK-VI is considered more reliable than that of SK-IV.

To test the robustness of the signal efficiency correction derived from the Am/Be calibration data and simulation, we applied an alternative neutron signal detection algorithm to both the calibration and atmospheric neutrino datasets in SK-VI. This algorithm uses a likelihood-based vertex fitter \cite{bonsai} to reconstruct the Gd$(n,\gamma)$ vertex independently of the neutrino interaction vertex, making it more resilient to uncertainties in outgoing neutron kinematics that affect the detection efficiency in the baseline algorithm described in Section~\ref{sec:ntagalgo}. Using this reference algorithm, we obtained $\langle N \rangle_\text{overall} = 2.49 \pm 0.06 \pm 0.05$ for SK-VI, consistent with the baseline result.

\begin{figure*}[p!]
\includegraphics[width=0.9\linewidth]{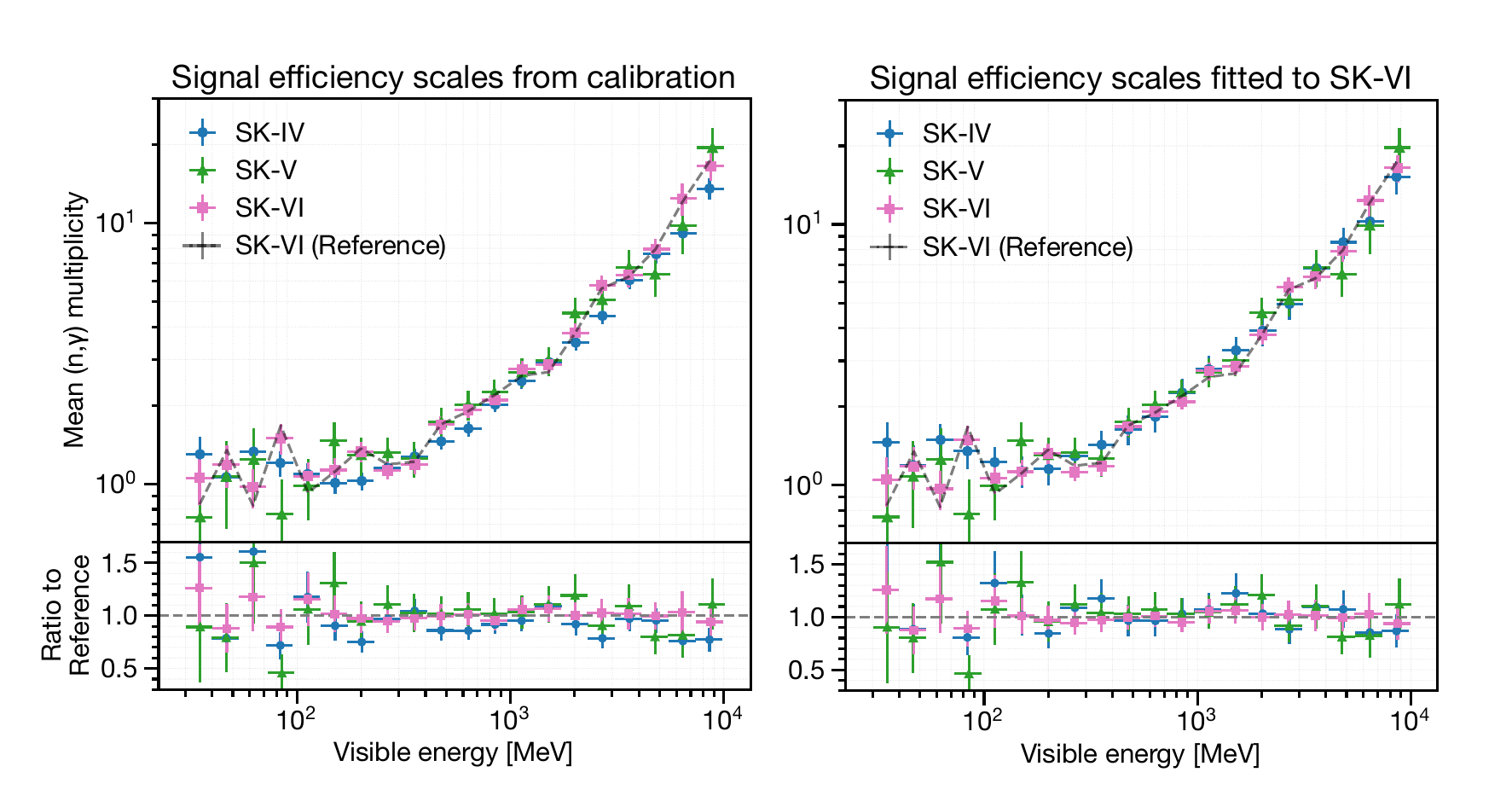}
\caption{\label{fig:abscale_phasecomp} Average $(n,\gamma)$ multiplicities plotted against neutrino event visible energy across SK phases. The left plot uses signal efficiency scales from Am/Be calibration (Section \ref{sec:ambe}), while the right plot adjusts SK-IV and SK-V scales to match $\langle N \rangle _\text{total}$ of SK-VI. Error bars include statistical and systematic uncertainties, with the right plot also accounting for scale differences. The dashed ``SK-VI (Reference)'' line shows SK-VI data using a neutron-energy-independent reference algorithm.}
\end{figure*}

Relying on the signal efficiency scale obtained for SK-VI, we adjust the correction factors for all phases such that $\langle N \rangle_\text{overall}$ aligns with the SK-VI reference value of 2.49. The difference between the calibration-based and phase-consistency-based correction factors is treated as an independent systematic uncertainty in the signal efficiency scale and is included in the uncertainty budget shown in Figure \ref{fig:fracerr}. This yields a final $\langle N \rangle_\text{overall}$ of $2.49 \pm 0.03\text{ (stat)} \pm 0.26\text{ (syst)}$, corresponding to a total systematic uncertainty of approximately 10\%. Figure \ref{fig:abscale_phasecomp} illustrates the impact of the two signal efficiency correction approaches: one based on Am/Be neutron source calibration and the other based on phase consistency.

\section{Effective metrics for model evaluation}
\label{appendix:slopeintercept}

\begin{figure*}[htbp!]
\includegraphics[width=\linewidth]{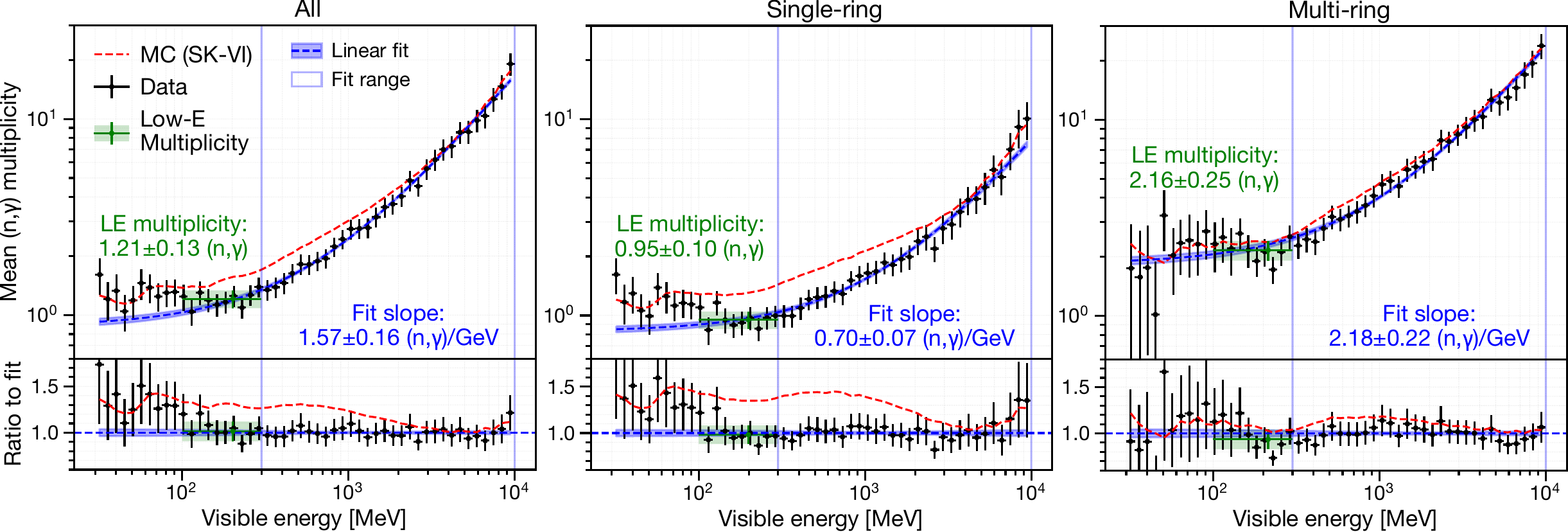}
\caption{\label{fig:lincomb} Average $(n,\gamma)$ multiplicity as a function of visible energy for neutrino events: all events (left), single-ring events (center), and multi-ring events (right). Red dashed lines show predictions from the full MC simulation in the SK-VI baseline setup. Blue dashed lines represent fitted linear functions with 1$\sigma$ prediction intervals shaded in blue, with the fit covering the energy range [0.3, 10] GeV. Green points represent low-energy multiplicity in the [0.1, 0.3] GeV range. Error bars include both statistical and systematic uncertainties.}
\end{figure*}

To separately evaluate the contributions from FSI within the target nucleus and downstream secondary interactions, we define two effective metrics.  The first metric is the low-energy (low-E) multiplicity as the average $(n,\gamma)$ multiplicity in the [0.1, 0.3] GeV visible energy range, where the contribution of low-energy nucleon FSI is significant (see Figure \ref{fig:n_source} in Appendix \ref{appendix:n_source}). Events below 0.1 GeV are excluded due to large systematic uncertainties. The second metric is the slope of the linear increase in average $(n,\gamma)$ multiplicity as a function of visible energy, obtained by performing a linear fit over the range [0.3, 10] GeV, as shown in Figure \ref{fig:lincomb}.

\begin{figure*}[htbp!]
    \includegraphics[width=\linewidth]{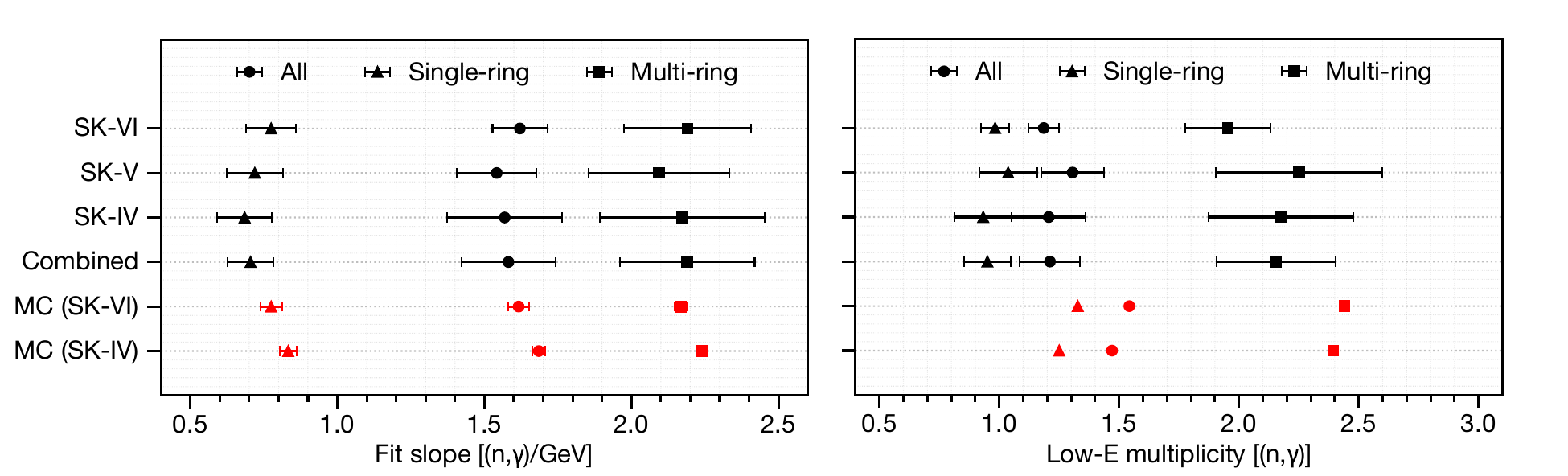}
    \caption{\label{fig:slopeconsistency} Measured slopes (left) and low-energy multiplicities (right) across different SK phases and event types. Data error bars (black) represent both statistical and systematic uncertainties, while MC error bars in the left panel indicate statistical uncertainties. In the right panel, MC predictions correspond to the `true' average $(n,\gamma)$ multiplicity.}
\end{figure*}

Figure \ref{fig:slopeconsistency} compares the measured slopes and low-energy multiplicities across different SK phases and event topologies. For the slope fits shown in the figure, the binning scheme was adjusted to ensure sufficient statistics (typically $\gtrsim 30$ neutrino events per bin). The measured values are consistent across SK phases within the assigned uncertainties. Both metrics are higher for multi-ring events than for single-ring events, as expected due to the higher fraction of deep inelastic scattering (DIS) interactions in the multi-ring sample (see Figure~\ref{fig:intbd}). While the observed slopes agree with baseline model predictions, the measured low-energy multiplicities were lower than expected.

\begin{figure*}[htbp!]
\includegraphics[width=\linewidth]{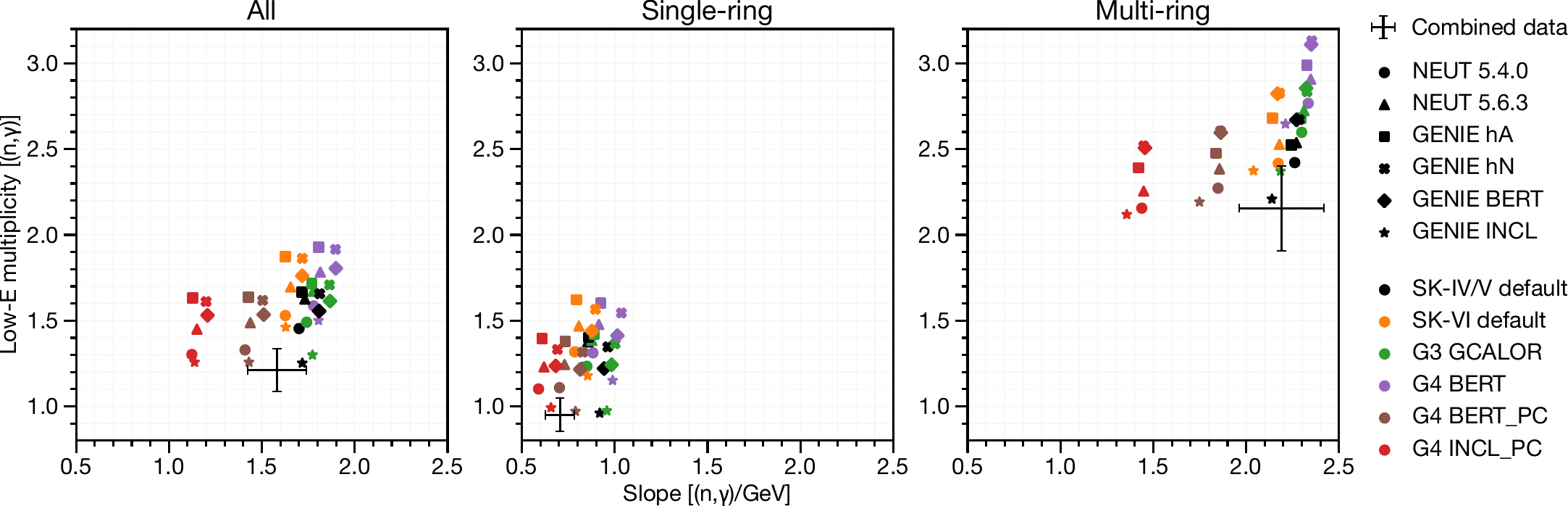}
\caption{\label{fig:model_scatter} Scatter plots of fitted slopes and low-energy multiplicities in model predictions and data (black crosses, error bars include both statistical and systematic uncertainties), shown for all events (left), single-ring events (middle), and multi-ring events (right). Shapes represent neutrino event generator options used, while colors indicate the secondary hadron-nucleus interaction model options used.}
\end{figure*}

Figure \ref{fig:model_scatter} compares the measured slopes and low-energy multiplicities with model predictions. The predictions show three distinct groups of slopes: the \texttt{G4 INCL\_PC} model gives the smallest slopes, \texttt{G4 BERT\_PC} predicts intermediate slopes, and other variants of the Bertini cascade model produce the largest slopes. The slope observed in the single-ring data better matches the \texttt{G4 INCL\_PC} predictions, while the slope in the multi-ring data is closer to the \texttt{G4 BERT\_PC} predictions. This distinction is most evident in the bottom panel of Figure \ref{fig:nmult_comp}, where \texttt{G4 INCL\_PC} matches well with sub-GeV single-ring data, and \texttt{G4 BERT\_PC} is a better fit for multi-GeV multi-ring data. For low-energy multiplicities, \texttt{NEUT 5.4.0}, \texttt{NEUT 5.6.3}, and \texttt{GENIE INCL} are preferred, as they predict lower neutron production. In contrast, models such as \texttt{GENIE hA}, \texttt{hN}, and \texttt{BERT} overestimate low-energy multiplicities, producing values significantly higher than the estimated $1\sigma$ uncertainty. These trends in model predictions and agreement with data remained robust under variations in the visible energy ranges used to define both metrics.

\section{Sources of neutron production}
\label{appendix:n_source}

\begin{figure*}[htbp!]
    \centering
    \includegraphics[width=0.5\linewidth]{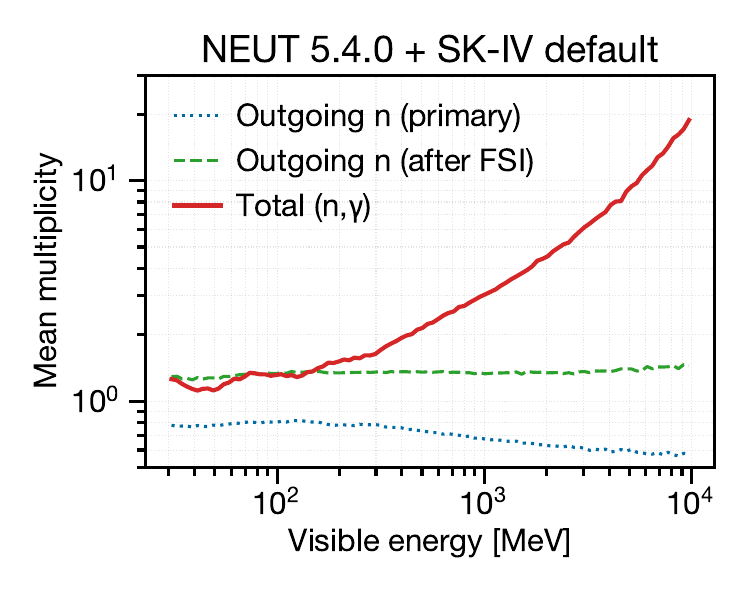}
    \caption{Mean multiplicity of outgoing neutrons from the primary neutrino interaction (blue, dotted) and subsequent FSIs (green, dashed) and resulting total $(n,\gamma)$ reactions (red, solid), in the baseline SK-IV simulation setup.}
    \label{fig:n_source}
\end{figure*}

Here, we clarify the sources of neutron production in our simulation. Figure \ref{fig:n_source} shows the baseline SK-IV neutron production as a function of visible energy. The number of outgoing neutrons from neutrino-nucleus interactions remains nearly constant across visible energies. At lower energies, the outgoing neutrons have small momenta, with each neutron corresponding closely to a single $(n,\gamma)$ reaction. At higher energies, larger hadron momentum results in a linear increase in secondary neutron production. This indicates that, for the two metrics defined in Appendix \ref{appendix:slopeintercept}, the low-energy multiplicity is sensitive to nucleon transport modeling, while the slope is primarily determined by the secondary hadron-nucleus interaction model.

\begin{figure*}[htbp!]
\includegraphics[width=0.8\linewidth]{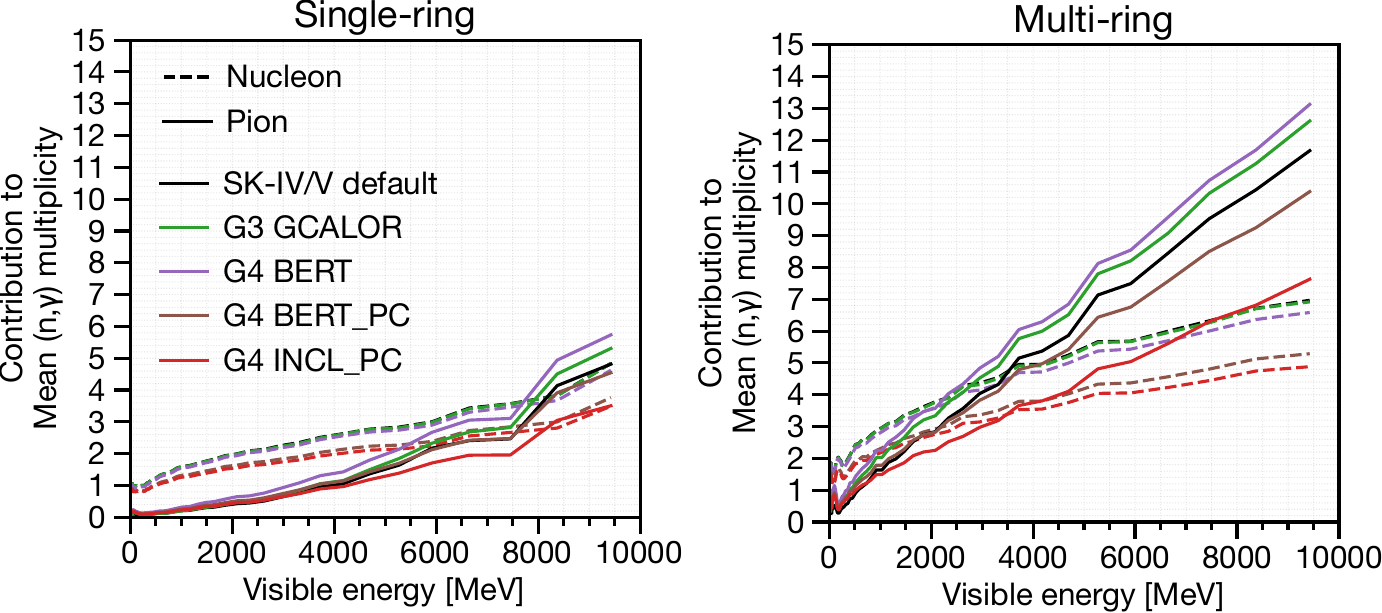}
\caption{\label{fig:n_contribution} Average contribution of outgoing nucleons (dashed lines) and pions (solid lines) to the observed $(n,\gamma)$ multiplicities as a function of visible energy, shown separately for single-ring events (left) and multi-ring events (right). The line colors represent different secondary hadron-nucleus interaction models used in the predictions.}
\end{figure*}

\begin{figure*}[htbp!]
\includegraphics[width=0.9
\linewidth]{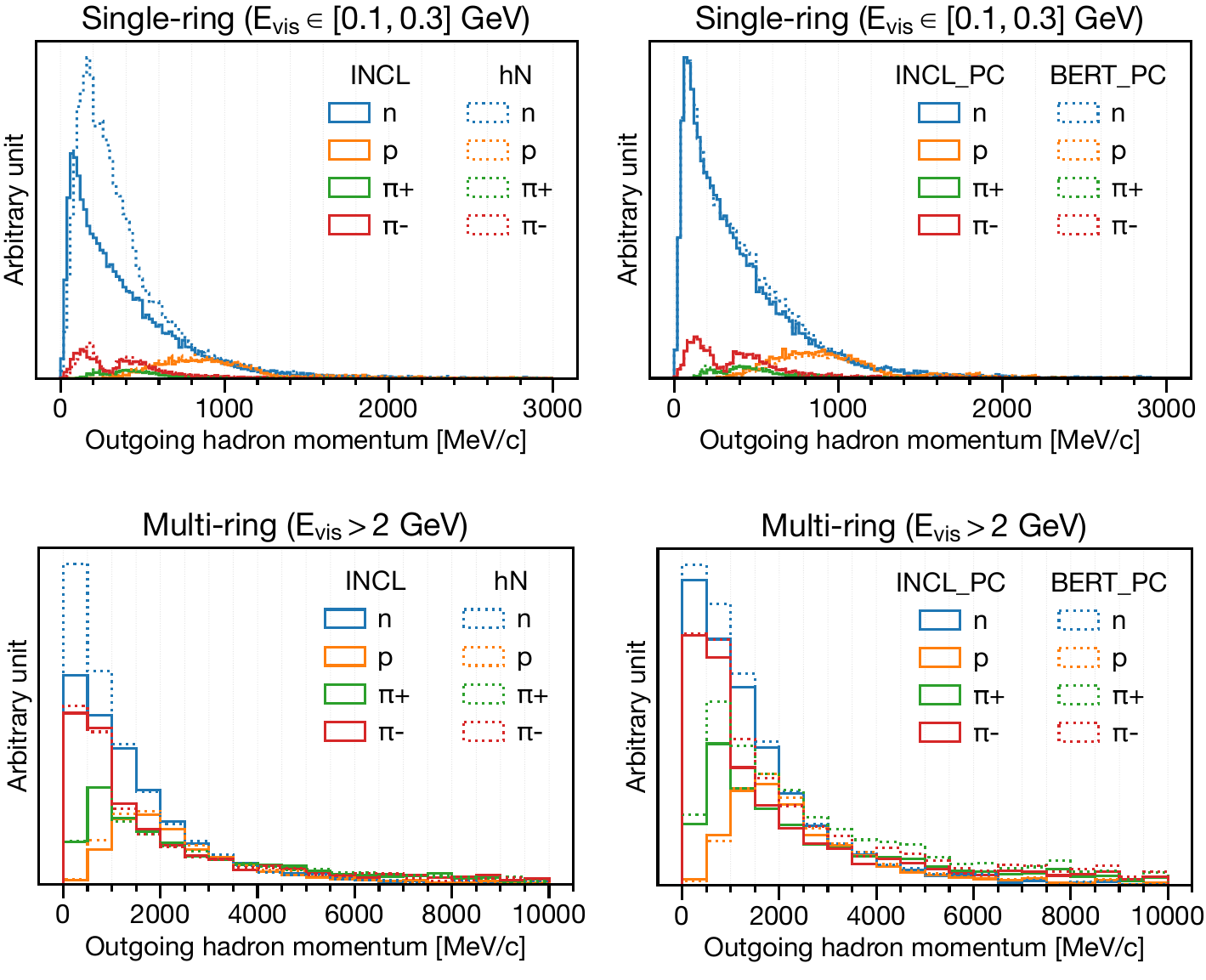}
\caption{\label{fig:n_source_by_particle} Momentum distributions of outgoing hadrons predicted by neutrino event generators, weighted by the average $(n,\gamma)$ multiplicity per projectile hadron momentum bin as predicted by secondary hadron-nucleus interaction models. The area under each histogram bin represents the average contribution of a specific hadron at a given momentum to the observable $(n,\gamma)$ multiplicity. The top panels show predictions for single-ring events with visible energy ($\text{E}_\text{vis}$) in the range [0.1, 0.3] GeV, while the bottom panels correspond to multi-ring events with visible energy above 2 GeV. The left plots compare two different cascade models for final-state interactions (\texttt{GENIE INCL} and \texttt{GENIE hN}) coupled with the secondary hadron-nucleus interaction model \texttt{G4 INCL\_PC}, whereas the right plots compare two secondary interaction models (\texttt{G4 INCL\_PC} and \texttt{G4 BERT\_PC}) coupled with the neutrino event generator option \texttt{GENIE hN}.}
\end{figure*}

Figure \ref{fig:n_contribution} shows the contributions of nucleons and pions to the $(n,\gamma)$ multiplicity as a function of visible energy, for different secondary hadron-nucleus interaction models. A general trend across all models is that, in the single-ring sample, most $(n,\gamma)$ reactions are induced by outgoing nucleons from neutrino interactions, while in the multi-ring sample, outgoing pions dominate in the few-GeV range. Figure \ref{fig:n_source_by_particle} further breaks down the contributions to the $(n,\gamma)$ multiplicity by outgoing particle momentum, for two FSI models (\texttt{GENIE INCL} and \texttt{GENIE hN}) and two secondary interaction models (\texttt{G4 INCL\_PC} and \texttt{G4 BERT\_PC}). In single-ring atmospheric neutrino events with visible energy in the range [0.1, 0.3] GeV, the total $(n,\gamma)$ multiplicity is primarily driven by neutrons below 1 GeV/c (from neutrino-nucleus interactions), with negligible variation across secondary interaction models. However, in multi-ring events with visible energies above 2 GeV, the total contribution from outgoing pions at all momenta becomes comparable to that from neutrons.

\clearpage

\bibliographystyle{apsrev4-2}
\bibliography{bibliography}

\providecommand{\noopsort}[1]{}\providecommand{\singleletter}[1]{#1}%
\begin{thebibliography}{77}%
\makeatletter
\providecommand \@ifxundefined [1]{%
 \@ifx{#1\undefined}
}%
\providecommand \@ifnum [1]{%
 \ifnum #1\expandafter \@firstoftwo
 \else \expandafter \@secondoftwo
 \fi
}%
\providecommand \@ifx [1]{%
 \ifx #1\expandafter \@firstoftwo
 \else \expandafter \@secondoftwo
 \fi
}%
\providecommand \natexlab [1]{#1}%
\providecommand \enquote  [1]{``#1''}%
\providecommand \bibnamefont  [1]{#1}%
\providecommand \bibfnamefont [1]{#1}%
\providecommand \citenamefont [1]{#1}%
\providecommand \href@noop [0]{\@secondoftwo}%
\providecommand \href [0]{\begingroup \@sanitize@url \@href}%
\providecommand \@href[1]{\@@startlink{#1}\@@href}%
\providecommand \@@href[1]{\endgroup#1\@@endlink}%
\providecommand \@sanitize@url [0]{\catcode `\\12\catcode `\$12\catcode
  `\&12\catcode `\#12\catcode `\^12\catcode `\_12\catcode `\%12\relax}%
\providecommand \@@startlink[1]{}%
\providecommand \@@endlink[0]{}%
\providecommand \url  [0]{\begingroup\@sanitize@url \@url }%
\providecommand \@url [1]{\endgroup\@href {#1}{\urlprefix }}%
\providecommand \urlprefix  [0]{URL }%
\providecommand \Eprint [0]{\href }%
\providecommand \doibase [0]{https://doi.org/}%
\providecommand \selectlanguage [0]{\@gobble}%
\providecommand \bibinfo  [0]{\@secondoftwo}%
\providecommand \bibfield  [0]{\@secondoftwo}%
\providecommand \translation [1]{[#1]}%
\providecommand \BibitemOpen [0]{}%
\providecommand \bibitemStop [0]{}%
\providecommand \bibitemNoStop [0]{.\EOS\space}%
\providecommand \EOS [0]{\spacefactor3000\relax}%
\providecommand \BibitemShut  [1]{\csname bibitem#1\endcsname}%
\let\auto@bib@innerbib\@empty
%</preamble>
\bibitem [{\citenamefont {Abratenko}\ \emph
  {et~al.}(2024{\natexlab{a}})\citenamefont {Abratenko} \emph
  {et~al.}}]{uboone}%
  \BibitemOpen
  \bibfield  {author} {\bibinfo {author} {\bibfnamefont {P.}~\bibnamefont
  {Abratenko}} \emph {et~al.} (\bibinfo {collaboration} {MicroBooNE
  collaboration}),\ }\href {https://doi.org/10.1103/PhysRevD.110.013006}
  {\bibfield  {journal} {\bibinfo  {journal} {Phys. Rev. D}\ }\textbf {\bibinfo
  {volume} {110}},\ \bibinfo {pages} {013006} (\bibinfo {year}
  {2024}{\natexlab{a}})}\BibitemShut {NoStop}%
\bibitem [{\citenamefont {Elkins}\ \emph {et~al.}(2019)\citenamefont {Elkins}
  \emph {et~al.}}]{mv_neutron_2019}%
  \BibitemOpen
  \bibfield  {author} {\bibinfo {author} {\bibfnamefont {M.}~\bibnamefont
  {Elkins}} \emph {et~al.} (\bibinfo {collaboration} {MINERvA collaboration}),\
  }\href {https://doi.org/10.1103/PhysRevD.100.052002} {\bibfield  {journal}
  {\bibinfo  {journal} {Phys. Rev. D}\ }\textbf {\bibinfo {volume} {100}},\
  \bibinfo {pages} {052002} (\bibinfo {year} {2019})}\BibitemShut {NoStop}%
\bibitem [{\citenamefont {Abratenko}\ \emph
  {et~al.}(2024{\natexlab{b}})\citenamefont {Abratenko} \emph
  {et~al.}}]{uboone_neutron}%
  \BibitemOpen
  \bibfield  {author} {\bibinfo {author} {\bibfnamefont {P.}~\bibnamefont
  {Abratenko}} \emph {et~al.} (\bibinfo {collaboration} {MicroBooNE
  collaboration}),\ }\href {https://doi.org/10.1140/epjc/s10052-024-13423-z}
  {\bibfield  {journal} {\bibinfo  {journal} {Eur. Phys. J. C}\ }\textbf
  {\bibinfo {volume} {84}},\ \bibinfo {pages} {1052} (\bibinfo {year}
  {2024}{\natexlab{b}})}\BibitemShut {NoStop}%
\bibitem [{\citenamefont {Ankowski}\ \emph {et~al.}(2015)\citenamefont
  {Ankowski}, \citenamefont {Coloma}, \citenamefont {Huber}, \citenamefont
  {Mariani},\ and\ \citenamefont {Vagnoni}}]{missing_energy_bias_to_cp}%
  \BibitemOpen
  \bibfield  {author} {\bibinfo {author} {\bibfnamefont {A.~M.}\ \bibnamefont
  {Ankowski}}, \bibinfo {author} {\bibfnamefont {P.}~\bibnamefont {Coloma}},
  \bibinfo {author} {\bibfnamefont {P.}~\bibnamefont {Huber}}, \bibinfo
  {author} {\bibfnamefont {C.}~\bibnamefont {Mariani}},\ and\ \bibinfo {author}
  {\bibfnamefont {E.}~\bibnamefont {Vagnoni}},\ }\href
  {https://doi.org/10.1103/PhysRevD.92.091301} {\bibfield  {journal} {\bibinfo
  {journal} {Phys. Rev. D}\ }\textbf {\bibinfo {volume} {92}},\ \bibinfo
  {pages} {091301} (\bibinfo {year} {2015})}\BibitemShut {NoStop}%
\bibitem [{\citenamefont {Cowan}\ \emph {et~al.}(1956)\citenamefont {Cowan},
  \citenamefont {Reines}, \citenamefont {Harrison}, \citenamefont {Kruse},\
  and\ \citenamefont {McGuire}}]{reines_cowan}%
  \BibitemOpen
  \bibfield  {author} {\bibinfo {author} {\bibfnamefont {C.~L.}\ \bibnamefont
  {Cowan}}, \bibinfo {author} {\bibfnamefont {F.}~\bibnamefont {Reines}},
  \bibinfo {author} {\bibfnamefont {F.~B.}\ \bibnamefont {Harrison}}, \bibinfo
  {author} {\bibfnamefont {H.~W.}\ \bibnamefont {Kruse}},\ and\ \bibinfo
  {author} {\bibfnamefont {A.~D.}\ \bibnamefont {McGuire}},\ }\href
  {https://doi.org/10.1126/science.124.3212.103} {\bibfield  {journal}
  {\bibinfo  {journal} {Science}\ }\textbf {\bibinfo {volume} {124}},\ \bibinfo
  {pages} {103} (\bibinfo {year} {1956})}\BibitemShut {NoStop}%
\bibitem [{\citenamefont {Wester}\ \emph {et~al.}(2024)\citenamefont {Wester}
  \emph {et~al.}}]{sk_osc_2024}%
  \BibitemOpen
  \bibfield  {author} {\bibinfo {author} {\bibfnamefont {T.}~\bibnamefont
  {Wester}} \emph {et~al.} (\bibinfo {collaboration} {Super-Kamiokande
  collaboration}),\ }\href {https://doi.org/10.1103/PhysRevD.109.072014}
  {\bibfield  {journal} {\bibinfo  {journal} {Phys. Rev. D}\ }\textbf {\bibinfo
  {volume} {109}},\ \bibinfo {pages} {072014} (\bibinfo {year}
  {2024})}\BibitemShut {NoStop}%
\bibitem [{\citenamefont {Takenaka}\ \emph {et~al.}(2020)\citenamefont
  {Takenaka} \emph {et~al.}}]{sk_pdk_2020}%
  \BibitemOpen
  \bibfield  {author} {\bibinfo {author} {\bibfnamefont {A.}~\bibnamefont
  {Takenaka}} \emph {et~al.} (\bibinfo {collaboration} {Super-Kamiokande
  collaboration}),\ }\href {https://doi.org/10.1103/PhysRevD.102.112011}
  {\bibfield  {journal} {\bibinfo  {journal} {Phys. Rev. D}\ }\textbf {\bibinfo
  {volume} {102}},\ \bibinfo {pages} {112011} (\bibinfo {year}
  {2020})}\BibitemShut {NoStop}%
\bibitem [{\citenamefont {Agostinelli}\ \emph {et~al.}(2003)\citenamefont
  {Agostinelli} \emph {et~al.}}]{geant4_1}%
  \BibitemOpen
  \bibfield  {author} {\bibinfo {author} {\bibfnamefont {S.}~\bibnamefont
  {Agostinelli}} \emph {et~al.} (\bibinfo {collaboration} {Geant4
  collaboration}),\ }\href
  {https://doi.org/https://doi.org/10.1016/S0168-9002(03)01368-8} {\bibfield
  {journal} {\bibinfo  {journal} {Nucl. Instrum. Methods A}\ }\textbf {\bibinfo
  {volume} {506}},\ \bibinfo {pages} {250} (\bibinfo {year}
  {2003})}\BibitemShut {NoStop}%
\bibitem [{\citenamefont {Allison}\ \emph {et~al.}(2006)\citenamefont {Allison}
  \emph {et~al.}}]{geant4_2}%
  \BibitemOpen
  \bibfield  {author} {\bibinfo {author} {\bibfnamefont {J.}~\bibnamefont
  {Allison}} \emph {et~al.},\ }\href {https://doi.org/10.1109/TNS.2006.869826}
  {\bibfield  {journal} {\bibinfo  {journal} {IEEE Trans. Nucl. Sci.}\ }\textbf
  {\bibinfo {volume} {53}},\ \bibinfo {pages} {270} (\bibinfo {year}
  {2006})}\BibitemShut {NoStop}%
\bibitem [{\citenamefont {Allison}\ \emph {et~al.}(2016)\citenamefont {Allison}
  \emph {et~al.}}]{geant4_3}%
  \BibitemOpen
  \bibfield  {author} {\bibinfo {author} {\bibfnamefont {J.}~\bibnamefont
  {Allison}} \emph {et~al.},\ }\href
  {https://doi.org/https://doi.org/10.1016/j.nima.2016.06.125} {\bibfield
  {journal} {\bibinfo  {journal} {Nucl. Instrum. Methods A}\ }\textbf {\bibinfo
  {volume} {835}},\ \bibinfo {pages} {186} (\bibinfo {year}
  {2016})}\BibitemShut {NoStop}%
\bibitem [{\citenamefont {Cugnon}(1982)}]{inc_review}%
  \BibitemOpen
  \bibfield  {author} {\bibinfo {author} {\bibfnamefont {J.}~\bibnamefont
  {Cugnon}},\ }\href
  {https://doi.org/https://doi.org/10.1016/0375-9474(82)90200-7} {\bibfield
  {journal} {\bibinfo  {journal} {Nucl. Phys. A}\ }\textbf {\bibinfo {volume}
  {387}},\ \bibinfo {pages} {191} (\bibinfo {year} {1982})}\BibitemShut
  {NoStop}%
\bibitem [{\citenamefont {David}\ \emph {et~al.}(2011)\citenamefont {David}
  \emph {et~al.}}]{iaea_benchmark}%
  \BibitemOpen
  \bibfield  {author} {\bibinfo {author} {\bibfnamefont {J.-C.}\ \bibnamefont
  {David}} \emph {et~al.},\ }\href {https://doi.org/10.15669/pnst.2.942}
  {\bibfield  {journal} {\bibinfo  {journal} {Prog. Nucl. Sci. Technol.}\
  }\textbf {\bibinfo {volume} {2}},\ \bibinfo {pages} {942} (\bibinfo {year}
  {2011})}\BibitemShut {NoStop}%
\bibitem [{\citenamefont {Akutsu}(2019)}]{t2k_neutron_2019}%
  \BibitemOpen
  \bibfield  {author} {\bibinfo {author} {\bibfnamefont {R.}~\bibnamefont
  {Akutsu}},\ }\href
  {https://www-sk.icrr.u-tokyo.ac.jp/sk/_pdf/articles/PhD_thesis.pdf} {\bibinfo
  {type} {{Ph.D.} thesis}},\ \bibinfo  {school} {{The Univ.\ of Tokyo}}
  (\bibinfo {year} {2019})\BibitemShut {NoStop}%
\bibitem [{\citenamefont {Aharmim}\ \emph {et~al.}(2019)\citenamefont {Aharmim}
  \emph {et~al.}}]{sno_neutron_2019}%
  \BibitemOpen
  \bibfield  {author} {\bibinfo {author} {\bibfnamefont {B.}~\bibnamefont
  {Aharmim}} \emph {et~al.} (\bibinfo {collaboration} {SNO collaboration}),\
  }\href {https://doi.org/10.1103/PhysRevD.99.112007} {\bibfield  {journal}
  {\bibinfo  {journal} {Phys. Rev. D}\ }\textbf {\bibinfo {volume} {99}},\
  \bibinfo {pages} {112007} (\bibinfo {year} {2019})}\BibitemShut {NoStop}%
\bibitem [{\citenamefont {Olivier}\ \emph {et~al.}(2023)\citenamefont {Olivier}
  \emph {et~al.}}]{mv_neutron_2023}%
  \BibitemOpen
  \bibfield  {author} {\bibinfo {author} {\bibfnamefont {A.}~\bibnamefont
  {Olivier}} \emph {et~al.} (\bibinfo {collaboration} {MINERvA
  collaboration}),\ }\href {https://doi.org/10.1103/PhysRevD.108.112010}
  {\bibfield  {journal} {\bibinfo  {journal} {Phys. Rev. D}\ }\textbf {\bibinfo
  {volume} {108}},\ \bibinfo {pages} {112010} (\bibinfo {year}
  {2023})}\BibitemShut {NoStop}%
\bibitem [{\citenamefont {Abe}\ \emph {et~al.}(2023)\citenamefont {Abe} \emph
  {et~al.}}]{kl_ncqe}%
  \BibitemOpen
  \bibfield  {author} {\bibinfo {author} {\bibfnamefont {S.}~\bibnamefont
  {Abe}} \emph {et~al.} (\bibinfo {collaboration} {KamLAND Collaboration}),\
  }\href {https://doi.org/10.1103/PhysRevD.107.072006} {\bibfield  {journal}
  {\bibinfo  {journal} {Phys. Rev. D}\ }\textbf {\bibinfo {volume} {107}},\
  \bibinfo {pages} {072006} (\bibinfo {year} {2023})}\BibitemShut {NoStop}%
\bibitem [{\citenamefont {Sakai}\ \emph {et~al.}(2024)\citenamefont {Sakai}
  \emph {et~al.}}]{sakai}%
  \BibitemOpen
  \bibfield  {author} {\bibinfo {author} {\bibfnamefont {S.}~\bibnamefont
  {Sakai}} \emph {et~al.} (\bibinfo {collaboration} {Super-Kamiokande
  Collaboration}),\ }\href {https://doi.org/10.1103/PhysRevD.109.L011101}
  {\bibfield  {journal} {\bibinfo  {journal} {Phys. Rev. D}\ }\textbf {\bibinfo
  {volume} {109}},\ \bibinfo {pages} {L011101} (\bibinfo {year}
  {2024})}\BibitemShut {NoStop}%
\bibitem [{\citenamefont {Chadwick}\ \emph {et~al.}(2011)\citenamefont
  {Chadwick} \emph {et~al.}}]{endf7}%
  \BibitemOpen
  \bibfield  {author} {\bibinfo {author} {\bibfnamefont {M.~B.}\ \bibnamefont
  {Chadwick}} \emph {et~al.},\ }\href
  {https://doi.org/10.1016/j.nds.2011.11.002} {\bibfield  {journal} {\bibinfo
  {journal} {Nucl. Data Sheets}\ }\textbf {\bibinfo {volume} {112}},\ \bibinfo
  {pages} {2887 } (\bibinfo {year} {2011})}\BibitemShut {NoStop}%
\bibitem [{\citenamefont {Abe}\ \emph {et~al.}(2022{\natexlab{a}})\citenamefont
  {Abe} \emph {et~al.}}]{sk_1st_gdloading}%
  \BibitemOpen
  \bibfield  {author} {\bibinfo {author} {\bibfnamefont {K.}~\bibnamefont
  {Abe}} \emph {et~al.} (\bibinfo {collaboration} {Super-Kamiokande
  collaboration}),\ }\href {https://doi.org/10.1016/j.nima.2021.166248}
  {\bibfield  {journal} {\bibinfo  {journal} {Nucl. Instrum. Methods A}\
  }\textbf {\bibinfo {volume} {1027}},\ \bibinfo {pages} {166248} (\bibinfo
  {year} {2022}{\natexlab{a}})}\BibitemShut {NoStop}%
\bibitem [{\citenamefont {Abe}\ \emph {et~al.}(2024)\citenamefont {Abe} \emph
  {et~al.}}]{sk_2nd_gdloading}%
  \BibitemOpen
  \bibfield  {author} {\bibinfo {author} {\bibfnamefont {K.}~\bibnamefont
  {Abe}} \emph {et~al.},\ }\href
  {https://doi.org/https://doi.org/10.1016/j.nima.2024.169480} {\bibfield
  {journal} {\bibinfo  {journal} {Nucl. Instrum. Methods A}\ }\textbf {\bibinfo
  {volume} {1065}},\ \bibinfo {pages} {169480} (\bibinfo {year}
  {2024})}\BibitemShut {NoStop}%
\bibitem [{\citenamefont {Fukuda}\ \emph {et~al.}(2003)\citenamefont {Fukuda}
  \emph {et~al.}}]{sk_detector}%
  \BibitemOpen
  \bibfield  {author} {\bibinfo {author} {\bibfnamefont {S.}~\bibnamefont
  {Fukuda}} \emph {et~al.} (\bibinfo {collaboration} {Super-Kamiokande
  collaboration}),\ }\href
  {https://doi.org/https://doi.org/10.1016/S0168-9002(03)00425-X} {\bibfield
  {journal} {\bibinfo  {journal} {Nucl. Instrum. Methods A}\ }\textbf {\bibinfo
  {volume} {501}},\ \bibinfo {pages} {418} (\bibinfo {year}
  {2003})}\BibitemShut {NoStop}%
\bibitem [{\citenamefont {Abe}\ \emph {et~al.}(2014)\citenamefont {Abe} \emph
  {et~al.}}]{sk_calib}%
  \BibitemOpen
  \bibfield  {author} {\bibinfo {author} {\bibfnamefont {K.}~\bibnamefont
  {Abe}} \emph {et~al.},\ }\href {https://doi.org/10.1016/j.nima.2013.11.081}
  {\bibfield  {journal} {\bibinfo  {journal} {Nucl. Instrum. Methods A}\
  }\textbf {\bibinfo {volume} {737}},\ \bibinfo {pages} {253–272} (\bibinfo
  {year} {2014})}\BibitemShut {NoStop}%
\bibitem [{\citenamefont {Yamada}\ \emph {et~al.}(2010)\citenamefont {Yamada}
  \emph {et~al.}}]{sk_qbee}%
  \BibitemOpen
  \bibfield  {author} {\bibinfo {author} {\bibfnamefont {S.}~\bibnamefont
  {Yamada}} \emph {et~al.},\ }\href {https://doi.org/10.1109/TNS.2009.2034854}
  {\bibfield  {journal} {\bibinfo  {journal} {IEEE Trans. Nucl. Sci.}\ }\textbf
  {\bibinfo {volume} {57}},\ \bibinfo {pages} {428} (\bibinfo {year}
  {2010})}\BibitemShut {NoStop}%
\bibitem [{\citenamefont {Andreopoulos}\ \emph {et~al.}(2010)\citenamefont
  {Andreopoulos} \emph {et~al.}}]{genie_main}%
  \BibitemOpen
  \bibfield  {author} {\bibinfo {author} {\bibfnamefont {C.}~\bibnamefont
  {Andreopoulos}} \emph {et~al.},\ }\href
  {https://doi.org/10.1016/j.nima.2009.12.009} {\bibfield  {journal} {\bibinfo
  {journal} {Nucl. Instrum. Methods A}\ }\textbf {\bibinfo {volume} {614}},\
  \bibinfo {pages} {87} (\bibinfo {year} {2010})}\BibitemShut {NoStop}%
\bibitem [{\citenamefont {Andreopoulos}\ \emph {et~al.}()\citenamefont
  {Andreopoulos} \emph {et~al.}}]{genie_manual}%
  \BibitemOpen
  \bibfield  {author} {\bibinfo {author} {\bibfnamefont {C.}~\bibnamefont
  {Andreopoulos}} \emph {et~al.},\ }\href@noop {} {}\Eprint
  {https://arxiv.org/abs/1510.05494} {arXiv:1510.05494} \BibitemShut {NoStop}%
\bibitem [{\citenamefont {Tena-Vidal}\ \emph {et~al.}(2021)\citenamefont
  {Tena-Vidal} \emph {et~al.}}]{genie_v3tune}%
  \BibitemOpen
  \bibfield  {author} {\bibinfo {author} {\bibfnamefont {J.}~\bibnamefont
  {Tena-Vidal}} \emph {et~al.} (\bibinfo {collaboration} {GENIE
  collaboration}),\ }\href {https://doi.org/10.1103/PhysRevD.104.072009}
  {\bibfield  {journal} {\bibinfo  {journal} {Phys. Rev. D}\ }\textbf {\bibinfo
  {volume} {104}},\ \bibinfo {pages} {072009} (\bibinfo {year}
  {2021})}\BibitemShut {NoStop}%
\bibitem [{\citenamefont {Shiozawa}(1999)}]{apfit}%
  \BibitemOpen
  \bibfield  {author} {\bibinfo {author} {\bibfnamefont {M.}~\bibnamefont
  {Shiozawa}},\ }\href
  {https://doi.org/https://doi.org/10.1016/S0168-9002(99)00359-9} {\bibfield
  {journal} {\bibinfo  {journal} {Nucl. Instrum. Methods A}\ }\textbf {\bibinfo
  {volume} {433}},\ \bibinfo {pages} {240} (\bibinfo {year}
  {1999})}\BibitemShut {NoStop}%
\bibitem [{\citenamefont {Honda}\ \emph {et~al.}(2011)\citenamefont {Honda},
  \citenamefont {Kajita}, \citenamefont {Kasahara},\ and\ \citenamefont
  {Midorikawa}}]{honda_2011}%
  \BibitemOpen
  \bibfield  {author} {\bibinfo {author} {\bibfnamefont {M.}~\bibnamefont
  {Honda}}, \bibinfo {author} {\bibfnamefont {T.}~\bibnamefont {Kajita}},
  \bibinfo {author} {\bibfnamefont {K.}~\bibnamefont {Kasahara}},\ and\
  \bibinfo {author} {\bibfnamefont {S.}~\bibnamefont {Midorikawa}},\ }\href
  {https://doi.org/10.1103/PhysRevD.83.123001} {\bibfield  {journal} {\bibinfo
  {journal} {Phys. Rev. D}\ }\textbf {\bibinfo {volume} {83}},\ \bibinfo
  {pages} {123001} (\bibinfo {year} {2011})}\BibitemShut {NoStop}%
\bibitem [{\citenamefont {Workman}\ \emph {et~al.}(2022)\citenamefont {Workman}
  \emph {et~al.}}]{pdg_nu_ev_gen_rev}%
  \BibitemOpen
  \bibfield  {author} {\bibinfo {author} {\bibfnamefont {R.~L.}\ \bibnamefont
  {Workman}} \emph {et~al.} (\bibinfo {collaboration} {Particle Data Group}),\
  }\href {https://doi.org/10.1093/ptep/ptac097} {\bibfield  {journal} {\bibinfo
   {journal} {Prog. Theor. Exp. Phys.}\ }\textbf {\bibinfo {volume} {083C01}},\
  \bibinfo {pages} {729} (\bibinfo {year} {2022})}\BibitemShut {NoStop}%
\bibitem [{\citenamefont {Hayato}\ and\ \citenamefont
  {Pickering}(2021)}]{neut}%
  \BibitemOpen
  \bibfield  {author} {\bibinfo {author} {\bibfnamefont {Y.}~\bibnamefont
  {Hayato}}\ and\ \bibinfo {author} {\bibfnamefont {L.}~\bibnamefont
  {Pickering}},\ }\href {https://doi.org/10.1140/epjs/s11734-021-00287-7}
  {\bibfield  {journal} {\bibinfo  {journal} {Eur. Phys. J. Spec. Top.}\
  }\textbf {\bibinfo {volume} {230}},\ \bibinfo {pages} {4469} (\bibinfo {year}
  {2021})}\BibitemShut {NoStop}%
\bibitem [{\citenamefont {Salcedo}\ \emph {et~al.}(1988)\citenamefont
  {Salcedo}, \citenamefont {Oset}, \citenamefont {Vicente-Vacas},\ and\
  \citenamefont {Garcia-Recio}}]{salcedo_oset}%
  \BibitemOpen
  \bibfield  {author} {\bibinfo {author} {\bibfnamefont {L.~L.}\ \bibnamefont
  {Salcedo}}, \bibinfo {author} {\bibfnamefont {E.}~\bibnamefont {Oset}},
  \bibinfo {author} {\bibfnamefont {M.~J.}\ \bibnamefont {Vicente-Vacas}},\
  and\ \bibinfo {author} {\bibfnamefont {C.}~\bibnamefont {Garcia-Recio}},\
  }\href {https://doi.org/10.1016/0375-9474(88)90310-7} {\bibfield  {journal}
  {\bibinfo  {journal} {Nucl. Phys. A}\ }\textbf {\bibinfo {volume} {484}},\
  \bibinfo {pages} {557} (\bibinfo {year} {1988})}\BibitemShut {NoStop}%
\bibitem [{\citenamefont {Guerra}\ \emph {et~al.}(2019)\citenamefont {Guerra}
  \emph {et~al.}}]{neut_pi_fsi_tune}%
  \BibitemOpen
  \bibfield  {author} {\bibinfo {author} {\bibfnamefont {E.~S.~P.}\
  \bibnamefont {Guerra}} \emph {et~al.},\ }\href
  {https://doi.org/10.1103/PhysRevD.99.052007} {\bibfield  {journal} {\bibinfo
  {journal} {Phys. Rev. D}\ }\textbf {\bibinfo {volume} {99}},\ \bibinfo
  {pages} {052007} (\bibinfo {year} {2019})}\BibitemShut {NoStop}%
\bibitem [{\citenamefont {Bertini}(1963)}]{bertini}%
  \BibitemOpen
  \bibfield  {author} {\bibinfo {author} {\bibfnamefont {H.~W.}\ \bibnamefont
  {Bertini}},\ }\href {https://doi.org/10.1103/PhysRev.131.1801} {\bibfield
  {journal} {\bibinfo  {journal} {Phys. Rev.}\ }\textbf {\bibinfo {volume}
  {131}},\ \bibinfo {pages} {1801} (\bibinfo {year} {1963})}\BibitemShut
  {NoStop}%
\bibitem [{\citenamefont {de~Perio}()}]{pdperio}%
  \BibitemOpen
  \bibfield  {author} {\bibinfo {author} {\bibfnamefont {P.}~\bibnamefont
  {de~Perio}},\ }\href {http://dx.doi.org/10.1063/1.3661590} {}\bibinfo {note}
  {\href{http://dx.doi.org/10.1063/1.3661590}{\textit{NEUT Pion FSI}}, in
  \textit{AIP Conf. Proc.}, ed. S. K. Singh, J. G. Morfin, M. Sakuda, and K. D.
  Purohit (AIP, Melville, NY, 2011)}\BibitemShut {NoStop}%
\bibitem [{\citenamefont {Ankowski}\ \emph {et~al.}(2012)\citenamefont
  {Ankowski}, \citenamefont {Benhar}, \citenamefont {Mori}, \citenamefont
  {Yamaguchi},\ and\ \citenamefont {Sakuda}}]{160_occ_prob}%
  \BibitemOpen
  \bibfield  {author} {\bibinfo {author} {\bibfnamefont {A.~M.}\ \bibnamefont
  {Ankowski}}, \bibinfo {author} {\bibfnamefont {O.}~\bibnamefont {Benhar}},
  \bibinfo {author} {\bibfnamefont {T.}~\bibnamefont {Mori}}, \bibinfo {author}
  {\bibfnamefont {R.}~\bibnamefont {Yamaguchi}},\ and\ \bibinfo {author}
  {\bibfnamefont {M.}~\bibnamefont {Sakuda}},\ }\href
  {https://doi.org/10.1103/PhysRevLett.108.052505} {\bibfield  {journal}
  {\bibinfo  {journal} {Phys. Rev. Lett.}\ }\textbf {\bibinfo {volume} {108}},\
  \bibinfo {pages} {052505} (\bibinfo {year} {2012})}\BibitemShut {NoStop}%
\bibitem [{\citenamefont {Ejiri}(1993)}]{16o_knockout_br}%
  \BibitemOpen
  \bibfield  {author} {\bibinfo {author} {\bibfnamefont {H.}~\bibnamefont
  {Ejiri}},\ }\href {https://doi.org/10.1103/PhysRevC.48.1442} {\bibfield
  {journal} {\bibinfo  {journal} {Phys. Rev. C}\ }\textbf {\bibinfo {volume}
  {48}},\ \bibinfo {pages} {1442} (\bibinfo {year} {1993})}\BibitemShut
  {NoStop}%
\bibitem [{\citenamefont {Zeitnitz}\ and\ \citenamefont
  {Gabriel}(1994)}]{gcalor}%
  \BibitemOpen
  \bibfield  {author} {\bibinfo {author} {\bibfnamefont {C.}~\bibnamefont
  {Zeitnitz}}\ and\ \bibinfo {author} {\bibfnamefont {T.~A.}\ \bibnamefont
  {Gabriel}},\ }\href {https://doi.org/10.1016/0168-9002(94)90613-0} {\bibfield
   {journal} {\bibinfo  {journal} {Nucl. Instrum. Meth. A}\ }\textbf {\bibinfo
  {volume} {349}},\ \bibinfo {pages} {106} (\bibinfo {year}
  {1994})}\BibitemShut {NoStop}%
\bibitem [{\citenamefont {Brun}\ \emph {et~al.}(1987)\citenamefont {Brun},
  \citenamefont {Bruyant}, \citenamefont {Maire}, \citenamefont {McPherson},\
  and\ \citenamefont {Zanarini}}]{geant3}%
  \BibitemOpen
  \bibfield  {author} {\bibinfo {author} {\bibfnamefont {R.}~\bibnamefont
  {Brun}}, \bibinfo {author} {\bibfnamefont {F.}~\bibnamefont {Bruyant}},
  \bibinfo {author} {\bibfnamefont {M.}~\bibnamefont {Maire}}, \bibinfo
  {author} {\bibfnamefont {A.~C.}\ \bibnamefont {McPherson}},\ and\ \bibinfo
  {author} {\bibfnamefont {P.}~\bibnamefont {Zanarini}},\ }\href
  {https://cds.cern.ch/record/1119728/files/CERN-DD-EE-84-1.pdf} {\emph
  {\bibinfo {title} {{GEANT3}}}},\ \bibinfo {type} {Tech. Rep.}\ (\bibinfo
  {institution} {CERN, CERN-DD-EE-84-1},\ \bibinfo {year} {1987})\BibitemShut
  {NoStop}%
\bibitem [{\citenamefont {Bertini}(1969)}]{bertini_validation}%
  \BibitemOpen
  \bibfield  {author} {\bibinfo {author} {\bibfnamefont {H.~W.}\ \bibnamefont
  {Bertini}},\ }\href {https://doi.org/10.1103/PhysRev.188.1711} {\bibfield
  {journal} {\bibinfo  {journal} {Phys. Rev.}\ }\textbf {\bibinfo {volume}
  {188}},\ \bibinfo {pages} {1711} (\bibinfo {year} {1969})}\BibitemShut
  {NoStop}%
\bibitem [{\citenamefont {Rose}(1991)}]{endf6}%
  \BibitemOpen
  \bibfield  {author} {\bibinfo {author} {\bibfnamefont {P.~F.}\ \bibnamefont
  {Rose}},\ }\href {https://doi.org/10.2172/10132931} {\emph {\bibinfo {title}
  {ENDF-201: ENDF/B-VI Summary Documentation}}},\ \bibinfo {type} {Tech. Rep.}\
  (\bibinfo  {institution} {BNL, BNL-NCS-17541},\ \bibinfo {year}
  {1991})\BibitemShut {NoStop}%
\bibitem [{\citenamefont {Mendoza}\ and\ \citenamefont
  {Cano-Ott}(2018)}]{neutronhp}%
  \BibitemOpen
  \bibfield  {author} {\bibinfo {author} {\bibfnamefont {E.}~\bibnamefont
  {Mendoza}}\ and\ \bibinfo {author} {\bibfnamefont {D.}~\bibnamefont
  {Cano-Ott}},\ }\href
  {http://inis.iaea.org/search/search.aspx?orig_q=RN:50006622} {\emph {\bibinfo
  {title} {Update of the Evaluated Neutron Cross Section Libraries for the
  {Geant4} Code}}},\ \bibinfo {type} {Tech. Rep.}\ (\bibinfo  {institution}
  {IAEA, INDC(NDS)-0758},\ \bibinfo {year} {2018})\BibitemShut {NoStop}%
\bibitem [{\citenamefont {Hagiwara}\ \emph {et~al.}(2019)\citenamefont
  {Hagiwara} \emph {et~al.}}]{annrigd_157}%
  \BibitemOpen
  \bibfield  {author} {\bibinfo {author} {\bibfnamefont {K.}~\bibnamefont
  {Hagiwara}} \emph {et~al.},\ }\href {https://doi.org/10.1093/ptep/ptz002}
  {\bibfield  {journal} {\bibinfo  {journal} {Prog. Theor. Exp. Phys.}\
  }\textbf {\bibinfo {volume} {2019}},\ \bibinfo {pages} {023D01} (\bibinfo
  {year} {2019})}\BibitemShut {NoStop}%
\bibitem [{\citenamefont {Tanaka}\ \emph {et~al.}(2020)\citenamefont {Tanaka}
  \emph {et~al.}}]{annrigd_155}%
  \BibitemOpen
  \bibfield  {author} {\bibinfo {author} {\bibfnamefont {T.}~\bibnamefont
  {Tanaka}} \emph {et~al.},\ }\href {https://doi.org/10.1093/ptep/ptaa015}
  {\bibfield  {journal} {\bibinfo  {journal} {Prog. Theor. Exp. Phys.}\
  }\textbf {\bibinfo {volume} {2020}},\ \bibinfo {pages} {043D02} (\bibinfo
  {year} {2020})}\BibitemShut {NoStop}%
\bibitem [{\citenamefont {Barger}\ \emph {et~al.}(1980)\citenamefont {Barger},
  \citenamefont {Whisnant}, \citenamefont {Pakvasa},\ and\ \citenamefont
  {Phillips}}]{barger}%
  \BibitemOpen
  \bibfield  {author} {\bibinfo {author} {\bibfnamefont {V.}~\bibnamefont
  {Barger}}, \bibinfo {author} {\bibfnamefont {K.}~\bibnamefont {Whisnant}},
  \bibinfo {author} {\bibfnamefont {S.}~\bibnamefont {Pakvasa}},\ and\ \bibinfo
  {author} {\bibfnamefont {R.~J.~N.}\ \bibnamefont {Phillips}},\ }\href
  {https://doi.org/10.1103/PhysRevD.22.2718} {\bibfield  {journal} {\bibinfo
  {journal} {Phys. Rev. D}\ }\textbf {\bibinfo {volume} {22}},\ \bibinfo
  {pages} {2718} (\bibinfo {year} {1980})}\BibitemShut {NoStop}%
\bibitem [{\citenamefont {Jiang}\ \emph {et~al.}(2019)\citenamefont {Jiang}
  \emph {et~al.}}]{mjiang}%
  \BibitemOpen
  \bibfield  {author} {\bibinfo {author} {\bibfnamefont {M.}~\bibnamefont
  {Jiang}} \emph {et~al.} (\bibinfo {collaboration} {Super-Kamiokande
  collaboration}),\ }\href {https://doi.org/10.1093/ptep/ptz015} {\bibfield
  {journal} {\bibinfo  {journal} {Prog. Theor. Exp. Phys.}\ }\textbf {\bibinfo
  {volume} {2019}},\ \bibinfo {pages} {053F01} (\bibinfo {year}
  {2019})}\BibitemShut {NoStop}%
\bibitem [{\citenamefont {Dziewonski}\ and\ \citenamefont
  {Anderson}(1981)}]{prem}%
  \BibitemOpen
  \bibfield  {author} {\bibinfo {author} {\bibfnamefont {A.~M.}\ \bibnamefont
  {Dziewonski}}\ and\ \bibinfo {author} {\bibfnamefont {D.~L.}\ \bibnamefont
  {Anderson}},\ }\href
  {https://doi.org/https://doi.org/10.1016/0031-9201(81)90046-7} {\bibfield
  {journal} {\bibinfo  {journal} {Phys. Earth Planet. Inter.}\ }\textbf
  {\bibinfo {volume} {25}},\ \bibinfo {pages} {297} (\bibinfo {year}
  {1981})}\BibitemShut {NoStop}%
\bibitem [{\citenamefont {Abe}\ \emph {et~al.}(2022{\natexlab{b}})\citenamefont
  {Abe} \emph {et~al.}}]{sk4_ntag}%
  \BibitemOpen
  \bibfield  {author} {\bibinfo {author} {\bibfnamefont {K.}~\bibnamefont
  {Abe}} \emph {et~al.} (\bibinfo {collaboration} {Super-Kamiokande
  collaboration}),\ }\href {https://doi.org/10.1088/1748-0221/17/10/p10029}
  {\bibfield  {journal} {\bibinfo  {journal} {J. Instrum.}\ }\textbf {\bibinfo
  {volume} {17}}\bibinfo  {number} { (10)},\ \bibinfo {pages}
  {P10029}}\BibitemShut {NoStop}%
\bibitem [{\citenamefont {Han}(2021)}]{han_mthesis}%
  \BibitemOpen
\bibfield  {number} {  }\bibfield  {author} {\bibinfo {author} {\bibfnamefont
  {S.}~\bibnamefont {Han}},\ }\href
  {https://www-sk.icrr.u-tokyo.ac.jp/sk/_pdf/articles/mthesis-Han.pdf}
  {\bibinfo {type} {{Master's} thesis}},\ \bibinfo  {school} {{The Univ.\ of
  Tokyo}} (\bibinfo {year} {2021})\BibitemShut {NoStop}%
\bibitem [{\citenamefont {Chollet}\ \emph {et~al.}(2015)\citenamefont {Chollet}
  \emph {et~al.}}]{keras}%
  \BibitemOpen
  \bibfield  {author} {\bibinfo {author} {\bibfnamefont {F.}~\bibnamefont
  {Chollet}} \emph {et~al.},\ }\href@noop {} {\bibinfo {title}
  {\textit{Keras}}},\ \bibinfo {howpublished} {\url{https://keras.io}}
  (\bibinfo {year} {2015})\BibitemShut {NoStop}%
\bibitem [{\citenamefont {He}\ \emph {et~al.}()\citenamefont {He},
  \citenamefont {Zhang}, \citenamefont {Ren},\ and\ \citenamefont
  {Sun}}]{he_normal}%
  \BibitemOpen
  \bibfield  {author} {\bibinfo {author} {\bibfnamefont {K.}~\bibnamefont
  {He}}, \bibinfo {author} {\bibfnamefont {X.}~\bibnamefont {Zhang}}, \bibinfo
  {author} {\bibfnamefont {S.}~\bibnamefont {Ren}},\ and\ \bibinfo {author}
  {\bibfnamefont {J.}~\bibnamefont {Sun}},\ }\href@noop {} {}\Eprint
  {https://arxiv.org/abs/1502.01852} {arXiv:1502.01852} \BibitemShut {NoStop}%
\bibitem [{\citenamefont {Kingma}\ and\ \citenamefont {Ba}()}]{adam}%
  \BibitemOpen
  \bibfield  {author} {\bibinfo {author} {\bibfnamefont {D.~P.}\ \bibnamefont
  {Kingma}}\ and\ \bibinfo {author} {\bibfnamefont {J.}~\bibnamefont {Ba}},\
  }\href {https://arxiv.org/abs/1412.6980} {}\Eprint
  {https://arxiv.org/abs/1412.6980} {arXiv:1412.6980} \BibitemShut {NoStop}%
\bibitem [{\citenamefont {Ito}\ \emph {et~al.}(2023)\citenamefont {Ito},
  \citenamefont {Wada}, \citenamefont {Yano}, \citenamefont {Hino},
  \citenamefont {Ommura}, \citenamefont {Harada}, \citenamefont {Minamino},\
  and\ \citenamefont {Ishitsuka}}]{ambe_intensity}%
  \BibitemOpen
  \bibfield  {author} {\bibinfo {author} {\bibfnamefont {H.}~\bibnamefont
  {Ito}}, \bibinfo {author} {\bibfnamefont {K.}~\bibnamefont {Wada}}, \bibinfo
  {author} {\bibfnamefont {T.}~\bibnamefont {Yano}}, \bibinfo {author}
  {\bibfnamefont {Y.}~\bibnamefont {Hino}}, \bibinfo {author} {\bibfnamefont
  {Y.}~\bibnamefont {Ommura}}, \bibinfo {author} {\bibfnamefont
  {M.}~\bibnamefont {Harada}}, \bibinfo {author} {\bibfnamefont
  {A.}~\bibnamefont {Minamino}},\ and\ \bibinfo {author} {\bibfnamefont
  {M.}~\bibnamefont {Ishitsuka}},\ }\href
  {https://doi.org/https://doi.org/10.1016/j.nima.2023.168701} {\bibfield
  {journal} {\bibinfo  {journal} {Nucl. Instrum. Methods A}\ }\textbf {\bibinfo
  {volume} {1057}},\ \bibinfo {pages} {168701} (\bibinfo {year}
  {2023})}\BibitemShut {NoStop}%
\bibitem [{\citenamefont {Han}(2024)}]{han_thesis}%
  \BibitemOpen
  \bibfield  {author} {\bibinfo {author} {\bibfnamefont {S.}~\bibnamefont
  {Han}},\ }\href
  {https://www-sk.icrr.u-tokyo.ac.jp/sk/_pdf/articles/2024/han_dthesis.pdf}
  {\bibinfo {type} {{Ph.D.} thesis}},\ \bibinfo  {school} {{The Univ.\ of
  Tokyo}} (\bibinfo {year} {2024})\BibitemShut {NoStop}%
\bibitem [{\citenamefont {Hino}\ \emph {et~al.}(2024)\citenamefont {Hino} \emph
  {et~al.}}]{hino_g4}%
  \BibitemOpen
  \bibfield  {author} {\bibinfo {author} {\bibfnamefont {Y.}~\bibnamefont
  {Hino}} \emph {et~al.},\ }\href {https://doi.org/10.1093/ptep/ptae193}
  {\bibfield  {journal} {\bibinfo  {journal} {Prog. Theor. Exp. Phys.}\
  }\textbf {\bibinfo {volume} {2025}},\ \bibinfo {pages} {013C01} (\bibinfo
  {year} {2024})}\BibitemShut {NoStop}%
\bibitem [{\citenamefont {Pritychenko}\ and\ \citenamefont
  {Mughabghab}(2012)}]{endf_paper}%
  \BibitemOpen
  \bibfield  {author} {\bibinfo {author} {\bibfnamefont {B.}~\bibnamefont
  {Pritychenko}}\ and\ \bibinfo {author} {\bibfnamefont {S.~F.}\ \bibnamefont
  {Mughabghab}},\ }\href
  {https://doi.org/https://doi.org/10.1016/j.nds.2012.11.007} {\bibfield
  {journal} {\bibinfo  {journal} {Nucl. Data Sheets}\ }\textbf {\bibinfo
  {volume} {113}},\ \bibinfo {pages} {3120} (\bibinfo {year}
  {2012})}\BibitemShut {NoStop}%
\bibitem [{\citenamefont {Hastie}\ and\ \citenamefont
  {Tibshirani}(1986)}]{gam}%
  \BibitemOpen
  \bibfield  {author} {\bibinfo {author} {\bibfnamefont {T.~J.}\ \bibnamefont
  {Hastie}}\ and\ \bibinfo {author} {\bibfnamefont {R.~J.}\ \bibnamefont
  {Tibshirani}},\ }\href {https://doi.org/10.1214/ss/1177013604} {\bibfield
  {journal} {\bibinfo  {journal} {Statist. Sci.}\ }\textbf {\bibinfo {volume}
  {1}},\ \bibinfo {pages} {297} (\bibinfo {year} {1986})}\BibitemShut {NoStop}%
\bibitem [{\citenamefont {Servén}\ \emph {et~al.}(2018)\citenamefont
  {Servén}, \citenamefont {Brummitt},\ and\ \citenamefont {Abedi}}]{pygam}%
  \BibitemOpen
  \bibfield  {author} {\bibinfo {author} {\bibfnamefont {D.}~\bibnamefont
  {Servén}}, \bibinfo {author} {\bibfnamefont {C.}~\bibnamefont {Brummitt}},\
  and\ \bibinfo {author} {\bibfnamefont {H.}~\bibnamefont {Abedi}},\ }\href
  {https://doi.org/10.5281/zenodo.1476122} {\bibinfo {title} {dswah/py{GAM}:
  v0.8.0}} (\bibinfo {year} {2018})\BibitemShut {NoStop}%
\bibitem [{\citenamefont {Wright}\ and\ \citenamefont
  {Kelsey}(2015)}]{g4_bertini}%
  \BibitemOpen
  \bibfield  {author} {\bibinfo {author} {\bibfnamefont {D.}~\bibnamefont
  {Wright}}\ and\ \bibinfo {author} {\bibfnamefont {M.}~\bibnamefont
  {Kelsey}},\ }\href
  {https://doi.org/https://doi.org/10.1016/j.nima.2015.09.058} {\bibfield
  {journal} {\bibinfo  {journal} {Nucl. Instrum. Methods A}\ }\textbf {\bibinfo
  {volume} {804}},\ \bibinfo {pages} {175} (\bibinfo {year}
  {2015})}\BibitemShut {NoStop}%
\bibitem [{\citenamefont {Mancusi}\ \emph {et~al.}(2014)\citenamefont
  {Mancusi}, \citenamefont {Boudard}, \citenamefont {Cugnon}, \citenamefont
  {David}, \citenamefont {Kaitaniemi},\ and\ \citenamefont {Leray}}]{g4_incl}%
  \BibitemOpen
  \bibfield  {author} {\bibinfo {author} {\bibfnamefont {D.}~\bibnamefont
  {Mancusi}}, \bibinfo {author} {\bibfnamefont {A.}~\bibnamefont {Boudard}},
  \bibinfo {author} {\bibfnamefont {J.}~\bibnamefont {Cugnon}}, \bibinfo
  {author} {\bibfnamefont {J.-C.}\ \bibnamefont {David}}, \bibinfo {author}
  {\bibfnamefont {P.}~\bibnamefont {Kaitaniemi}},\ and\ \bibinfo {author}
  {\bibfnamefont {S.}~\bibnamefont {Leray}},\ }\href
  {https://doi.org/10.1103/PhysRevC.90.054602} {\bibfield  {journal} {\bibinfo
  {journal} {Phys. Rev. C}\ }\textbf {\bibinfo {volume} {90}},\ \bibinfo
  {pages} {054602} (\bibinfo {year} {2014})}\BibitemShut {NoStop}%
\bibitem [{\citenamefont {Kelic}\ \emph {et~al.}()\citenamefont {Kelic},
  \citenamefont {Ricciardi},\ and\ \citenamefont {Schmidt}}]{abla07}%
  \BibitemOpen
  \bibfield  {author} {\bibinfo {author} {\bibfnamefont {A.}~\bibnamefont
  {Kelic}}, \bibinfo {author} {\bibfnamefont {M.~V.}\ \bibnamefont
  {Ricciardi}},\ and\ \bibinfo {author} {\bibfnamefont {K.-H.}\ \bibnamefont
  {Schmidt}},\ }\href@noop {} {}\Eprint {https://arxiv.org/abs/0906.4193}
  {arXiv:0906.4193} \BibitemShut {NoStop}%
\bibitem [{\citenamefont {Quesada~Molina}\ \emph {et~al.}(2011)\citenamefont
  {Quesada~Molina}, \citenamefont {Ivanchenko}, \citenamefont {Ivanchenko},
  \citenamefont {Cortés-Giraldo}, \citenamefont {Folger}, \citenamefont
  {Howard},\ and\ \citenamefont {Wright}}]{g4preco}%
  \BibitemOpen
  \bibfield  {author} {\bibinfo {author} {\bibfnamefont {J.}~\bibnamefont
  {Quesada~Molina}}, \bibinfo {author} {\bibfnamefont {V.}~\bibnamefont
  {Ivanchenko}}, \bibinfo {author} {\bibfnamefont {A.}~\bibnamefont
  {Ivanchenko}}, \bibinfo {author} {\bibfnamefont {M.}~\bibnamefont
  {Cortés-Giraldo}}, \bibinfo {author} {\bibfnamefont {G.}~\bibnamefont
  {Folger}}, \bibinfo {author} {\bibfnamefont {A.}~\bibnamefont {Howard}},\
  and\ \bibinfo {author} {\bibfnamefont {D.}~\bibnamefont {Wright}},\ }\href
  {https://doi.org/10.15669/pnst.2.936} {\bibfield  {journal} {\bibinfo
  {journal} {Prog. Nucl. Sci. Technol.}\ }\textbf {\bibinfo {volume} {2}},\
  \bibinfo {pages} {936} (\bibinfo {year} {2011})}\BibitemShut {NoStop}%
\bibitem [{\citenamefont {Grichine}(2009)}]{glauber}%
  \BibitemOpen
  \bibfield  {author} {\bibinfo {author} {\bibfnamefont {V.~M.}\ \bibnamefont
  {Grichine}},\ }\href@noop {} {\bibfield  {journal} {\bibinfo  {journal} {Eur.
  Phys. J. C}\ }\textbf {\bibinfo {volume} {62}},\ \bibinfo {pages} {399}
  (\bibinfo {year} {2009})}\BibitemShut {NoStop}%
\bibitem [{\citenamefont {Weisskopf}(1937)}]{weisskopf}%
  \BibitemOpen
  \bibfield  {author} {\bibinfo {author} {\bibfnamefont {V.}~\bibnamefont
  {Weisskopf}},\ }\href {https://doi.org/10.1103/PhysRev.52.295} {\bibfield
  {journal} {\bibinfo  {journal} {Phys. Rev.}\ }\textbf {\bibinfo {volume}
  {52}},\ \bibinfo {pages} {295} (\bibinfo {year} {1937})}\BibitemShut
  {NoStop}%
\bibitem [{\citenamefont {Dostrovsky}\ \emph {et~al.}(1959)\citenamefont
  {Dostrovsky}, \citenamefont {Fraenkel},\ and\ \citenamefont
  {Friedlander}}]{dostrovsky}%
  \BibitemOpen
  \bibfield  {author} {\bibinfo {author} {\bibfnamefont {I.}~\bibnamefont
  {Dostrovsky}}, \bibinfo {author} {\bibfnamefont {Z.}~\bibnamefont
  {Fraenkel}},\ and\ \bibinfo {author} {\bibfnamefont {G.}~\bibnamefont
  {Friedlander}},\ }\href {https://doi.org/10.1103/PhysRev.116.683} {\bibfield
  {journal} {\bibinfo  {journal} {Phys. Rev.}\ }\textbf {\bibinfo {volume}
  {116}},\ \bibinfo {pages} {683} (\bibinfo {year} {1959})}\BibitemShut
  {NoStop}%
\bibitem [{\citenamefont {Chatterjee}\ \emph {et~al.}(1981)\citenamefont
  {Chatterjee}, \citenamefont {Murthy},\ and\ \citenamefont
  {Gupta}}]{chatterjee}%
  \BibitemOpen
  \bibfield  {author} {\bibinfo {author} {\bibfnamefont {A.}~\bibnamefont
  {Chatterjee}}, \bibinfo {author} {\bibfnamefont {K.~H.~N.}\ \bibnamefont
  {Murthy}},\ and\ \bibinfo {author} {\bibfnamefont {S.~K.}\ \bibnamefont
  {Gupta}},\ }\href {https://doi.org/10.1007/BF02848235} {\bibfield  {journal}
  {\bibinfo  {journal} {Pramana}\ }\textbf {\bibinfo {volume} {16}},\ \bibinfo
  {pages} {391} (\bibinfo {year} {1981})}\BibitemShut {NoStop}%
\bibitem [{\citenamefont {Kalbach}(2002)}]{kalbach}%
  \BibitemOpen
  \bibfield  {author} {\bibinfo {author} {\bibfnamefont {C.}~\bibnamefont
  {Kalbach}},\ }\href@noop {} {\emph {\bibinfo {title} {{PRECO-2000}, Exciton
  Model Preequilibrium Code System with Direct Reactions}}},\ \bibinfo {type}
  {Tech. Rep.}\ (\bibinfo  {institution} {NEA},\ \bibinfo {year}
  {2002})\BibitemShut {NoStop}%
\bibitem [{\citenamefont {Boudard}\ and\ \citenamefont {Cugnon}(2008)}]{incl4}%
  \BibitemOpen
  \bibfield  {author} {\bibinfo {author} {\bibfnamefont {A.}~\bibnamefont
  {Boudard}}\ and\ \bibinfo {author} {\bibfnamefont {J.}~\bibnamefont
  {Cugnon}},\ }\href
  {http://www-nds.iaea.org/reports-new/indc-reports/indc-nds/indc-nds-0530.pdf}
  {\emph {\bibinfo {title} {{INCL4} - The {Liège INC} Model for High-Energy
  Hadron-Nucleus Reactions: A Short Description of the {INCL}4.2 and {INCL}4.4
  Versions}}},\ \bibinfo {type} {Tech. Rep.}\ (\bibinfo  {institution} {IAEA,
  INDC(NDS)-0530},\ \bibinfo {year} {2008})\BibitemShut {NoStop}%
\bibitem [{\citenamefont {Boudard}\ \emph {et~al.}(2013)\citenamefont
  {Boudard}, \citenamefont {Cugnon}, \citenamefont {David}, \citenamefont
  {Leray},\ and\ \citenamefont {Mancusi}}]{incl_pb}%
  \BibitemOpen
  \bibfield  {author} {\bibinfo {author} {\bibfnamefont {A.}~\bibnamefont
  {Boudard}}, \bibinfo {author} {\bibfnamefont {J.}~\bibnamefont {Cugnon}},
  \bibinfo {author} {\bibfnamefont {J.-C.}\ \bibnamefont {David}}, \bibinfo
  {author} {\bibfnamefont {S.}~\bibnamefont {Leray}},\ and\ \bibinfo {author}
  {\bibfnamefont {D.}~\bibnamefont {Mancusi}},\ }\href
  {https://doi.org/10.1103/PhysRevC.87.014606} {\bibfield  {journal} {\bibinfo
  {journal} {Phys. Rev. C}\ }\textbf {\bibinfo {volume} {87}},\ \bibinfo
  {pages} {014606} (\bibinfo {year} {2013})}\BibitemShut {NoStop}%
\bibitem [{\citenamefont {Ershova}\ \emph {et~al.}(2022)\citenamefont {Ershova}
  \emph {et~al.}}]{ershova}%
  \BibitemOpen
  \bibfield  {author} {\bibinfo {author} {\bibfnamefont {A.}~\bibnamefont
  {Ershova}} \emph {et~al.},\ }\href
  {https://doi.org/10.1103/PhysRevD.106.032009} {\bibfield  {journal} {\bibinfo
   {journal} {Phys. Rev. D}\ }\textbf {\bibinfo {volume} {106}},\ \bibinfo
  {pages} {032009} (\bibinfo {year} {2022})}\BibitemShut {NoStop}%
\bibitem [{\citenamefont {Mancusi}\ \emph {et~al.}(2017)\citenamefont {Mancusi}
  \emph {et~al.}}]{mancusi}%
  \BibitemOpen
  \bibfield  {author} {\bibinfo {author} {\bibfnamefont {D.}~\bibnamefont
  {Mancusi}} \emph {et~al.},\ }\href
  {https://doi.org/10.1140/epja/i2017-12263-0} {\bibfield  {journal} {\bibinfo
  {journal} {Eur. Phys. J. A}\ }\textbf {\bibinfo {volume} {53}},\ \bibinfo
  {pages} {80} (\bibinfo {year} {2017})}\BibitemShut {NoStop}%
\bibitem [{\citenamefont {Aoust}\ and\ \citenamefont
  {Cugnon}(2006)}]{incl_pion}%
  \BibitemOpen
  \bibfield  {author} {\bibinfo {author} {\bibfnamefont {T.}~\bibnamefont
  {Aoust}}\ and\ \bibinfo {author} {\bibfnamefont {J.}~\bibnamefont {Cugnon}},\
  }\href {https://doi.org/10.1103/PhysRevC.74.064607} {\bibfield  {journal}
  {\bibinfo  {journal} {Phys. Rev. C}\ }\textbf {\bibinfo {volume} {74}},\
  \bibinfo {pages} {064607} (\bibinfo {year} {2006})}\BibitemShut {NoStop}%
\bibitem [{\citenamefont {Dytman}\ \emph {et~al.}(2021)\citenamefont {Dytman},
  \citenamefont {Hayato}, \citenamefont {Raboanary}, \citenamefont {Sobczyk},
  \citenamefont {Tena-Vidal},\ and\ \citenamefont
  {Vololoniaina}}]{dytman_validation}%
  \BibitemOpen
  \bibfield  {author} {\bibinfo {author} {\bibfnamefont {S.}~\bibnamefont
  {Dytman}}, \bibinfo {author} {\bibfnamefont {Y.}~\bibnamefont {Hayato}},
  \bibinfo {author} {\bibfnamefont {R.}~\bibnamefont {Raboanary}}, \bibinfo
  {author} {\bibfnamefont {J.~T.}\ \bibnamefont {Sobczyk}}, \bibinfo {author}
  {\bibfnamefont {J.}~\bibnamefont {Tena-Vidal}},\ and\ \bibinfo {author}
  {\bibfnamefont {N.}~\bibnamefont {Vololoniaina}},\ }\href
  {https://doi.org/10.1103/PhysRevD.104.053006} {\bibfield  {journal} {\bibinfo
   {journal} {Phys. Rev. D}\ }\textbf {\bibinfo {volume} {104}},\ \bibinfo
  {pages} {053006} (\bibinfo {year} {2021})}\BibitemShut {NoStop}%
\bibitem [{\citenamefont {Ashida}\ \emph {et~al.}(2024)\citenamefont {Ashida}
  \emph {et~al.}}]{ashida_n16o}%
  \BibitemOpen
  \bibfield  {author} {\bibinfo {author} {\bibfnamefont {Y.}~\bibnamefont
  {Ashida}} \emph {et~al.},\ }\href
  {https://doi.org/10.1103/PhysRevC.109.014620} {\bibfield  {journal} {\bibinfo
   {journal} {Phys. Rev. C}\ }\textbf {\bibinfo {volume} {109}},\ \bibinfo
  {pages} {014620} (\bibinfo {year} {2024})}\BibitemShut {NoStop}%
\bibitem [{\citenamefont {Tano}\ \emph {et~al.}(2024)\citenamefont {Tano} \emph
  {et~al.}}]{tano_n16o}%
  \BibitemOpen
  \bibfield  {author} {\bibinfo {author} {\bibfnamefont {T.}~\bibnamefont
  {Tano}} \emph {et~al.},\ }\href {https://doi.org/10.1093/ptep/ptae159}
  {\bibfield  {journal} {\bibinfo  {journal} {Prog. Theor. Exp. Phys.}\
  }\textbf {\bibinfo {volume} {2024}},\ \bibinfo {pages} {113D01} (\bibinfo
  {year} {2024})}\BibitemShut {NoStop}%
\bibitem [{\citenamefont {Hino}\ \emph {et~al.}()\citenamefont {Hino},
  \citenamefont {Ashida}, \citenamefont {Tano},\ and\ \citenamefont
  {Koshio}}]{e525}%
  \BibitemOpen
  \bibfield  {author} {\bibinfo {author} {\bibfnamefont {Y.}~\bibnamefont
  {Hino}}, \bibinfo {author} {\bibfnamefont {Y.}~\bibnamefont {Ashida}},
  \bibinfo {author} {\bibfnamefont {T.}~\bibnamefont {Tano}},\ and\ \bibinfo
  {author} {\bibfnamefont {Y.}~\bibnamefont {Koshio}},\ }\href
  {https://arxiv.org/abs/2506.14388} {}\Eprint
  {https://arxiv.org/abs/2506.14388} {arXiv:2506.14388} \BibitemShut {NoStop}%
\bibitem [{\citenamefont {{Super-Kamiokande
  collaboration}}(2025)}]{data_release}%
  \BibitemOpen
  \bibfield  {author} {\bibinfo {author} {\bibnamefont {{Super-Kamiokande
  collaboration}}},\ }\href {https://doi.org/10.5281/zenodo.15392411}
  {10.5281/zenodo.15392411} (\bibinfo {year} {2025})\BibitemShut {NoStop}%
\bibitem [{\citenamefont {Smy}()}]{bonsai}%
  \BibitemOpen
  \bibfield  {author} {\bibinfo {author} {\bibfnamefont {M.}~\bibnamefont
  {Smy}},\ }\href
  {https://inspirehep.net/files/c1d5dd7b462ea94e7a87d5c6959d41c9} {}\bibinfo
  {note}
  {\href{https://inspirehep.net/files/c1d5dd7b462ea94e7a87d5c6959d41c9}{\textit{Low
  Energy Event Reconstruction and Selection in Super-Kamiokande-III}}, in
  \textit{Proc. 30th Int. Cosmic Ray Conf.}, ed. R. Caballero, J. C. D’Olivo,
  G. Medina-Tanco, L. Nellen, F. A. Sánchez, J. F. Valdés-Galicia (UNAM,
  Mexico City, Mexico, 2008)}\BibitemShut {NoStop}%
\end{thebibliography}%

\end{document}